\documentclass[onecolumn,trackchanges]{aastex62}

\usepackage{rotating}
\usepackage{float}
\usepackage{graphicx,amsmath}
\usepackage{xcolor}
\usepackage{tikz}
\usetikzlibrary{shapes,arrows}
\usepackage{hyperref}

\received{18 December 2018}
\revised{22 March 2019}
\accepted{26 March 2019}

\submitjournal{ApJ}
\shorttitle{Models of twisted flux tube}
\shortauthors{Sen \& Mangalam}

\begin{document}

\title{Open and closed magnetic configurations of twisted flux tubes}

\correspondingauthor{A. Mangalam}
\email{mangalam@iiap.res.in}

\author[0000-0003-1546-381X]{Samrat Sen}
\affiliation{Indian Institute of Astrophysics, Sarjapur road, Koramangala 2nd Block, Bangalore-560034, India}
\email{samrat@iiap.res.in}

\author[0000-0001-9282-0011]{A. Mangalam}
\affiliation{Indian Institute of Astrophysics, Sarjapur road, Koramangala 2nd Block, Bangalore-560034, India}
\bibliographystyle{aasjournal}

\begin{abstract}
We construct two classes of magnetohydrostatic (MHS) equilibria for an axisymmetric vertical flux tube spanning from the photosphere to the lower part of the transition region within a realistic stratified solar atmosphere subject to solar gravity. We assume a general quadratic expression  of the magnetic flux function for the gas pressure and poloidal current and solve the Grad-Shafranov equation analytically. The solution is a combination of a homogeneous and a particular part where the former is separable by a Coulomb function in $r$ and exponential in $z$, while the particular part is an open configuration that has no $z$ dependence.  We also present another open field solution by using a self-similar formulation with two different profile functions and incorporating stratified solar gravity to maintain the magnetohydrostatic equilibria, which is a modification of earlier self-similar models with a twist. We study the admitted parameter space that is consistent with the conditions in the solar atmosphere and derive magnetic and the thermodynamic structures inside the flux tube that are reasonably consistent with the photospheric magnetic bright points (MBPs) for both open and closed field Coulomb function and self-similar models as estimated from observations and simulations. The obtained open and closed field flux tube solutions can be used as the background conditions for the numerical simulations for the study of the wave propagation through the flux tubes. The solutions can also be used to construct realistic magnetic canopies. 
\end{abstract}

\keywords {magnetohydrodynamics (MHD) -- Sun: activity -- Sun: magnetic fields -- Sun: photosphere -- Sun: transition region}

\section{Introduction}
The small scale magnetic structure in the solar photosphere plays an important role in several phenomena like the evolution of active regions \citep{muller1987, 2000ApJ...541.1059A, 2007ApJ...666L.137C}, heating of corona through the dissipation of waves \citep{1998A&A...337L...9R, 2017NatSR...743147S} and reconnection between flux tubes \citep{1986ApJ...311.1001V, 1994A&A...283..232M}. Magnetic flux tubes span from the  photosphere to the higher atmosphere and are observed in the form of small scale magnetic structures. The topological rearrangement of these flux tubes due to the motion of the photospheric footpoints gives rise to the magnetic reconnection leading to the energy release in the solar corona \citep{1988ApJ...330..474P, 2005ESASP.596E..14P, 2013ApJ...769...59T}. Therefore, the modeling of the proverbial flux tube is one of the key aspects to understand various phenomena on the solar surface and its outer atmosphere.

Several attempts have been made earlier to construct the model of flux tubes for both twisted and untwisted magnetic fields. \cite{Schluter_Temesvary_1958} studied a two dimensional (2D) axisymmetric flux tube model without twist for sunspots  using self--similar structure, where a self-similar parameter was defined as a combination of $r$ and $z$, and the relative vertical magnetic field strength at any arbitrary point w.r.t. the magnetic field strength at the axis is scaled with a Gaussian profile function of the self-similar parameter. This model is valid for open field lines where the magnetic lines of force rise from a horizontal plane and do not return in the model domain. \cite{1971SoPh...16..398Y} implemented a twist in the self--similar structure to model the  sunspots. In this model, an empirical form of the azimuthal magnetic field strength $B_\phi(r,z)$ was taken from the data obtained from observations \citep{1965IAUS...22..267S}. By solving for the  variation of the pitch angle and  gradient of the pitch angle, the thermodynamic quantities with the depth were calculated. Motivated by the model and the self--similar structure proposed by \cite{Schluter_Temesvary_1958}, \cite{1980SoPh...77..63O} assumed a quadratic form of the flux function for the gas pressure to model a closed field flux tube, where the magnetic lines of force rise and return to the same horizontal plane.\\
\cite{1986A&A...170..126S} have numerically studied a 2D model of open single flux tube with a twist using the standard boundary conditions including sheet current to study the magnetic field line structure within and outside the flux tube. The magnetic and thermodynamic structure for both single and multiple flux tubes which span from  photosphere to corona have been studied for the case of untwisted magnetic field \citep{2013MNRAS.435..689G, 2014ApJ...789...42G},  where an empirical form of the magnetic field components is motivated by a  self-similar construction. A numerical model of flux tubes has been studied by \citet{2015A&A...577A.126M}, where an empirical form of magnetic flux function has been assumed; this was followed by a model to study the propagation of the MHD waves through the flux tubes with an azimuthal velocity perturbation. The steady structure of the 2D flux tube was used as a background initial condition to study the propagation of the MHD waves. For example, \cite{2009A&A...508..951V} assumed an empirical form of gas pressure for investigating the wave propagation and energy transport through the flux tube. Other interesting results of wave behavior in the solar atmosphere have been presented by several authors. \cite{2009SoPh..258..219F} have studied the propagation of the acoustic wave through the solar atmosphere due to the periodic drivers at the photosphere, and \cite{2010A&A...515A.107S} have modeled the wave propagation through the photospheric magnetic bright points (MBPs).\\
In this work, we have constructed two different models of flux tubes with twisted magnetic field for open and closed field lines by solving Grad--Shafranov equation (GSE) \citep{1958conf...21...190, 1958jetp...710...33}. Here, we have assumed a quadratic form of the flux function for the gas pressure and poloidal current which has been used to study the equilibrium solution of terrestrial plasma \citep{doi:10.1063/1.1756167}, and we extend it to solar flux tubes. As the MHD waves follow the magnetic field lines, it is important to model flux tubes with open field lines, so that MHD waves propagate through the flux tube and dissipate in the upper atmosphere, which is a key aspect of the coronal heating. A key aspect of this paper is to show that the closed field model, reported in \cite{2018AdSpR..61..617S} (SM18 hereafter), is a special case of the open field model with a twisted field line. The flux tube we build is axisymmetric in structure and spans vertically upward from photosphere to the transition region. The case of a linear form of the flux function for the gas pressure and poloidal current, an equilibrium solution near the magnetic axis of a plasma torus has been reported by \cite{1968JETP...26..2S}. SM18 have studied the homogeneous solution of GSE which is a special case of the general solution of the quadratic case to model a flux tube with closed field lines with a twist. Here, we present the full solution of the GSE including both homogeneous and the particular parts to model a twisted open field flux tube. The other model we have built is a self--similar magnetic structure with twist, with a generalized Gaussian (or power law) incorporated into the magnetic shape functions; the gas pressure and poloidal current  are taken to be quadratic functions of the flux function. The self--similar flux tube model expands with height which spans from the photosphere to transition region. After building the solutions semi--analytically and applying appropriate boundary conditions (BCs), we calculate the magnetic field structure and thermodynamic quantities inside the flux tube. As magnetic bright points (MBPs) observed in the photosphere \citep{muller1987, 2007ApJ...666L.137C, 2010ApJ...723L.164L, 2010A&A...515A.107S} are likely to be flux tubes, we compare our model with the existing observations and simulations of MBPs.

The paper is organized as follows. In \S \ref{GSE}, we apply the GSE to the cylindrical flux tube case and describe the common BCs which are physically realistic and used in modeling of our flux tubes. In \S \ref{coulomb_solution}, we present the Coulomb function model for open and closed fields, the appropriate BCs, and show how the open field Coulomb model generalizes the Coulomb field closed model. The solution of the self-similar model and the appropriate BCs are presented in \S \ref{ss_sol}. In \S \ref{result}, the results of our simulations and the variation of the magnetic and thermodynamic profile functions are presented for Coulomb function and self-similar models, and in \S \ref{compare}, the results obtained from the models are applied to the existing observations of MBPs and the simulations for other solar flux tubes. In \S \ref{discussions}, we have compared between the Coulomb function and self-similar models and find the regime of the validity; we have also discussed the advancement made and how the models for open and closed field flux tubes are useful for building realistic structures. Finally, in \S \ref{summary}, we summarize and highlight the major points of the work and conclude how the work may be useful for future numerical studies. A glossary of all the symbols used throughout the paper is provided in Table \ref{glos_tab}.

\begin{longrotatetable}
\begin{table}[h!]
\large
\begin{center}
\scalebox{0.65}
{
\begin{tabular}{|c l | c l|}
\hline \hline
{\bf Symbols common to all the models } & & & \\
\hline
$r$ & Radial coordinate in cylindrical geometry & $\phi$ & Azimuthal coordinate in cylindrical geometry\\
$z$ & Vertical coordinate in cylindrical geometry & $B_r$ & Radial component of the magnetic field strength \\
$B_\phi$ & Azimuthal component of the magnetic field strength & $B_z$ & Vertical component of the magnetic field strength\\
$B_0$ & Magnetic field strength at the center of the flux tube & $I_p$ & Poloidal current\\
$\Psi$ & Total flux function & $\Psi_b$ & Flux at the boundary \\
$k$ & Scale factor of the pressure scale height & $p$ & Gas pressure at arbitrary point\\
$p_T$ & Total pressure inside flux tube & $p_0$ & Gas pressure at the photosphere\\
$p_e$ & Pressure out side the flux tube in the solar atmosphere & $p_t$ & Pressure at the transition region \\ 
$z_t$ & Height of the transition region from the photosphere & $\rho$ & Gas density inside the flux tube\\
$T$ & Temperature inside the flux tube & $g$ & Acceleration due to gravity at the photosphere\\
$R_g$ & Universal gas constant & $\mu_e$ & Effective molar mass \\
$\bar{\mu}$ & Mean effective molar mass &  &  \\
 
\hline \hline

{\bf Coulomb function model } & & & \\
\hline
$a,\alpha,b,\beta,\kappa$ &  Parameters of the GSE & $\psi_h$ & Homogeneous solution of GSE, same as $\psi_C^C$\\
$s$ & Radial part of the homogeneous solution & $Z$ & Vertical part of homogeneous solution\\
$\psi_p$ & Particular solution of GSE & $\psi^X_C$ & General Coulomb solution of the GSE\\
$\psi^O_C$ & Coulomb function open field solution & $\psi^C_C$ & Coulomb function closed field solution \\
$j_\phi$ & Sheet current along azimuthal direction at flux tube boundary & $j_z$ & Sheet current along vertical direction at flux tube boundary \\ 
$p_{20}$ & Gas pressure at the center  of the flux tube & $R$ & Radius of the flux tube  \\
$r_b$ & Cut-off radius inside the flux tube  &  & \\
 \hline \hline
 
{\bf Self-similar model} & & & \\ \hline
$\xi$& Self-similar parameter  & $D_X$ & Shape function of the flux function \\
$B'_{z0}$ & Vertical gradient of magnetic field at the center of the flux tube & $\bar{\chi}$ & Dimensionless twist parameter of the field lines \\
$\bar{f}$ & Dimensionless shape function parameter & $p_c$ & Gas pressure at the center of the flux tube \\ 
$\psi^O_S$ & Self-similar open field flux & $R_{90}$ &  Radius of the flux tube where $90$\% of the flux is enclosed \\ 
\hline \hline  
{\bf Generalized Gaussian flux tube model } & && \\
\hline
 $\psi_G$ & Flux function  & $D_G$ & Shape function of the flux function \\
$n_G$ & Power index of shape function profile  & $R_G$ & Radius of the flux tube \\
\hline \hline
{\bf Power law flux tube model } & && \\
\hline
$\psi_P$ & Flux function  & $D_P$ & Shape function of the flux function \\
$n_P$ & Power index of shape function profile  & $R_P$ & Radius of the flux tube \\
\hline
\end{tabular}
}
\end{center}
\caption{Glossary of symbols used in the different flux tube models.}
\label{glos_tab}
\end{table}
\end{longrotatetable}

\section{Grad--Shafranov equation for the cylindrical flux tube }\label{GSE}

In a magnetic medium of field strength ${\bf B}$, with gas (or plasma) pressure $p$ and mass density $\rho$, the magnetohydrostatic (MHS) pressure balance equation is given by
\begin{equation}\label{mhs}
-\nabla p+\frac{1}{4\pi}(\nabla \times {\bf B})\times {\bf B}+\rho {\bf g}=0, 
\end{equation}      
where ${\bf g}$ denotes the acceleration due to gravity at the solar surface. The individual components of ${\bf B}$ can be expressed in terms of the poloidal flux function, $\Psi(r,z)=\int_0^r r' B_z(r',z) {\rm d}r'$, in the following way
\begin{align}\label{mc}
B_r=-\frac{1}{r}\frac{\partial \Psi}{\partial z} ;\quad B_z=\frac{1}{r}\frac{\partial \Psi}{\partial r} ; \quad B_\phi=\frac{I_p}{r},
\end{align}
where $I_p$ represents the poloidal current. These forms of $B_r,B_\phi$ and $B_z$ automatically ensure the solenoidal condition for ${\bf B}$. Using the axisymmetric condition we split the MHS equilibrium eqn (\ref{mhs}) into $r$ and $z$ direction and plug in the forms of magnetic field components from eqn (\ref{mc}), to find two scalar partial differential equations
\begin{subequations}
\begin{align}\label{gsec}
\frac{\partial \Psi}{\partial r}\frac{\partial^2\Psi}{\partial z^2}+\frac{\partial \Psi}{\partial r} \frac{\partial^2\Psi}{\partial r^2}-\frac{1}{r}\bigg(\frac{\partial \Psi}{\partial r}\bigg)^2+\frac{1}{2}\frac{\partial I_p^2}{\partial r}=-4\pi r^2\frac{\partial p}{\partial r} \\ \label{gsec2}
-\frac{\partial p}{\partial z}+\frac{1}{4\pi}\bigg[\frac{1}{r}\frac{\partial \Psi}{\partial z}\bigg(\frac{1}{r^2}\frac{\partial \Psi}{\partial r}-\frac{1}{r}\frac{\partial^2\Psi}{\partial r^2}\bigg)-\frac{1}{r^2}\frac{\partial \Psi}{\partial z}\frac{\partial^2\Psi}{\partial z^2}-\frac{1}{2r^2}\frac{\partial I_p^2}{\partial z}\bigg]-\rho g=0,
\end{align}
\end{subequations}  
where we assume the form of the gas pressure to be
\begin{align}\label{p}
p(r,z)=p_1(\Psi)+p_2(z);
\end{align}
this form is required in order to have a non-zero density (see SM18). The $\phi$ part of eqn (\ref{mhs}) gives $\displaystyle{\nabla \Psi \times \nabla I_p =0}$, which implies $I_p=I_p(\Psi)$. We have the following form from eqns (\ref{gsec}, \ref{p}) for the GSE to be given by
\begin{align}\label{rc}
\frac{\partial^2\Psi}{\partial r^2}-\frac{1}{r}\frac{\partial \Psi}{\partial r}+\frac{\partial^2\Psi}{\partial z^2}=-\frac{1}{2}\frac{\partial I_p^2(\Psi)}{\partial \Psi}-4\pi r^2\frac{\partial p_1(\Psi)}{\partial \Psi}.
\end{align}
From eqns (\ref{gsec2}, \ref{p}) we find
\begin{align}\label{zc}
-\frac{\partial p_2}{\partial z}-\frac{\partial p_1(\Psi)}{\partial z}+\frac{1}{4\pi}\bigg[\frac{1}{r^2}\frac{\partial \Psi}{\partial z}\bigg(\frac{1}{r}\frac{\partial \Psi}{\partial r}-\frac{\partial^2\Psi}{\partial r^2}\bigg)-\frac{1}{r^2}\frac{\partial \Psi}{\partial z}\frac{\partial^2\Psi}{\partial z^2}-\frac{1}{2r^2}\frac{\partial I_p^2(\Psi)}{\partial z}\bigg]-\rho g=0.
\end{align}
Following SM$18$, by multiplying $\displaystyle{4 \pi r^2\frac{\partial z}{\partial \Psi}}$ on both sides of eqn (\ref{zc}) and using eqn (\ref{rc}), we obtain
\begin{align}\label{density}
\rho(z)=-\frac{1}{g}\frac{{\rm d} p_2(z)}{{\rm d} z}.
\end{align}
We will see later that the prescription of $p_2(z)$ will lead $\rho$ to be a positive quantity, and hence the density within the flux tube is independent of the radial distance $r$ but varies with height $z$. The temperature, $T$, inside the flux tube is calculated by the ideal gas law according to the following form
\begin{align}\label{temperature}
T(r,z)=\frac{\bar{\mu}\ p(r,z)}{R_{g}\ \rho(z)},
\end{align}
where, $R_{g}=8.314$ J mol$^{-1}$ K$^{-1}$ represents the gas constant and 
\begin{align}\label{mu}
\bar{\mu}=\frac{1}{z_t}\int_0^{z_t}\mu_e(z){\rm d}z=1.12
\end{align}
is the mean effective molar mass from photosphere to transition region given by the empirical relation, $\mu_e(z)=1.288 \bigg[ 1-0.535 ( \frac{z}{2.152})^3 \bigg]$ \citep{2015AstL...41..211S} in the domain of $0<z<2.152$ Mm. The formulary of the derived functions for the Coulomb function helical flux tube model are summarized in the Table \ref{func_models}. In \S \ref{coulomb_solution} and \S \ref{ss_sol}, we reduce the GSE for different models of flux tubes having open or closed field line structures. A flowchart of the solutions of the two  different flux tube models obtained is shown in Fig. \ref{flowchart}.\\
Before we solve for the various cylindrical structures, we discuss the boundary conditions below that are crucial to the models, applicable to both open and closed field flux tubes. The magnetic field lines that rise from a horizontal plane and do not return to the same plane within the domain of interest, are called open field lines (see Figs. [\ref{gse_openfield_topology}, \ref{gausstopo_topology}, \ref{powerlawtopo_topology}]). On the other hand, the field lines that rise and return to the same horizontal plane are called the closed field lines (see Fig. \ref{gse_closedfield_topology}). We take an idealized case in which the flux tube is embedded in a magnetic field free region where there is no current outside the flux tube. We apply the following standard BCs which are used by several authors [e.g. \cite{2000JApA...21..299M, 2015AstL...41..211S, 2018AdSpR..61..617S}], that [$B_r(r=0,z)=0$, $B_\phi(r=0,z)=0$] which implies that the magnetic field line is vertical at the axis of the flux tube. At the boundary, the radial component vanishes i.e. $B_r(r=R,z)=0$. We also use the BCs that the total pressure at the boundary of the flux tube matches with the external pressure and the radial average of the internal gas pressure at the transition region ($z=z_t$) is equal to $p_t$, where the pressure at the photosphere ($z=0$) outside the flux tube is taken to be $p_0=1.228\times 10^5$ dyne cm$^{-2}$ and at the transition region ($z_{t}=2$ Mm), it is $p_{t}=0.1488$ dyne cm$^{-2}$; these are taken from Avrett-Loeser model \citep{2008ApJS..175..229A}. We specify the appropriate BCs to model both open and closed field flux tubes below: 
\begin{subequations}
\begin{center}
\begin{align}\label{gse_bc1}
& \text {BC 1:}\quad  B_r(r=0,z) =0 \\ \label{gse_bc2}
& \text {BC 2:}\quad  B_\phi(r=0,z)=0 \\ \label{gse_bc3}
& \text {BC 3:}\quad  B_r(R,z)=0 \\ \label{gse_bc4}
& \text {BC 4:}\quad  p_T(R,z) =p_e(z) \\ \label{gse_bc5}
& \text {BC 5:}\quad  \frac{1}{R}\int_0^R p(r,z_t) {\rm d}r =p_t. 
\end{align}
\end{center}
\end{subequations}
The BCs that distinguishes between the closed and open field flux tubes is the following
\begin{align}\label{Bphi_BC}
B_\phi(R,z)
\begin{cases}
=0; \quad \text {closed field}\\
\neq 0; \quad \text {open field},
\end{cases}
\end{align}
which reduces to the condition,
\begin{align}\label{psi_BC}
\Psi(R,z)=\Psi_b
\begin{cases}
=0; \quad \text {closed field}\\
\neq 0; \quad \text {open field},
\end{cases}
\end{align}
which is derived in \S \ref{coulomb_solution}. The open (general) solution is obtained in \S \ref{coulomb_solution} and it is reduced to the special case of the closed solution by taking $\Psi_b=0$ is presented in \S \ref{coulomb_solution}.

\begin{center}
\begin{sidewaysfigure}[hbt!]
\scalebox{0.8}
{

\tikzstyle{decision} = [diamond, draw,  
    text width=6em, text badly centered, node distance=5cm, inner sep=0pt]
\tikzstyle{block} = [rectangle, draw, 
    text width=6.5em, text centered, rounded corners, minimum height=2em]
\tikzstyle{line} = [draw, -latex']
\tikzstyle{cloud} = [draw, ellipse,node distance=3cm,
    minimum height=2em]

\begin{tikzpicture}[node distance = 2cm, auto]

\tikzstyle{block} = [rectangle, draw, 
    text width=9.2em, text centered,rounded corners, minimum height=1em]
\node [block] (c1) {\S \ref{GSE}\\ BCs:\\ $B_r(0,z)=0 $\\
$B_\phi(0,z)=0$\\
$B_r(R,z)=0$\\
$p_T(R,z)=p_e(z)$\\
$\frac{1}{R}\int_0^Rp(r,z_t){\rm d}r=p_t$};

\tikzstyle{block} = [rectangle, draw, 
    text width=7em, text centered,rounded corners, minimum height=1em]
\node [block, left of=c1, node distance=5cm] (c2) {\S \ref{coulomb_solution}\\ Cylindrical flux tube general solutions to GSE with $B_\phi  \neq 0$};

\tikzstyle{block} = [rectangle, draw, 
    text width=5em, text centered,rounded corners, minimum height=3em]
\node [block, left of=c2, below of=c2, node distance=5.5cm](l1){\S \ref{coulomb_solution}\\ Coulomb function model};

\tikzstyle{block} = [rectangle, draw, 
    text width=11em, text centered,rounded corners, minimum height=1em]
\node [block, right of=l1,node distance=12cm](r1){\S \ref{ss_sol}\\ Self-similar model\\ (Open field solution)\\ $\psi_b \neq 0$\\ $B_\phi(r \rightarrow \infty)=0$};

\tikzstyle{block} = [rectangle, draw, 
    text width=7em, text centered,rounded corners, minimum height=1em]
\node [block, below of=l1,node distance=3cm](l1b1){$\psi=\psi_h+\psi_p$\\$j_{\phi s} \neq 0$};

\tikzstyle{block} = [rectangle, draw, 
    text width=5.5em, text centered,rounded corners, minimum height=1em]
\node [block, left of=l1b1,node distance=4cm](l1b1l1){\S \ref{coulomb_solution}\\ Closed Field solution\\ $\psi_p=0$\\ $\psi_b=0$\\(SM18)};

\tikzstyle{block} = [rectangle, draw, 
    text width=7em, text centered,rounded corners, minimum height=1em]
\node [block, below of=l1b1l1,node distance=3.9cm](l1b2l1){$\psi(\varpi=1)=0$\\ $B_\phi(\varpi=1)=0$\\ $j_{zs}=0$};

\tikzstyle{block} = [rectangle, draw, 
    text width=5em, text centered,rounded corners, minimum height=1em]
\node [block, right of=l1b1,node distance=3.5cm](l1b1r1){\S \ref{coulomb_solution}\\ Open Field solution\\ $\psi_p \neq 0$\\ $\psi_b \neq 0$};

\tikzstyle{block} = [rectangle, draw, 
    text width=7em, text centered,rounded corners, minimum height=1em]
\node [block, below of=l1b1r1,node distance=3.9cm](l1b2r1){$\psi(\varpi=1)=\psi_b$\\ $B_\phi(\varpi=1) \neq 0$\\ $j_{zs} \neq 0$};

\tikzstyle{block} = [rectangle, draw, 
    text width=17em, text centered,rounded corners, minimum height=1em]
\node [block, left of=r1, below of=r1, node distance=3.3cm](l1b1r2){\S \ref{ss_sol}\\ Generalized Gaussian shape function \\$D_G(\xi)=D_{G0} \exp(-\xi^{n_G})$\\$(n_G>0)$\\
\citep{Schluter_Temesvary_1958, 1979SoPh...64..261O} for $n_G=2$};

\tikzstyle{block} = [rectangle, draw, 
    text width=12em, text centered,rounded corners, minimum height=1em]
\node [block, right of=l1b1r2,node distance=6.8cm](l1b1r3){\S \ref{ss_sol}\\ Power law shape function\\$D_P(\xi)=D_{P0} (1+\xi)^{-n_P}$\\$(n_P>2)$\\
(current paper)};

\draw [->,thick,line width=0.5mm,draw=blue!100](c2.south)to[out=-90,in=90]node [text width=1.9cm,above=0.17cm]{$R=$ finite}(l1.north);

\draw [->,thick,line width=0.5mm,draw=blue!100](c2.south)to[out=-90,in=90]node [text width=1.9cm,above=0.17cm]{$R \rightarrow \infty$}(r1.north);
\path [line] (c1)-- (c2);
\path [line] (l1b1)-- (l1b1l1);
\path [line] (l1b1)-- (l1b1r1);
\path [line] (l1b1l1)-- (l1b2l1);
\path [line] (l1b1r1)-- (l1b2r1);
\path [line] (l1)-- (l1b1);
\draw [->,thick,line width=0.5mm,draw=blue!88](r1.south)to[out=-90,in=90](l1b1r2.north);
\draw [->,thick,line width=0.5mm,draw=blue!88](r1.south)to[out=-90,in=90](l1b1r3.north);
\end{tikzpicture}
}
\caption{The two families of solutions presented in various sections are indicated above: the Coulomb function model and the self-similar model are shown along with the applicable boundary conditions.}
\label{flowchart}
\end{sidewaysfigure}
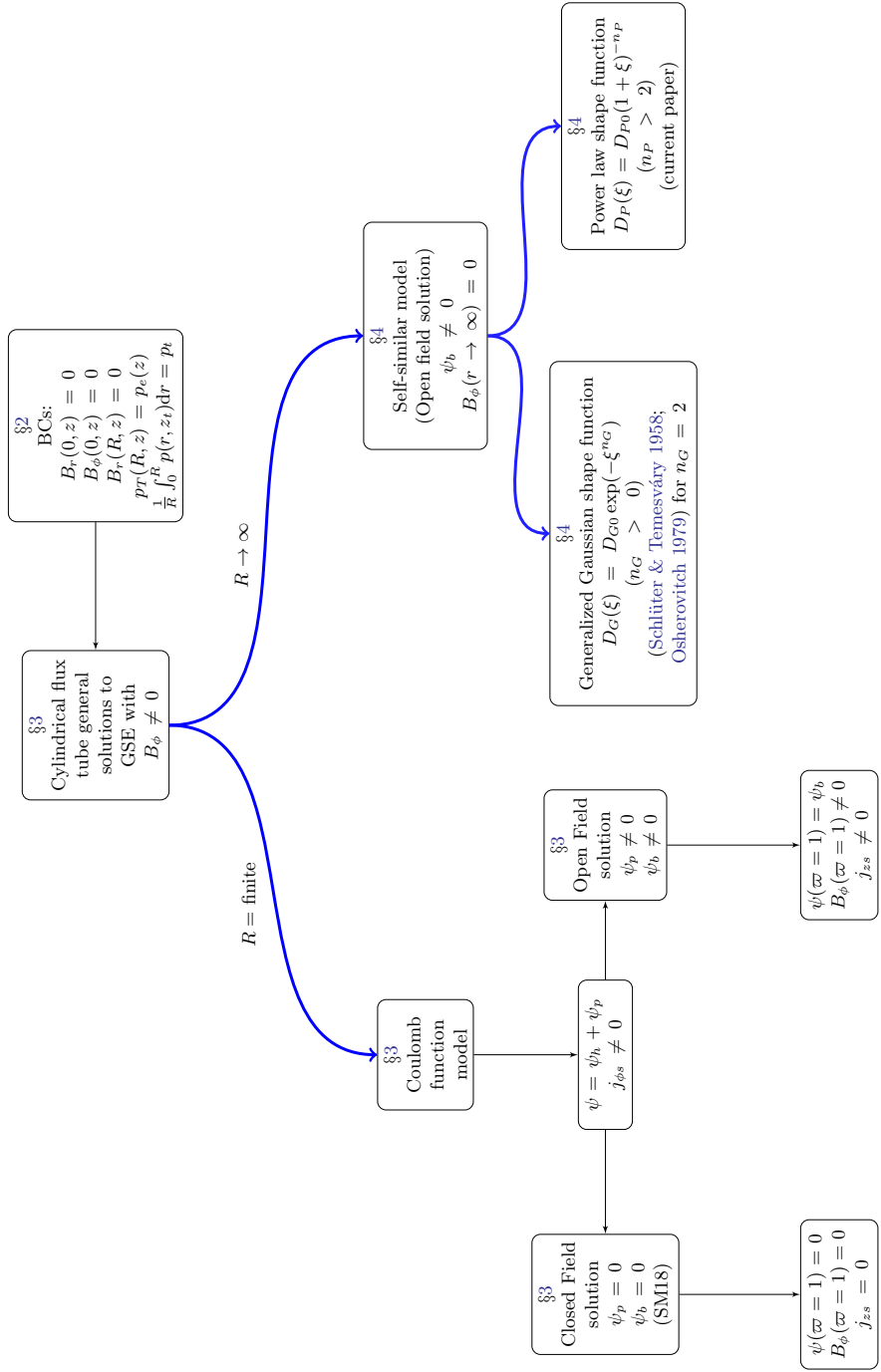
\end{center}

\section{Coulomb function solution of helical flux tube model} \label{coulomb_solution}
\begin{figure}[t]
\begin{center}
\includegraphics[scale=0.5]{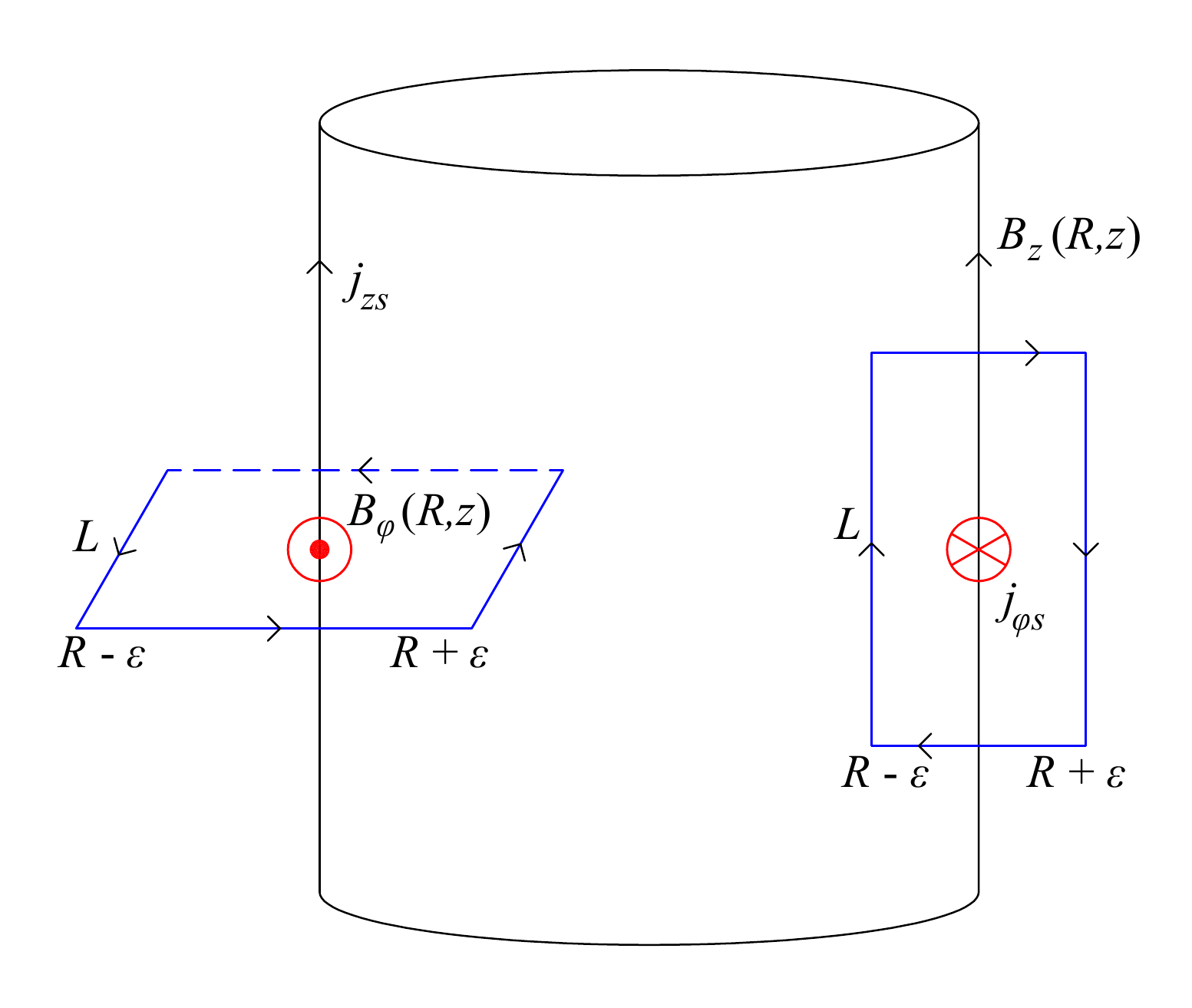}\hspace{1cm}
\caption{Geometry of the flux tube at the boundary showing sheet currents.}
\label{tube}
\end{center}
\end{figure}

For the magnetohydrostatic equilibria with uniform solar gravity and axisymmetric condition, we have split $p(r,z)=p_1(\Psi(r,z))+p_2(z)$ in order to have a non-zero density (SM18). If we take the form of $p_1(\Psi)$ as a linear function of $\Psi$, we found that the BCs 1--5 [eqn (\ref{gse_bc1})--(\ref{gse_bc5})], which are crucial for our model, will not be satisfied for arbitrary $R$ values. On the other hand, the quadratic function of $\Psi$, which is more general than the linear form, is the simplest allowed form for $p_1$ and $I_p$, satisfies all the BCs [eqns (\ref{gse_bc1})--(\ref{gse_bc5})], where $R$ becomes a free parameter, and can be chosen any value within the domain of our interest. Therefore, we assume $p_1(\Psi)$ and $I_p^2(\Psi)$ to be polynomials of $\Psi$ upto second order \citep{doi:10.1063/1.1756167},
\begin{subequations}
\begin{align}\label{p1}
& p_1(\Psi)=\frac{1}{4 \pi}\bigg(\frac{a'}{2} \Psi^2+b' \Psi\bigg), \\ \label{Ip}
& I_p^2(\Psi)=\alpha' \Psi^2+ 2\beta' \Psi+I_0^2,
\end{align}
\end{subequations}  
where the parameters $a', b',\alpha', \beta', I_0$ are to be determined by appropriate boundary conditions (BCs) and the function $p_2(z)$ will be evaluated later. Plugging eqns (\ref{p1}, \ref{Ip}) into eqn (\ref{rc}) we obtain a second order scalar linear partial differential equation
\begin{align}\label{gse}
\frac{\partial^2\Psi}{\partial r^2}-\frac{1}{r}\frac{\partial \Psi}{\partial r}+\frac{\partial^2\Psi}{\partial z^2}= -(a'r^2+\alpha')\Psi-(b'r^2+\beta').
\end{align} 
We define the dimensionless parameters (in the LHS) by introducing the scaling relations,
\begin{align}\nonumber
& \varpi=r/R, \ \ \ \tau=R/z_0, \ \ \ \psi_b= \frac{\Psi_b \sqrt{a'}}{B_0},\ \ \ a=\frac{a' R^4}{4}, \\ 
& \alpha=\frac{\alpha'}{4\sqrt{a'}}, \ \ \ b=\frac{b'}{B_0 \sqrt{a'}}, \ \ \ \beta=\frac{R^2 \beta'}{\Psi_b},
\end{align}
where $\Psi_b$, $R$, $B_0$ are the boundary flux, radius and the magnetic field strength at the center of the flux tube respectively, and ${\bar z}=z/z_0$, where $z_0$ is a constant.
To solve this equation, we split $\psi=\Psi/\Psi_b$ into homogeneous $\psi_h$ and particular part $\psi_p$, i.e. $\psi=\psi_h+\psi_p$. We plug $\psi$ into eqn (\ref{gse}) and separate  out the homogeneous and particular parts to obtain the following dimensionless equations
\begin{align}\label{hom}
& \frac{\partial^2\psi_h}{\partial \varpi^2}-\frac{1}{\varpi}\frac{\partial \psi_h}{\partial \varpi}+\tau^2 \frac{\partial^2\psi_h}{\partial {\bar z}^2}= -4\sqrt{a} (\sqrt{a}\varpi^2+2\alpha)\psi_h,  \\ \label{inhom}
& \frac{\partial^2\psi_p}{\partial \varpi^2}-\frac{1}{\varpi}\frac{\partial \psi_p}{\partial \varpi}= -(4a\varpi^2+8\sqrt{a}\alpha)\psi_p-\bigg(\frac{4ab}{\psi_b}\varpi^2+\beta\bigg).
\end{align} 
The solution of eqn (\ref{hom}) is separable and given by $\displaystyle{\psi_h(\varpi,{\bar z})=s(\varpi) Z({\bar z})}$ which has been shown in SM18 to be given by 
\begin{align}\label{S2}
s(\varpi)= c  F_0(-\alpha-\kappa^2,\sqrt{a}\varpi^2)  \qquad (\text{with $a>0$}),
\end{align}
where $\displaystyle{F_0(-\alpha-\kappa^2,\sqrt{a}\varpi^2)}$ represents the Coulomb function \citep{abramowitz+stegun} and $\displaystyle{\kappa=\frac{k R}{2\sqrt{2}a^{1/4}}},$ where the value of $k$ is evaluated later. The $z$--part solution of eqn (\ref{hom}) is given by  
\begin{align}\label{Z}
\displaystyle{Z({\bar z})=\exp\bigg(-\frac{2\sqrt{2}\kappa a^{1/4} {\bar z}}{\tau}\bigg)}.
\end{align}
The homogeneous solution takes the following form
\begin{align}\label{Ah}
\psi_h(\varpi,\bar{z})=c \exp\bigg(-\frac{2\sqrt{2}\kappa a^{1/4} {\bar z}}{\tau}\bigg)F_0(-\alpha-\kappa^2,\sqrt{a}\varpi^2).
\end{align}
The solution of the inhomogeneous eqn (\ref{inhom}) is given by a power series solution
\begin{align}\label{Ap}
\psi_p(\varpi)=-\frac{\beta}{8\alpha}+\frac{i \sqrt{a}\varpi^2}{2} e^{-i \sqrt{a}\varpi^2}\bigg(\frac{\beta}{8\alpha}-\frac{b}{\psi_b}\bigg)\sum_{n=0}^{\infty} \frac{F_2^1\big(n+2,1;n+2+i\alpha;\frac{1}{2}\big)(i\sqrt{a}\varpi^2)^n}{(n+1+i\alpha)n!}.
\end{align}
A similar but different homogeneous solution which is oscillatory along $z-$direction has been used for the cases of both $D-$shaped and toroidally diverted laboratory plasma \citep{doi:10.1063/1.1756167}. The general solution of the GS eqn (\ref{gse}) is given by $\displaystyle{\psi=\psi_h+\psi_p}$. Since $\psi(\varpi,{\bar z})$ and $\psi^*(\varpi,{\bar z})$, its complex conjugate, are the valid solutions of eqn (\ref{gse}), we construct a real solution of eqn (\ref{gse}) by redefining $\displaystyle{\frac{\psi_h(\varpi,\bar{z})+\psi_h^*(\varpi,{\bar z})}{2} \rightarrow \psi_h(\varpi,{\bar z})}$ and $\displaystyle{\frac{\psi_p(\varpi)+\psi_p^*(\varpi)}{2} \rightarrow \psi_p(\varpi)}$ which leads to \begin{displaymath}\frac{\psi(\varpi,{\bar z})+\psi^*(\varpi,{\bar z})}{2} \rightarrow \psi(\varpi,{\bar z})=\psi^X_C= \begin{cases}\psi_C^C=\psi_h \quad \text{(closed field)}\\ \psi_C^O=\psi_h+\psi_p \quad \text{(open field)}\end{cases}. \end{displaymath} The solution $\psi_h$ alone gives the closed field structure of flux tube (SM$18$), which we denote as $\psi^C_C$; the general solution is a combination of $\psi_h$ and $\psi_p$ and we denote the open field flux tube structure as $\psi^O_C$.\\
The total flux function $\psi^X_C(\varpi,{\bar z})$ is given by
\begin{align}\label{psi_tot}
\psi^X_C(\varpi,{\bar z})=s(\varpi) Z({\bar z})+\psi_p(\varpi),
\end{align} 
where $s(\varpi)$ and $\psi_p(\varpi)$ are given by eqns (\ref{S2}) and (\ref{Ap}) respectively. Now, $\psi^X_C(\varpi,{\bar z})$ has to be zero at the axis (i.e. $\varpi=0$) for all $\bar{z}$, to keep the field finite at the origin. Since $s(0)=0$, which satisfies the BC $1$ [eqn (\ref{gse_bc1})], we obtain from eqn (\ref{psi_tot}), $\psi_p(\varpi=0)=0$. From eqn (\ref{Ap}), we have $\displaystyle{\psi_p(\varpi=0)=-\frac{\beta}{8\alpha}}$. Therefore, we obtain $\beta=0$. From eqn (\ref{Ap}), $\psi_p(\varpi)$ reduces to
\begin{align}\label{Ap2}
\psi_p(\varpi)=\frac{i\sqrt{a}b\varpi^2}{4\psi_b}\bigg[  e^{i \sqrt{a}\varpi^2}\sum_{n=0}^{\infty} \frac{F_2^1\big(n+2,1;n+2-i\alpha;\frac{1}{2}\big)(-i\sqrt{a}\varpi^2)^n}{(n+1-i\alpha)n!} \\ \nonumber -e^{-i \sqrt{a}\varpi^2}\sum_{n=0}^{\infty} \frac{F_2^1\big(n+2,1;n+2+i\alpha;\frac{1}{2}\big)(i\sqrt{a}\varpi^2)^n}{(n+1+i\alpha)n!} \bigg].
\end{align}
From eqns (\ref{mc}) and (\ref{psi_tot}) we have 
\begin{align}\label{Bz2}
B_z(\varpi,{\bar z})=\frac{B_0\psi_b}{2\sqrt{a}\varpi}\bigg(\frac{\partial \psi_h}{\partial \varpi}+\frac{\partial \psi_p}{\partial \varpi}\bigg),
\end{align}
and we obtain the explicit form for
\begin{align} \label{coulomb_Bz_explicit}
B_z(\varpi,\bar{z})=&\frac{B_0}{8\varpi}\bigg[\bigg\lbrace1-\frac{ib}{2}\bigg(\frac{F_2^1(1,-i\alpha,2-i\alpha,-1)}{1-i\alpha}-\frac{F_2^1(1,i\alpha,2+i\alpha,-1)}{1+i\alpha}\bigg)\bigg \rbrace \\ \nonumber & \cdot \exp\bigg(-\frac{2\sqrt{2}\kappa a^{1/4} \bar{z}}{\tau}\bigg) \frac{{\rm d}}{{\rm d}\varpi}[F_0(-\alpha-\kappa^2,\sqrt{a}\varpi^2)+F^*_0(-\alpha-\kappa^2,\sqrt{a}\varpi^2)]\\ \nonumber &+ib \frac{{\rm d}}{{\rm d}\varpi}\bigg \lbrace \varpi^2 e^{i \sqrt{a}\varpi^2}\sum_{n=0}^\infty\frac{F_2^1(n+2,1,n+2-i\alpha,1/2)(-i\sqrt{a}\varpi^2)^n}{(n+1-i\alpha)n!}\\ \nonumber  &-\varpi^2 e^{-i \sqrt{a}\varpi^2}\sum_{n=0}^\infty\frac{F_2^1(n+2,1,n+2+i\alpha,1/2)(-i\sqrt{a}\varpi^2)^n}{(n+1+i\alpha)n!}\bigg \rbrace \bigg].
\end{align}
The steps to obtain eqn (\ref{coulomb_Bz_explicit}) are given in Appendix \ref{app_Bz}. From eqns (\ref{mc}, \ref{psi_tot}) we have
\begin{align}\label{Br2}
B_r(\varpi,{\bar z})=-\frac{B_0\psi_b \tau}{2\sqrt{a}} s(\varpi) Z'({\bar z}),
\end{align}
whose explicit form is given by
\begin{align}\label{coulomb_Br_explicit}
B_r(\varpi,\bar{z})=& \frac{B_0 \kappa}{2\sqrt{2}a^{1/4}\varpi}\exp\bigg(\frac{-2\sqrt{2}\kappa a^{1/4}\bar{z}}{\tau}\bigg)[F_0(-\alpha-\kappa^2,\sqrt{a}\varpi^2)+F^*_0(-\alpha-\kappa^2,\sqrt{a}\varpi^2)] \\ \nonumber & \cdot \bigg[1-\frac{ib}{2}\bigg(\frac{F_2^1(1,-i\alpha,2-i\alpha,-1)}{1-i \alpha}- \frac{F_2^1(1,i\alpha,2+i\alpha,-1)}{1+i \alpha}\bigg)\bigg].
\end{align}
From eqn (\ref{mc}) we obtain the toroidal component 
\begin{align}\label{B_phi2}
B_\phi(\varpi,\bar{z})=\frac{\sqrt{2}B_0\alpha^{1/2}\psi_b}{a^{1/4}\varpi}(\psi_h+\psi_p),
\end{align}
whose explicit form is given by
\begin{align}\label{coulomb_Bphi_explicit}
& B_\phi(\varpi,\bar{z}) = \frac{B_0 \alpha^{1/2}a^{-1/4}}{4\sqrt{2}}\bigg[\frac{1}{\varpi}\bigg\lbrace 1-\frac{ib}{2}\bigg(\frac{F_2^1(1,-i\alpha,2-i\alpha,-1)}{1-i\alpha}-\frac{F_2^1(1,i\alpha,2+i\alpha,-1)}{1+i\alpha}\bigg)\bigg\rbrace \\ \nonumber & \cdot \exp\bigg(-\frac{-2\sqrt{2}\kappa \bar{z}}{\tau}\bigg) [F_0(-\alpha-\kappa^2,\sqrt{a}\varpi^2)+F^*_0(-\alpha-\kappa^2,\sqrt{a}\varpi^2)] \\ \nonumber & +ib \varpi \bigg \lbrace e^{i \sqrt{a}\varpi^2}\sum_{n=0}^\infty\frac{F_2^1(n+2,1,n+2-i\alpha,1/2)(-i\sqrt{a}\varpi^2)^n}{(n+1-i\alpha)n!}\\ \nonumber  & -e^{-i \sqrt{a}\varpi^2}\sum_{n=0}^\infty\frac{F_2^1(n+2,1,n+2+i\alpha,1/2)(-i\sqrt{a}\varpi^2)^n}{(n+1+i\alpha)n!}\bigg \rbrace \bigg].
\end{align}
Applying the BC $3$ [eqn (\ref{gse_bc3})] and using eqn (\ref{Br2}), we find $\displaystyle{s(\varpi=1)=0}$. From BC $2$ [eqn (\ref{gse_bc2})] and eqn (\ref{Ip}) we find $I_0=0$. 
We assume that the external pressure from photosphere to transition region decreases exponentially as
\begin{align}\label{pex}
p_{e}(z)=p_0 \exp(-2kz),
\end{align}
where $k$ is the pressure scale height, which is determined by the relation, $\displaystyle{k=\frac{1}{2  z_{t}} \ln\bigg(\frac{p_0}{p_{t}}\bigg)} =3.405$ Mm$^{-1}$, where $p_0=1.22 \times 10^5$ dyne cm$^{-2}$, $p_t=0.148$ dyne cm$^{-2}$, and $z_t=2$ Mm. By matching the pressure scale heights inside and outside the flux tube, we see that $p_2(z)$ follows 
\begin{align}\label{p2}
p_2(z)=p_{20}\exp(-2kz),
\end{align}    
where $p_{20}$ is evaluated later. Finally, we have the expression of $p(r,z)$ from eqns (\ref{p}, \ref{p1}) to be given by
\begin{align}\label{prz}
p(r,z)=\frac{1}{4\pi}\bigg(\frac{a'}{2}\Psi^2+b'\Psi\bigg)+p_{20}\exp(-2kz),
\end{align}
whose explicit form is given by
\begin{align}\label{prz_explicit}
p(\varpi,\bar{z})=\displaystyle{  B_0^2\bigg[\bigg(\frac{\psi_b^2 s^2(\varpi)}{8\pi}+\bar{p}_{20}\bigg)Z^2(\bar{z})+\bigg(\frac{\psi_b^2 s(\varpi)\psi_p(\varpi)}{4\pi}+\frac{b\psi_b s(\varpi)}{2\sqrt{2a}}\bigg)Z(\bar{z})+\bigg(\frac{\psi_b^2 \psi_p^2}{8\pi}+\frac{b \psi_b \psi_p}{2\sqrt{2a}}\bigg)\bigg]},    
\end{align}
where $\bar{p}_{20}=p_{20}/B_0^2$, and $s(\varpi),\ Z(\bar{z}),\ \psi_p(\varpi)$ are given by eqns (\ref{S2}, \ref{Z}, \ref{Ap2}) respectively. We now calculate the total pressure at the boundary of the flux tube that includes the  contribution due to gas pressure and the magnetic forces due to the presence of the sheet currents $j_\phi$ (SM18) and $j_z$. The pressure and radial component of the MHS force balance eqn (\ref{mhs}) yields
\begin{align}\nonumber
-\frac{\partial p}{\partial r}\bigg|_{r=R}+\frac{1}{4\pi}\bigg(B_r\frac{\partial B_r}{\partial r}+B_z\frac{\partial B_r}{\partial z}\bigg)\bigg|_{r=R} -\frac{\partial}{\partial r}\bigg(\frac{B^2}{8\pi}\bigg)\bigg|_{r=R}\\\label{fbe} +j_\phi(R)B_z(R)-j_z(R) B_\phi(R)=0.
\end{align}  
The sheet currents $j_\phi$ and $j_z$ take the forms
\begin{subequations}
\begin{align}\label{jphis}
j_\phi(r)=j_{\phi s}\  \delta(r-R),\\ \label{jz}
j_z(r)=j_{z s}\  \delta(r-R).
\end{align}
\end{subequations}
Integrating  eqn (\ref{fbe}) w.r.t. $r$ from $r=R-\epsilon$ to $r=R+\epsilon$ where $\epsilon$ is an infinitesimal positive quantity we obtain
\begin{align}\nonumber
-\int_{R-\epsilon}^{R+\epsilon}\frac{\partial p}{\partial r}dr+\frac{1}{4\pi}\bigg(\int_{R-\epsilon}^{R+\epsilon}B_r\frac{\partial B_r}{\partial r}dr+\int_{R-\epsilon}^{R+\epsilon}B_z\frac{\partial B_r}{\partial z}dr\bigg)\\ \label{fbe2} -\int_{R-\epsilon}^{R+\epsilon}\frac{\partial}{\partial r}\bigg(\frac{B^2}{8\pi}\bigg)dr+ \int_{R-\epsilon}^{R+\epsilon}j_\phi(r)B_z(r){\rm d}r-\int_{R-\epsilon}^{R+\epsilon}j_z(r)B_\phi(r){\rm d}r=0,
\end{align}
which leads to
\begin{align}\nonumber
p_{i}(R,z)-p_{e}(z)+j_{\phi s}\ B_z(R)-j_{z s}\ B_\phi(R)\\ \label{fbe3} +\frac{1}{4\pi}\bigg[ B_r \frac{\partial B_r}{\partial r}+B_z\frac{\partial B_r}{\partial z} \bigg]_R+ \frac{B_i^2(R,z)-B_e^2(R,z)}{8\pi}=0,
\end{align}
where $[...]_R$ denotes the jump condition at the boundary and $\lbrace  B_i, p_i \rbrace$ and  $\lbrace
B_e, p_e \rbrace$ are the internal and external magnetic fields and gas pressures in the flux tube respectively. To calculate $j_{\phi s}$ and $j_{zs}$, we assume an infinitesimal current loop of vertical height $L$ and radial extent $R-\epsilon$ to $R+\epsilon$ placed at the boundary of the flux tube (see Fig. \ref{tube}) and by applying the line integral along the loop, we obtain 
\begin{subequations}
\begin{align}
B_z(R) L=4\pi L \int_{R-\epsilon}^{R+\epsilon}j_{\phi s}\delta(r-R){\rm d}r\\
-B_\phi(R) L=4\pi L \int_{R-\epsilon}^{R+\epsilon}j_{z s}\delta(r-R){\rm d}r,
\end{align}   
\end{subequations}
which implies
\begin{subequations}
\begin{align}\label{jphis2}
& j_{\phi s}=\frac{B_z(R)}{4\pi}\\ \label{jz2}
& j_{z s}=-\frac{B_\phi(R)}{4\pi}.
\end{align}
\end{subequations}
The total internal magnetic field is given by, $\displaystyle{B_i^2=B_r^2+B_\phi^2+B_z^2}$. Applying the BC $3$ [eqn (\ref{gse_bc3})] and $B_e=0$, we have from eqn (\ref{fbe3}),   
\begin{align}\label{fbe4}
p_{i}(R,z)-p_{e}(z)+\frac{3 B_z^2(R,z)}{8\pi}+\frac{3 B_\phi^2(R,z)}{8\pi}=0.
\end{align}
By expanding eqn (\ref{fbe4}) we obtain
\begin{align}\label{fbe5}
&\frac{\psi_b^2}{3}\bigg(1+\frac{6\alpha}{\sqrt{a}}\bigg)+\frac{2}{3}b \psi_b +\frac{\psi_b^2}{4a}\bigg[\frac{1}{\varpi^2}\bigg(s'(\varpi)\exp\bigg(-\frac{2\sqrt{2}\kappa a^{1/4}{\bar z}}{\tau}\bigg)+\psi'_p(\varpi)\bigg)^2\bigg]_{\varpi=1}\\ \nonumber &={\bar p} \exp\bigg(-\frac{4\sqrt{2}\kappa a^{1/4}{\bar z}}{\tau}\bigg),
\end{align}
where  $\displaystyle{\psi_b=\psi_p(\varpi=1)}$ and $\displaystyle{\bar{p}=\frac{8\pi (p_0-p_{20})}{3 B_0^2}}$. By equating the coefficients of $\displaystyle{\exp\bigg(-\frac{2\sqrt{2}\kappa a^{1/4} {\bar z}}{\tau}\bigg)}$, $\displaystyle{\exp\bigg(-\frac{4\sqrt{2}\kappa a^{1/4} {\bar z}}{\tau}\bigg)}$ and the constant quantity between both sides of eqn (\ref{fbe5}), we obtain 
\begin{subequations}
\begin{align}\label{c1}
& \bigg(\frac{1}{\varpi^2}s'(\varpi)\psi_p'(\varpi)\bigg)_{\varpi=1}=0,\\ \label{c2}
& \frac{\psi_b^2}{4a}\bigg(\frac{s'^2(\varpi)}{\varpi^2}\bigg)_{\varpi=1}={\bar p},   \\ \label{c3}
& \frac{\psi_b^2}{3}\bigg(1+\frac{6\alpha}{\sqrt{a}}\bigg)+\frac{2b \psi_b}{3}+\frac{\psi_b^2}{4a}\bigg[\frac{\psi^{'2}_p(\varpi)}{\varpi^2}\bigg]_{\varpi=1}=0.
\end{align}
\end{subequations}
From eqns (\ref{c1}, \ref{c2}) we find that $\bigg(\frac{\psi'_p(\varpi)}{\varpi}\bigg)_{\varpi=1}=0$, as $\bar{p} \neq 0$ and hence from eqn (\ref{c3})
\begin{align}\label{b_condition}
b=-\frac{\psi_b}{2}\bigg(1+\frac{6\alpha}{\sqrt{a}}\bigg).
\end{align}  
We now summarize the set of equations we need to solve numerically for the open field model to be given by
\begin{subequations}
\begin{align}\label{bc1}
&  s(\varpi=1)=0,\\ \label{bc2}
& \bigg(\frac{\psi'_p(\varpi)}{\varpi}\bigg)_{\varpi=1}=0,\\ \label{bc3}
& \bigg(\frac{s'^2(\varpi)}{\varpi^2}\bigg)_{\varpi=1}=\frac{4a{\bar p}}{\psi_b^2},\\ \label{bc5}
& b=-\frac{\psi_b}{2}\bigg(1+\frac{6\alpha}{\sqrt{a}}\bigg),\\ \label{bc4}
& \int_0^1 p(\varpi,{\bar z}_t) {\rm d}\varpi=p_t.
\end{align}
\end{subequations} 
 The explicit forms of the eqns (\ref{bc1}--\ref{bc4}) are given by eqns (\ref{bc1_explicit}--\ref{bc5_explicit}) in Appendix \ref{app_explicit_bcs}. The five eqns (\ref{bc1_explicit}--\ref{bc5_explicit}) consists of seven unknown variables, $\lbrace a, \alpha, b, \psi_b, \bar{p}, R, B_0 \rbrace$; so there is a unique solution to the Coulomb function open field model for a given pair of the unknown variables. The eqns (\ref{bc1_explicit}) and (\ref{bc2_explicit}) contain three variables $a$, $\alpha$ and $R$, and we use these two equations to obtain $a(R)$ and $\alpha(R)$. From eqn (\ref{bc5_explicit}), we calculate $\psi_b(R,B_0,b)$ and then find $b(R,B_0)$ from eqn (\ref{bc4_explicit}), and hence $\psi_b(R,B_0)$ and then evaluate $\bar{p}(R,B_0)$ from eqn (\ref{bc3_explicit}). As a result, the complete solution for the open field flux tube depends only on $R$ and $B_0$ which are the free parameters of the model. As per BC$1$--BC$5$ [eqns (\ref{gse_bc1}--\ref{gse_bc5})], which are used for the open field flux tube model, the magnetic field component at the boundary of the flux tube is given by
\begin{align}\label{Bphi_cfm}
B_\phi(\varpi=1,\bar{z})=\frac{\sqrt{2 \alpha}B_0 \psi_b^2}{a^{1/4}}.
\end{align}
If we demand additionally, that $B_\phi(\varpi=1,\bar{z})=0$, then, from eqn (\ref{Bphi_cfm}), $\psi_b=0$; also eqn (\ref{bc5}) gives $b=0$. Therefore from eqn (\ref{Ap2}), we obtain 
\begin{align}\label{cfs_ap0}
\psi_p(\varpi)=0.
\end{align}
This represents the solution of the homogeneous part $\psi_h$ of GSE, which has been discussed in SM$18$, that is applicable for closed field flux tube model. Therefore we need to solve eqns (\ref{bc1}, \ref{bc3}, \ref{bc4}), which are given by the explicit forms by eqns (\ref{bc1_explicit}, \ref{bc3_explicit}, \ref{bc5_explicit}) [with $b=\psi_b=0$], numerically to find the parameters $a$, $\alpha$ and $\bar{p}$ in terms of $\lbrace R, B_0\rbrace$, which are  the free parameters of the closed field model. The formulary of the derived functions for the Coulomb function helical flux tube model are summarized in the Table \ref{func_models}. We discuss the various configurations of Coulomb function open and closed field structure of flux tubes in \S \ref{CFM_results}.

\section{Self-similar model}\label{ss_sol}

The basic formulation of self-similar model of a flux tube is based on \cite{Schluter_Temesvary_1958} (ST$58$ hereafter). For an axially symmetric cylindrical geometry ($r, \phi, z$), where $\phi$ is ignorable, the magnetic field components are given by eqn (\ref{mc}). The coordinates $r$ and $z$ are combined together into a new dimensionless variable $\xi$ which is called the self-similar parameter and as a consequence, the flux function $\Psi$ can be expressed only as a function of $\xi$ i.e., $\Psi(r,z)=\Psi(\xi)$ (ST$58$). We define the dimensionless parameters (in the LHS) by introducing the scaling relations,
\begin{align} \nonumber
\varpi &=r/R,\ \ \
\tau =R/z_0,\ \ \
\psi =\Psi/\Psi_b,\ \ \
\psi_b =\frac{\Psi_b}{B_0 z_0^2},\\
\bar{p}_1 &=p_1/B_0^2,\ \ \
\bar{I}_p =\frac{I_p}{B_0 R},\ \ \
\bar{\chi} =\chi z_0^2,
\end{align} 
where, $\Psi_b$, $R$, $B_0$ are the boundary flux, radius and the magnetic field strength at the center of the flux tube respectively, and $\bar{z}=z/z_0$, where $z_0$ is a constant length. From ST$58$, the self-similar parameter $\xi$ is defined by
\begin{align}\label{ss_xi}
\xi= \zeta(\bar{z}) \varpi,
\end{align}  
which describes the radial size distribution of the flux tube with height from the base. Plugging in eqn (\ref{ss_xi}), we can rewrite the GS eqn (\ref{rc}) in the following form 
\begin{align}\label{ss_gse}
\frac{\psi_b^2}{2 \tau^2} \frac{{\rm d}}{{\rm d}\xi}\bigg(\frac{{\rm d}\psi}{{\rm d}\xi}\bigg)^2 \zeta'^2(\bar{z})+\frac{\psi_b^2}{\tau^2} \frac{1}{\xi}\bigg(\frac{{\rm d}\psi}{{\rm d}\xi}\bigg)^2 \zeta'(\bar{z})\zeta''(\bar{z})+\frac{\psi_b^2}{2 \tau^4} \frac{{\rm d}}{{\rm d}\xi}\bigg(\frac{1}{\xi}\frac{{\rm d}\psi}{{\rm d}\xi}\bigg)^2 \zeta^4(\bar{z})+\frac{1}{2 \xi^2}\frac{{\rm d}\bar{I}_p^2}{{\rm d}\xi}\zeta^2=-4\pi \frac{\partial \bar{p}_1}{\partial \xi},
\end{align}
and the $z-$part of GS eqn (\ref{gsec2}), gives the expression of $\rho$, eqn (\ref{density}), which is self consistent for both the Coulomb function and self-similar models. We define a quantity which is called the magnetic shape function given by
\begin{align}\label{shape_function}
D_X(\xi)=\frac{1}{\xi}\frac{{\rm d}\psi}{{\rm d}\xi}.
\end{align}
Plugging eqn (\ref{shape_function}) into eqn (\ref{ss_gse}) we obtain 
\begin{align}\label{ss_gse1}
\frac{\psi_b^2}{\tau^2}\xi D_X^2(\xi) \zeta \zeta''+\frac{\psi_b^2}{2\tau^2}\frac{{\rm d}}{{\rm d}\xi}[\xi^2 D_X^2(\xi)]\zeta'^2+\frac{\psi_b^2}{2\tau^4}\frac{{\rm d}}{{\rm d}\xi}\big(D_X^2(\xi)\big)\zeta^4+\frac{1}{2\xi^2}\frac{{\rm d}\bar{I}_p^2}{{\rm d}\xi}\zeta^2=-4\pi \frac{\partial \bar{p}_1}{\partial \xi},
\end{align}
and integrating eqn (\ref{ss_gse1}) w.r.t. $\xi$ from $0$ to $\infty$ we write
\begin{align}\label{ss_gse2}\nonumber
& \frac{\psi_b^2}{\tau^2}\zeta \zeta'' \int_0^{\infty}\xi D_X^2(\xi) {\rm d}\xi+\frac{\psi_b^2 \zeta'^2}{2 \tau^2}\big[ \xi^2 D_X^2(\xi)\big]_{\xi=0}^{\infty}+\frac{\psi_b^2 \zeta^4}{2 \tau^4}[D_X^2(\infty)-D_X^2(0)]\\ & +\frac{\zeta^2}{2}\int_0^{\infty} \frac{1}{\xi^2}\frac{{\rm d}\bar{I}_p^2}{{\rm d}\xi}{\rm d}\xi=-4\pi \int_0^{\infty}\frac{\partial \bar{p}_1}{\partial \xi}{\rm d}\xi.
\end{align} 
Following ST$58$, we define
\begin{align}\label{ss_basic}
y^2(\bar{z})=\frac{\psi_b D_0}{\tau}  \zeta^2(\bar{z}),
\end{align}
where, $\displaystyle{y(\bar{z}) \equiv \bigg(\frac{B_z(0,\bar{z})}{B_0}\bigg)^{1/2}}$, $B_0 \equiv B_z(0,0)$ and $D_0\equiv D_X(\xi=0)$. Next, using eqns (\ref{ss_gse2}) and (\ref{ss_basic}) we obtain 
\begin{align}\label{ss_gse3}\nonumber
& \frac{\psi_b}{\tau}\frac{y y''}{D_0} \int_0^{\infty}\xi D_X^2(\xi) {\rm d}\xi+\frac{\psi_b}{\tau} \frac{y'^2}{2D_0}\big[ \xi^2 D_X^2(\xi)\big]_{\xi=0}^{\infty}+\frac{y^4}{2D_0^2 \tau^2}[D_X^2(\infty)-D_0^2]\\ & + \frac{\tau}{\psi_b} \frac{y^2}{2 D_0}\int_0^{\infty} \frac{1}{\xi^2}\frac{{\rm d}\bar{I}_p^2}{{\rm d}\xi}{\rm d}\xi=-4\pi \int_0^{\infty}\frac{\partial \bar{p}_1}{\partial \xi}{\rm d}\xi.
\end{align}  
To solve the eqn (\ref{ss_gse3}), we need to specify the functional form of $p_1$, $I_p$ and $D_X(\xi)$ to study the flux tube model with twisted magnetic field under the similarity assumption. The functional form of  $p_1=\frac{f}{2}\Psi^2$ is taken from \cite{1980SoPh...77..63O}, where $f$ is the shape function parameter, and the poloidal current, $I_p$ defined by \cite{1971SoPh...16..398Y} and \cite{1979SoPh...64..261O}, and motivated from the observations of \cite{1965IAUS...22..267S}. Hence the form of gas pressure $p$ and poloidal current $I_p$ are taken to be 
\begin{align}\label{ss_p1}
p &=p_c \exp(-2kz)+\frac{f}{2} \Psi^2\\ \label{ss_Ip}
I_p^2&=\Psi^2_b \chi \xi^4 D_X^2(\xi),
\end{align}
for the positivity of $\rho(z)$ at all $z$ which is given by eqn (\ref{density}). Here, $p_2(z)=p_c \exp(-2kz)$ denotes the gas pressure at the flux tube axis, with $p_c$ is the pressure at the center of the flux tube, and $\displaystyle{\chi=\bigg(\frac{B_\phi}{r B_z}\bigg)^2},$ is a constant pitch angle parameter. We deviate from [\cite{1980SoPh...77..63O}; \cite{1971SoPh...16..398Y}] by employing the extra term, $p_c \exp(-2kz)$ with $p_1$ in eqn (\ref{ss_p1}), to maintain the hydrostatic vertical pressure balance condition under the influence of solar gravity, with two options for the shape function $D_X(\xi)$ specified by
\begin{align}\label{DgP}
D_X(\xi)=
  \begin{cases}
  D_G(\xi)=D_{G0} \exp(-\xi^{n_G}); \quad (n_G>0) :\quad \text{Generalized Gaussian}\\ 
 D_P(\xi)=D_{P0} (1+\xi)^{-n_P}; \quad (n_P>2) :\quad \text{Power law}
  \end{cases}
\end{align}
where
\begin{subequations} 
\begin{align} \label{DGO}
D_{G0}&=\frac{n_G}{\Gamma (2/n_G)}, \\
\label{DPO}
D_{P0}&=(n_P-1)(n_P-2),
\end{align}
\end{subequations}
We see that both the shape functions, eqn (\ref{DgP}) vanish asymptotically  at $\xi \rightarrow \infty$; hence from eqn (\ref{ss_gse3}) we obtain
\begin{align}\label{ss_int_eq}
\frac{\psi_b}{\tau}\frac{y y''}{D_0} \int_0^{\infty}\xi D_X^2(\xi) {\rm d}\xi-  \frac{y^4}{2 \tau^2}+\frac{\tau}{\psi_b}\frac{y^2}{D_0}\int_0^{\infty} \frac{1}{2\xi^2}\frac{{\rm d}\bar{I}_p^2}{{\rm d}\xi}{\rm d}\xi=-4\pi \int_0^{\infty}\frac{\partial \bar{p}_1}{\partial \xi}{\rm d}\xi.
\end{align}
Next, we evaluate the integrals of the eqn (\ref{ss_int_eq}) for both generalized Gaussian and power law shape functions. We will see later that, from eqn (\ref{flux_functionGP}), that the flux function $\psi_P(\xi)$ varies as $\xi^{2-n_P}$, in the domain $0< \xi< \infty$; therefore $\psi_P(\xi)$ will converge to a finite value at $\xi \rightarrow \infty$, if $n_p>2$. The results of the integrals are provided in Table  \ref{int_tab}. 
\begin{center}
\begin{table}[h!]\hspace{-20 mm}
  \resizebox{1.12 \textwidth}{!}{  
\begin{tabular}{|c |c | c|}\hline \hline
Functions  & Shape function $D_G(\xi) \quad (n_G>0)$  & Shape function $D_P(\xi) \quad (n_P>2)$  \\  \hline & &\\ 
$\displaystyle{\int_0^\infty \xi D_X^2(\xi){\rm d}\xi}$ & $\displaystyle{\frac{n_G}{\Gamma(2/n_G) 2^{2/n_G}}}$ & $\displaystyle{\frac{(n_P-1)(n_P-2)^2}{2(2n_P-1)}}$\\ \hline & &\\ 
$\displaystyle{\int_0^{\infty} \frac{1}{2\xi^2}\frac{{\rm d}\bar{I}_p^2}{{\rm d}\xi}{\rm d}\xi}$ & $\displaystyle{\frac{\bar{\chi} \psi_b^2 n_G}{2^{2/n_G} \tau^2 \Gamma(2/n_G)}}$ & $\displaystyle{\frac{\bar{\chi} \psi_b^2 (n_P-2)^2(n_P-1)}{2 \tau^2(2n_P-1)}}$\\ \hline & &\\ 
$\displaystyle{\int_0^{\infty}\frac{\partial \bar{p}_1}{\partial \xi}{\rm d}\xi}$ & $\displaystyle{\bar{f}\psi_b^2}/2$ & $\displaystyle{\bar{f}\psi_b^2/2}$\\ \hline  & &\\ 
$\lambda_X(n)$ & $\displaystyle{2^{2/n_G-1}}$ & $\displaystyle{\frac{2n_P-1}{n_P-2}}$ \\  \hline 
\end{tabular}
}
\caption{Expression of the integrals in eqn (\ref{ss_int_eq}) and $\lambda_X$ in eqn (\ref{ss_y_eq}), for generalized Gaussian ($X=G$), and power law ($X=P$) shape function, eqn (\ref{DgP}).}
\label{int_tab} 
\end{table}
\end{center}
Using the values of the integrals from Table \ref{int_tab} and redefining, $y'(\bar{z}=0)\equiv y'_0$, we reduce eqn (\ref{ss_int_eq}) to the following form 
\begin{align}\label{ss_y_eq}
\frac{{\rm d}y}{{\rm d}\bar{z}}=\bigg[\frac{\lambda_X(n)}{2 \psi_b \tau}(y^4-1)-2\bar{\chi}(y^2-1)-8\pi \psi_b \bar{f}\lambda_X(n) \tau \ln(y)+\frac{\bar{B'}^2_{z0}}{4}\bigg]^{1/2},
\end{align}
where, $\displaystyle{\bar{B'}_{z0}=\frac{B'_{z0}z_0}{B_0}}$, $\displaystyle{\bar{\chi}=\chi z_0^2}$ and $\displaystyle{\bar{f}=f z_0^4}$. The form of eqn (\ref{ss_y_eq}) is common for both generalized Gaussian ($X=G$) and power law ($X=P$) shape functions, where the functions $\lambda_X(n)$ for both shape functions are defined in Table \ref{int_tab}. Here we have used the notation $B'_{z0}=B'_z(0,0)$, which represents the vertical gradient of $B_z(0,z)$ at the center. From eqn (\ref{ss_y_eq}), we have the following integral relation
\begin{align}\label{w1_int}
\bar{z}(y)=\int_1^y \frac{{\rm d}y'}{G(y')},
\end{align} 
where the function $G(y)$ is given by 
\begin{align}\label{Gw}
G(y)=\bigg[\frac{\lambda_X(n)}{2 \psi_b \tau}(y^4-1)-2\bar{\chi}(y^2-1)-8\pi \psi_b \bar{f}\lambda_X(n) \tau \ln(y)+\frac{\bar{B'}^2_{z0}}{4}\bigg]^{1/2}.
\end{align}
We evaluate the integral (\ref{w1_int}) numerically which gives $\bar{z}=\bar{z}(y)$. Thereafter, inverting the function between $\bar{z}$ and $y$, we evaluate $y=y(\bar{z})$. From eqns (\ref{ss_xi}, \ref{ss_basic}) we obtain
\begin{align}\label{ss_para}
\xi=\sqrt{\frac{\tau}{\psi_b D_0}}\varpi y(\bar{z}).
\end{align}
Using the similarity assumption, $\displaystyle{B_z(r,z)=\frac{B_0 \psi_b}{\tau} \zeta^2(z) D_X(\xi)}$ (ST$58$), and eqns (\ref{mc}, \ref{ss_xi}, \ref{ss_para}) we calculate the magnetic field components, representing the most general self-similar solution, to be 
\begin{subequations}
\begin{align}\label{ss_Bz}
& B_z(\varpi,\bar{z})=\frac{B_0}{D_0}y^2(\bar{z}) D_X(\xi)\\ \label{ss_Br}
& B_r(\varpi,\bar{z})=-\frac{B_0 \varpi}{D_0}y(\bar{z}) y'(\bar{z})D_X(\xi)   \\ \label{ss_Bphi}
& B_\phi(\varpi,\bar{z})=\frac{\sqrt{\bar{\chi}}B_0}{D_0} \varpi y^2(\bar{z}) D_X(\xi).
\end{align}
\end{subequations}
The flux function for the self-similar model is obtained by integrating the shape function  
\begin{align}\label{flux_function}
\psi^O_S(\xi)=\int_0^\xi \xi' D_X(\xi') {\rm d}\xi'.
\end{align}
Employing eqns (\ref{shape_function}, \ref{DgP}), we obtain the open flux function for generalized Gaussian, $\psi_G$, and power law, $\psi_P$, models to be given by

\begin{align}\label{flux_functionGP}
\psi_S^O(\xi)=
\begin{cases}
\psi_G(\xi)= 1-\frac{\Gamma(2/n_G,\xi^{n_G})}{\Gamma(2/n_G)}  ; \quad (n_G>0): \quad  \text{Generalized Gaussian}\\ 
\psi_P(\xi)=1-(1+\xi)^{1-n_P}\big(1+\xi(n_P-1)\big) ; \quad (n_P>2): \quad \text{Power law}.
\end{cases}
\end{align}
From eqn (\ref{flux_functionGP}), it is seen that, $\psi_G(\xi)$ and $\psi_P(\xi)$ converges to unity for $\xi \rightarrow \infty$. The structure of the self-similar model of a flux tube is that the magnetic field decreases asymptotically in the radial direction to zero at infinity. The flux tube does not have any sharp boundary which can make a partition with the external solar atmosphere. In other words, the self-similar flux tube is embedded in a continuous magnetic medium which has the maximum field strength at the base of the axis of the flux tube and radius of the flux tube is infinity. We take the effective radius of the flux tube  as the distance from the axis on the $z=0$ plane, which makes a circular area where $90 \%$ of the total flux is enclosed. We call this radius as $R_{90}$. The total flux is zero at the axis and it increases asymptotically with $r$. The explicit forms of the magnetic field components, obtained from the eqns (\ref{ss_Bz}, \ref{ss_Br}, \ref{ss_Bphi}) by using eqns (\ref{DgP}, \ref{DGO}, \ref{DPO}, \ref{ss_para}) are:

\begin{align}\label{ss_Bz_explicit} 
B_z(\varpi,\bar{z})=
\begin{cases}
B_0 y^2(\bar{z}) \exp\bigg[ -\bigg( \sqrt{ \frac{\tau \Gamma(2/n_G)}{n_G \psi_b}} y(\bar{z}) \varpi\bigg)^{n_G}  \bigg], \quad (n_G>0): \quad \text{Generalized Gaussian}\\ 
B_0 y^2(\bar{z}) \bigg[1+\sqrt{\frac{\tau}{(n_P-1)(n_P-2)\psi_b}} y(\bar{z}) \varpi\bigg]^{-n_P}, \quad (n_P>2): \quad \text{Power law} 
\end{cases}
\end{align}

\begin{align}\label{ss_Br_explicit}
B_r(\varpi,\bar{z})=
\begin{cases}
-B_0 y(\bar{z})y'(\bar{z})\varpi \exp\bigg[ -\bigg( \sqrt{\frac{\tau \Gamma(2/n_G)}{n_G \psi_b}} y(\bar{z}) \varpi\bigg)^{n_G}  \bigg], \quad (n_G>0): \quad \text{Generalized Gaussian}\\
-B_0 y(\bar{z})y'(\bar{z})\varpi \bigg[1+\sqrt{\frac{\tau}{(n_P-1)(n_P-2)\psi_b}} y(\bar{z}) \varpi \bigg]^{-n_P}, \quad (n_P>2): \quad \text{Power law}
\end{cases}
\end{align}

\begin{align}\label{ss_Bphi_explicit}
B_\phi(\varpi,\bar{z})=
\begin{cases}
B_0 \sqrt{\bar{\chi}}  y^2(\bar{z})\varpi \exp\bigg[ -\bigg( \sqrt{\frac{\tau \Gamma(2/n_G)}{n_G \psi_b}} y(\bar{z}) \varpi\bigg)^{n_G}  \bigg], \quad (n_G>0): \quad \text{Generalized Gaussian}\\
B_0 \sqrt{\bar{\chi}} y^2(\bar{z}) \varpi \bigg[1+\sqrt{\frac{\tau}{(n_P-1)(n_P-2)\psi_b}} y(\bar{z}) \varpi\bigg]^{-n_P}, \quad (n_P>2): \quad \text{Power law}. 
\end{cases}
\end{align}
The magnetic field components $B_r(\varpi,\bar{z})$ and $B_\phi(\varpi,\bar{z})$ for the self-similar model follow the BCs ($1, 2, 3$) [eqns (\ref{gse_bc1}, \ref{gse_bc2}, \ref{gse_bc3})] for $R = \infty$. $B_z(\varpi,\bar{z})$ decreases monotonically with $\varpi$ and converges to zero at infinity. The total pressure far from the flux tube axis is only due to the gas pressure $p$. We use BC $4$ [eqn (\ref{gse_bc4})] at $z=0$, for $r \rightarrow \infty$, so that
\begin{align}\label{ssm_bc3}
p(r \rightarrow \infty,0)=p_0.
\end{align}
From eqn (\ref{flux_functionGP}), we see that the flux function for both generalized Gaussian and power law, converges to unity at $\varpi \rightarrow \infty$, i.e.
\begin{align}\label{psi_infinity}
\psi_S^O(\varpi \rightarrow\infty)= 1.
\end{align}
Using eqns (\ref{ss_p1}, \ref{ssm_bc3}), we obtain 
\begin{align}\label{ss_f}
\bar{f}=2(\bar{p}_0-\bar{p}_c),
\end{align}
and the explicit form of $p(\varpi, \bar{z})$ for both the generalized Gaussian and the power law models is given by 
\begin{align}\label{ss_explicit_prz}
p(\varpi,\bar{z})=\displaystyle{B_0^2\bigg(\frac{\bar{f}}{2} \psi^2+\bar{p}_c e^{-2\bar{k}\bar{z}}\bigg)},   
\end{align}
where $\bar{p}_0=p_0/B_0^2$ and $\bar{p}_c=p_c/B_0^2$. The formulary of the derived functions for the self-similar flux tube model are summarized in the Table \ref{func_models}. The flowchart of the solutions to the Coulomb function and self-similar models are shown in the Fig. \ref{flowchart}.

\section{Results obtained from the models}\label{result}
\subsection{Coulomb function helical flux tube model} \label{CFM_results}
\begin{figure}\hspace{10 mm}
\includegraphics[scale=0.6]{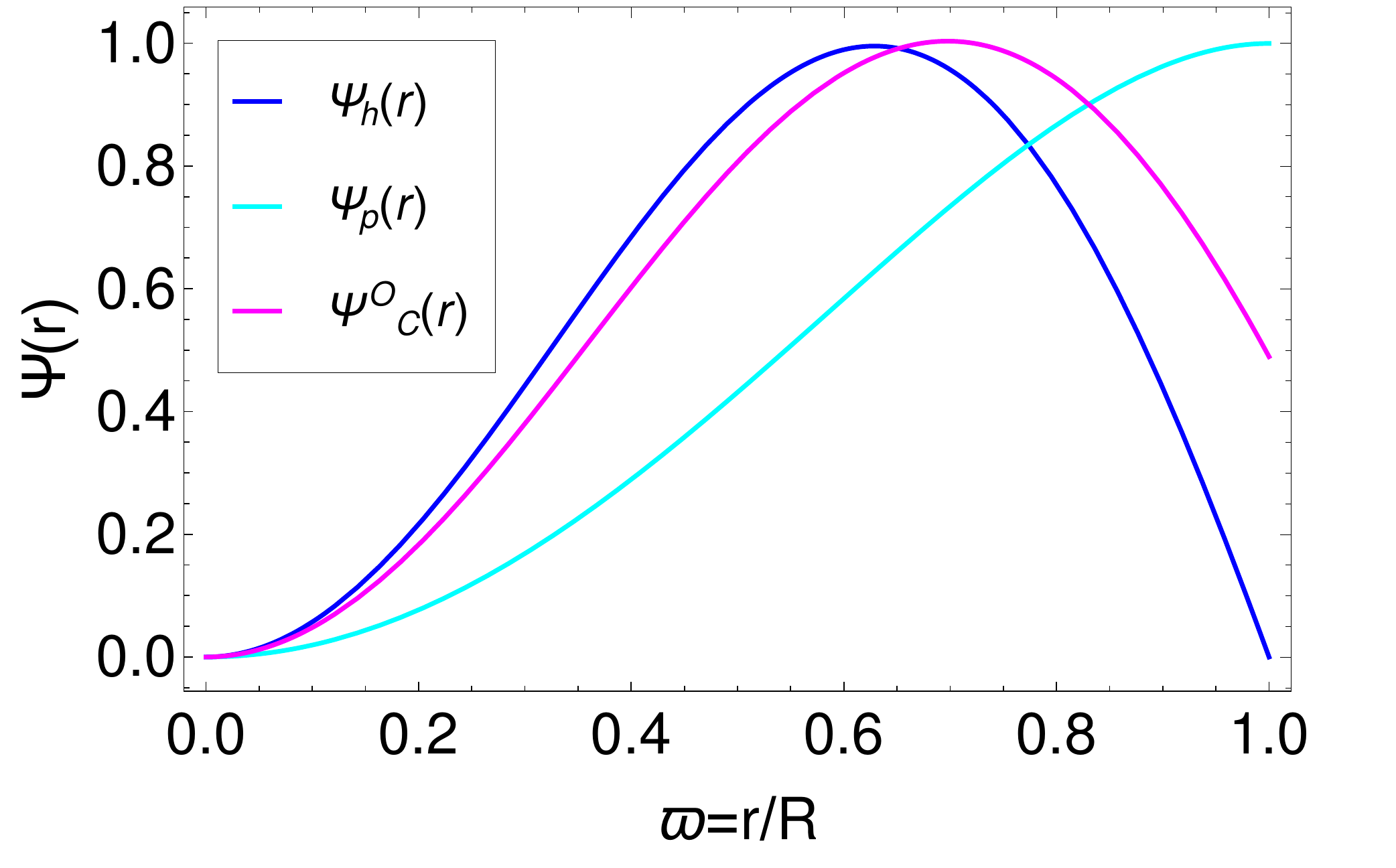}
\caption{The radial variation of the flux function, normalized with respect to the maximum value, obtained from Coulomb function open field model for run $C4$ in Table \ref{gse_parameter_tab}. The horizontal axis is scaled w.r.t. the total radius $R$.} 
\label{gse_Ar_plot}
\end{figure}

\begin{figure}\hspace{10 mm}
\includegraphics[scale=0.6]{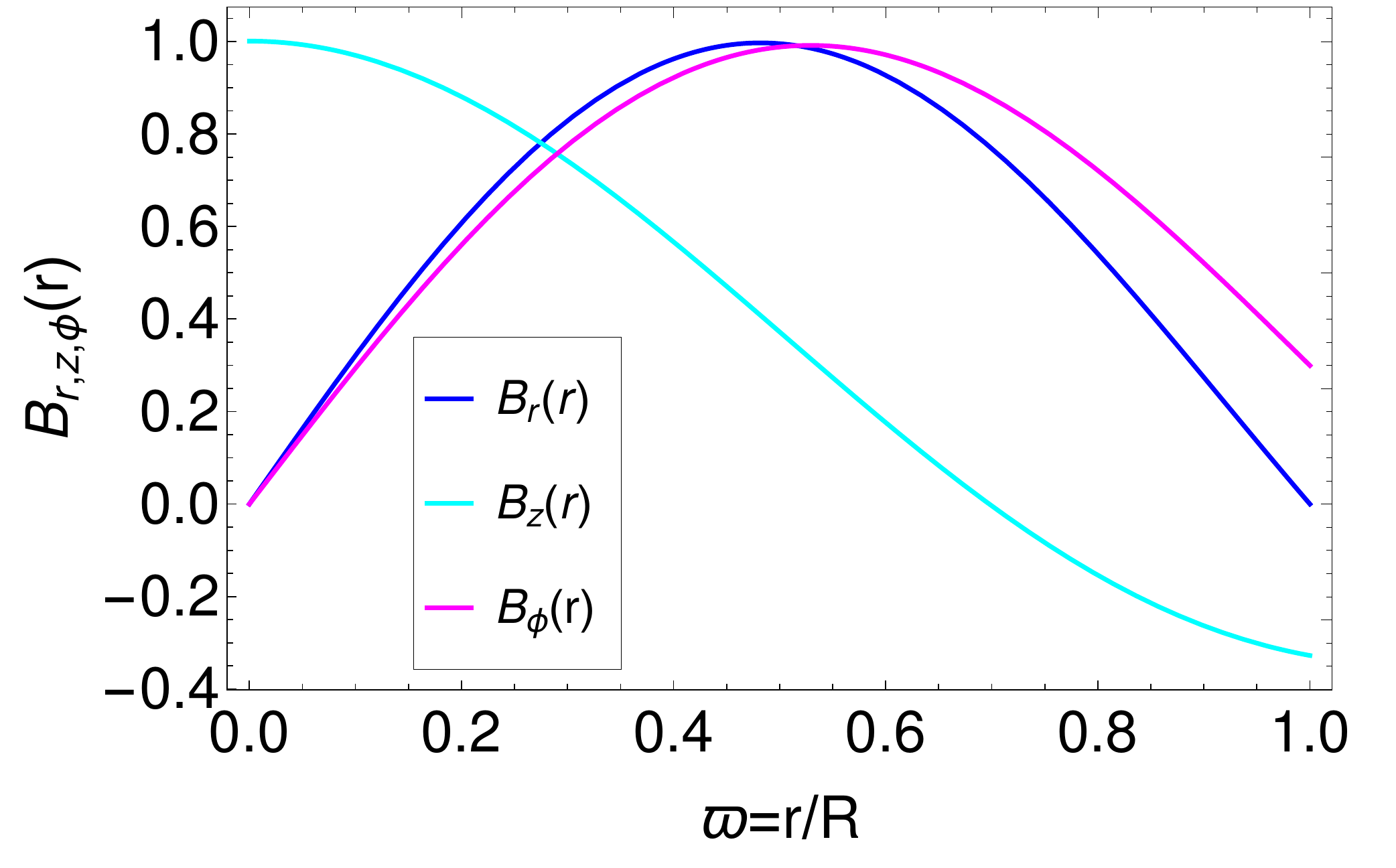}
\caption{The radial variation of $B_r$, $B_\phi$ and $B_z$, normalized with respect to the maximum values of $|B_r|$, $|B_\phi|$ and $|B_z|$ respectively, obtained from the Coulomb function open field  model, for run $C4$ in Table \ref{gse_parameter_tab}. The horizontal axis is scaled w.r.t. the total radius $R$.}  
\label{gse_B_plot}
\end{figure}

\begin{figure}[h!]\hspace{15 mm}
\includegraphics[scale=0.6]{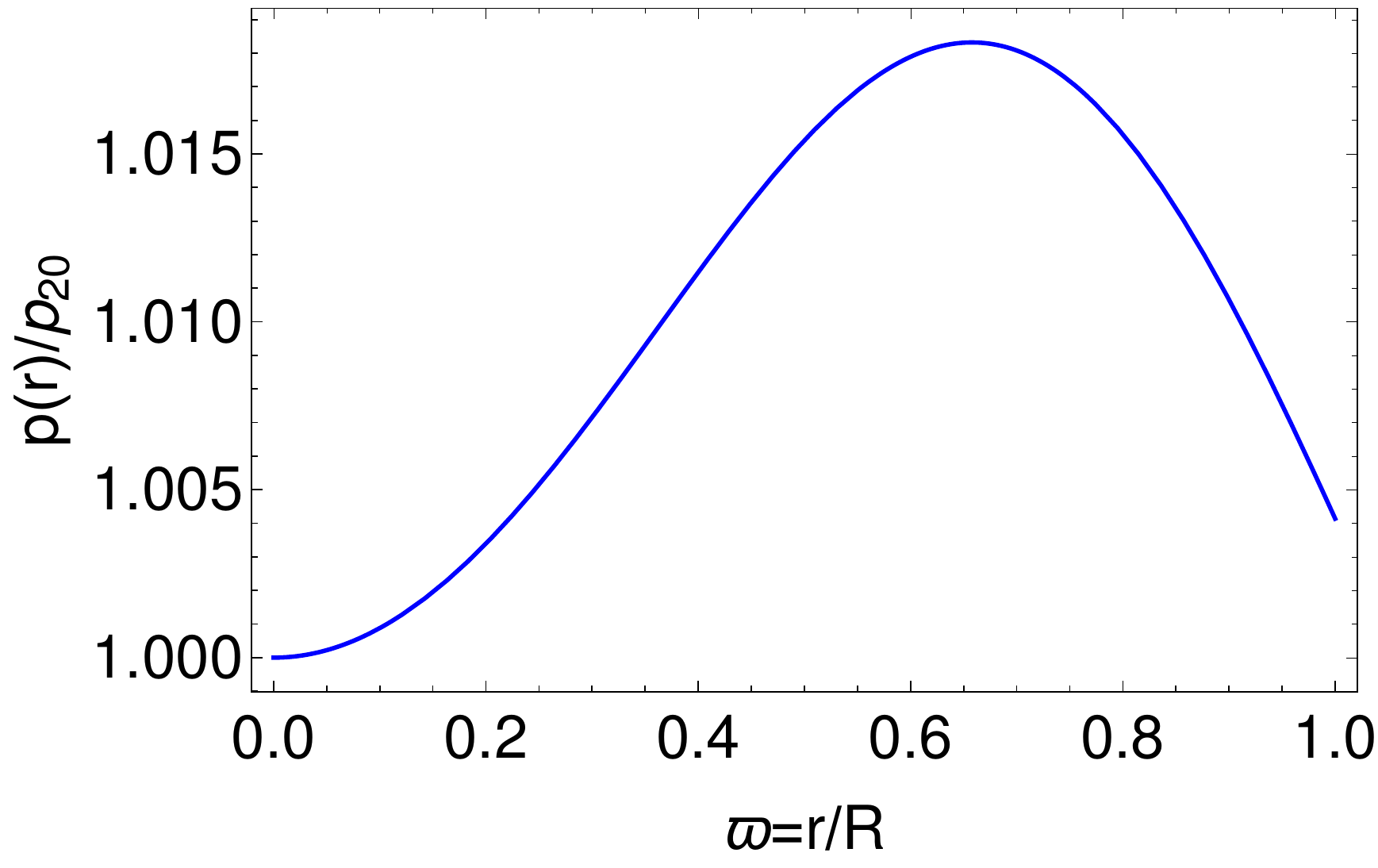}
\caption{The radial variation of $p$ normalized with the value at the center of the flux tube $p_{20}$, obtained from Coulomb function open field  model, for run $C4$ in Table \ref{gse_parameter_tab}. The horizontal axis is scaled with the total radius $R$.} 
\label{gse_pr_plot}
\end{figure}

\begin{figure}[h!]\hspace{13 mm}
\includegraphics[scale=0.45]{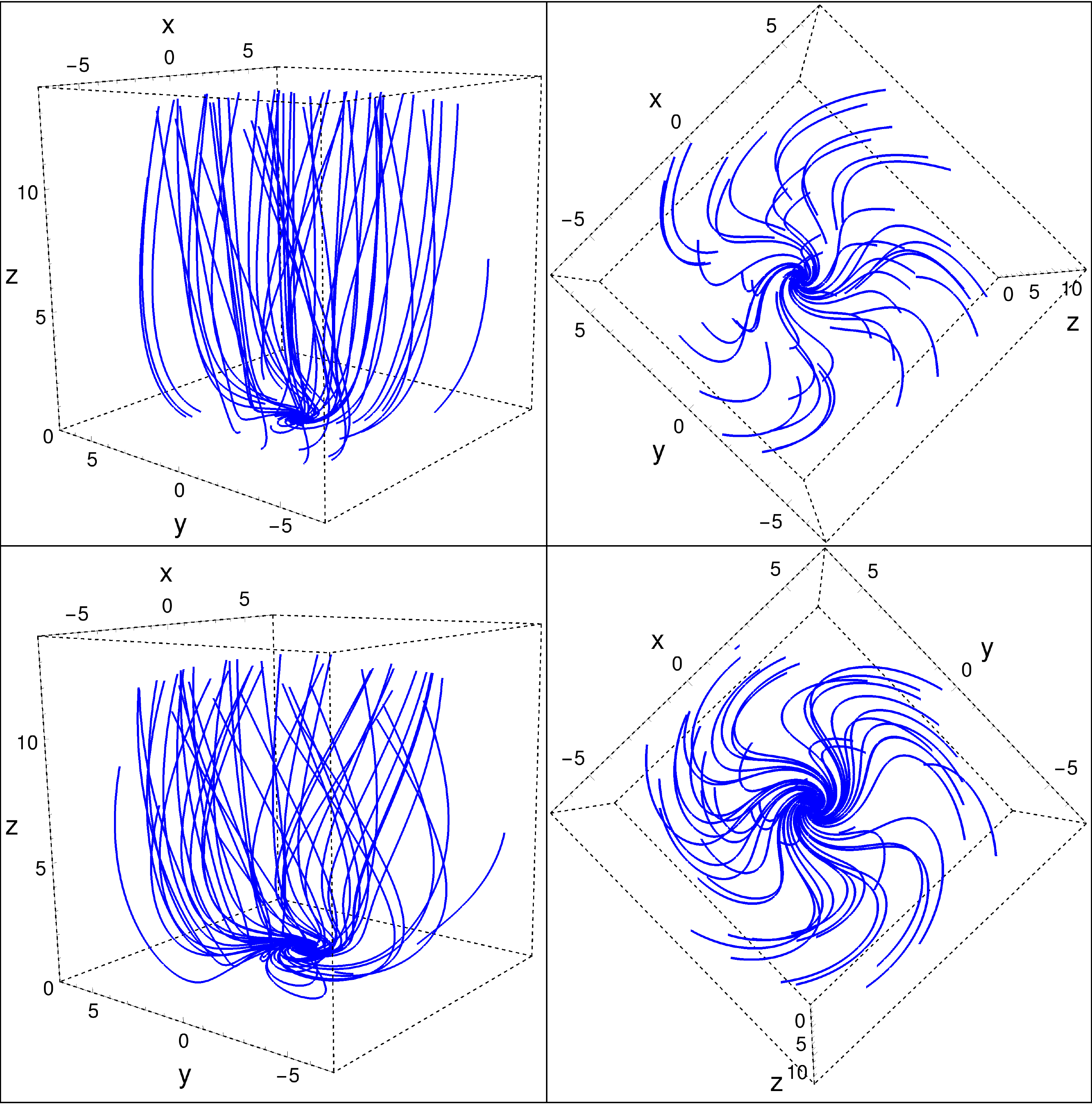}
\caption{The 3D configuration of $50$ different magnetic field lines for open field flux tube obtained from the Coulomb function helical flux tube model. The left and right columns show the side and top view of the configuration. The domain of the simulation box is $-7\leq x \leq 7$, $-7 \leq y \leq 7$ where the $x$ and $y$ axes are scaled in units of $20$ km. The vertical domain is $0 \leq z \leq 14$ where the $z$ axis is scaled in units of $150$ km. The field line configurations for the bottom and the top rows are simulated for the parameter sets of runs $C4$ and $C10$ respectively in Table {\ref{gse_parameter_tab}}.}
\label{gse_openfield_topology}
\end{figure}

\begin{figure}[h!]\hspace{13 mm}
\includegraphics[scale=0.52]{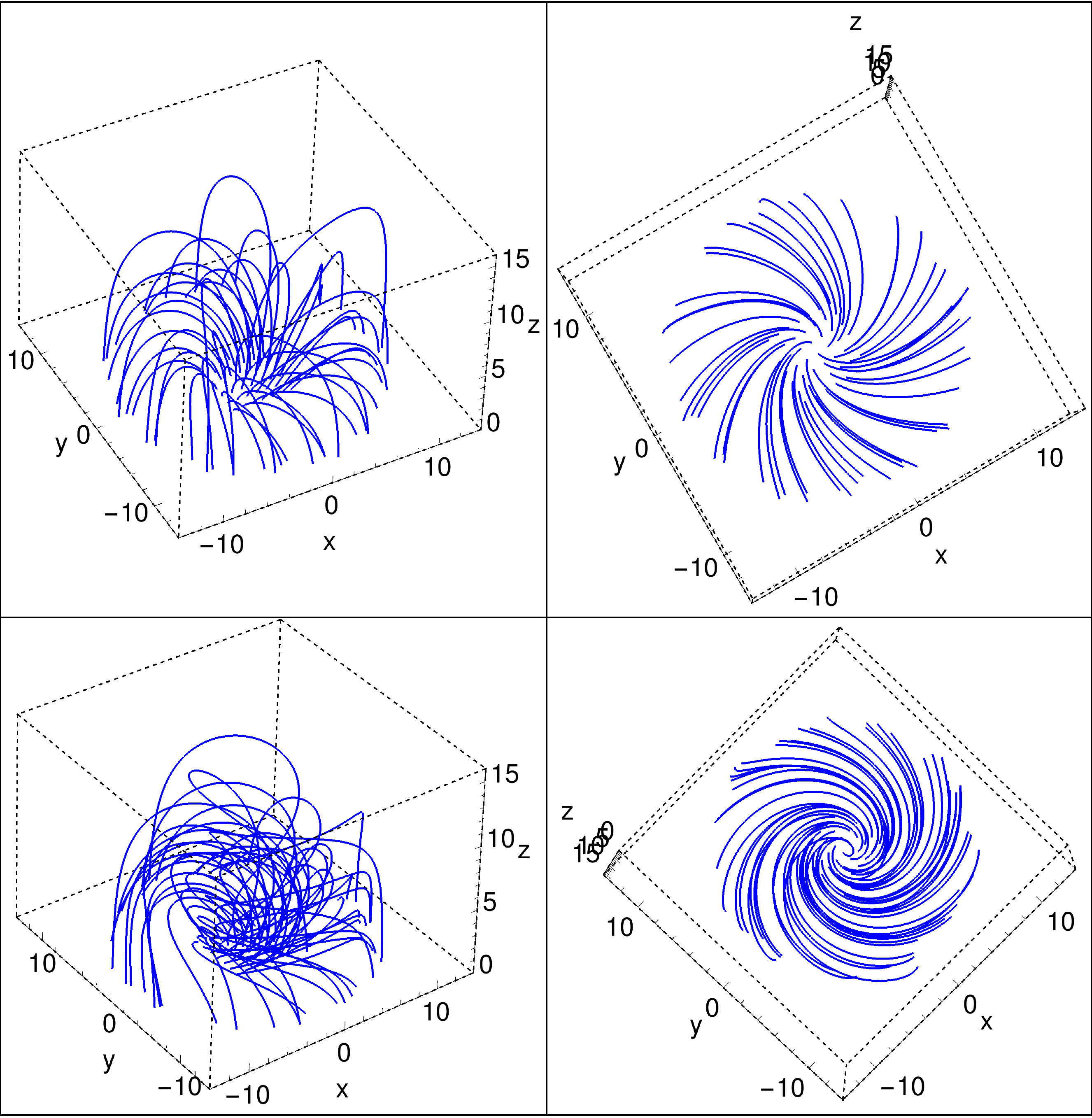}
\caption{The 3D configuration of $50$ different magnetic field lines for closed field flux tube obtained from the Coulomb function helical flux tube model. The left and right columns show the side and top view of the configuration. The domain of the simulation box is $-14\leq x \leq 14$, $-14 \leq y \leq 14$ where the $x$ and $y$ axes are scaled in units of $10$ km. The vertical domain is $0 \leq z \leq 15$ where the $z$ axis is scaled in units of $150$ km. The field line configurations for the bottom and the top rows are simulated for the parameter sets of runs $C4$ and $C10$ respectively in Table {\ref{gse_parameter_tab}}.}
\label{gse_closedfield_topology}
\end{figure}

\begin{figure}[h!]\hspace{30 mm}
\includegraphics[scale=0.65]{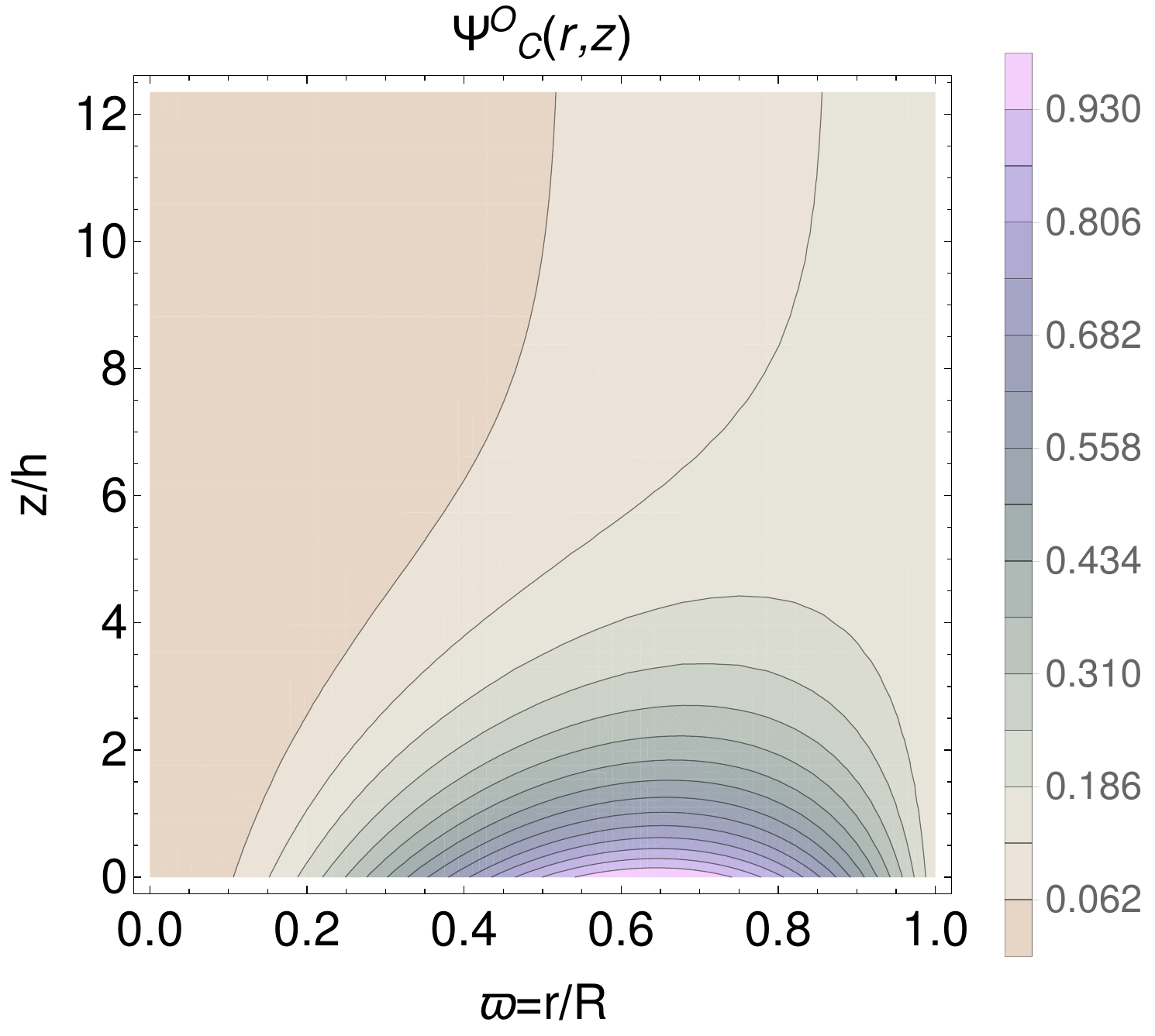}
\caption{A contour plot of the flux function corresponding to run $C4$ in Table \ref{gse_parameter_tab}, obtained from the Coulomb function open flux tube model. The horizontal axis is scaled to the radius $R$ and the vertical axis is scaled to the pressure scale height $h=162$ km. The contours have been normalized with respect to the maximum value of the flux function.}
\label{gse_Acp_plot}
\end{figure}

\begin{figure}[h!]\hspace{30 mm}
\includegraphics[scale=0.5]{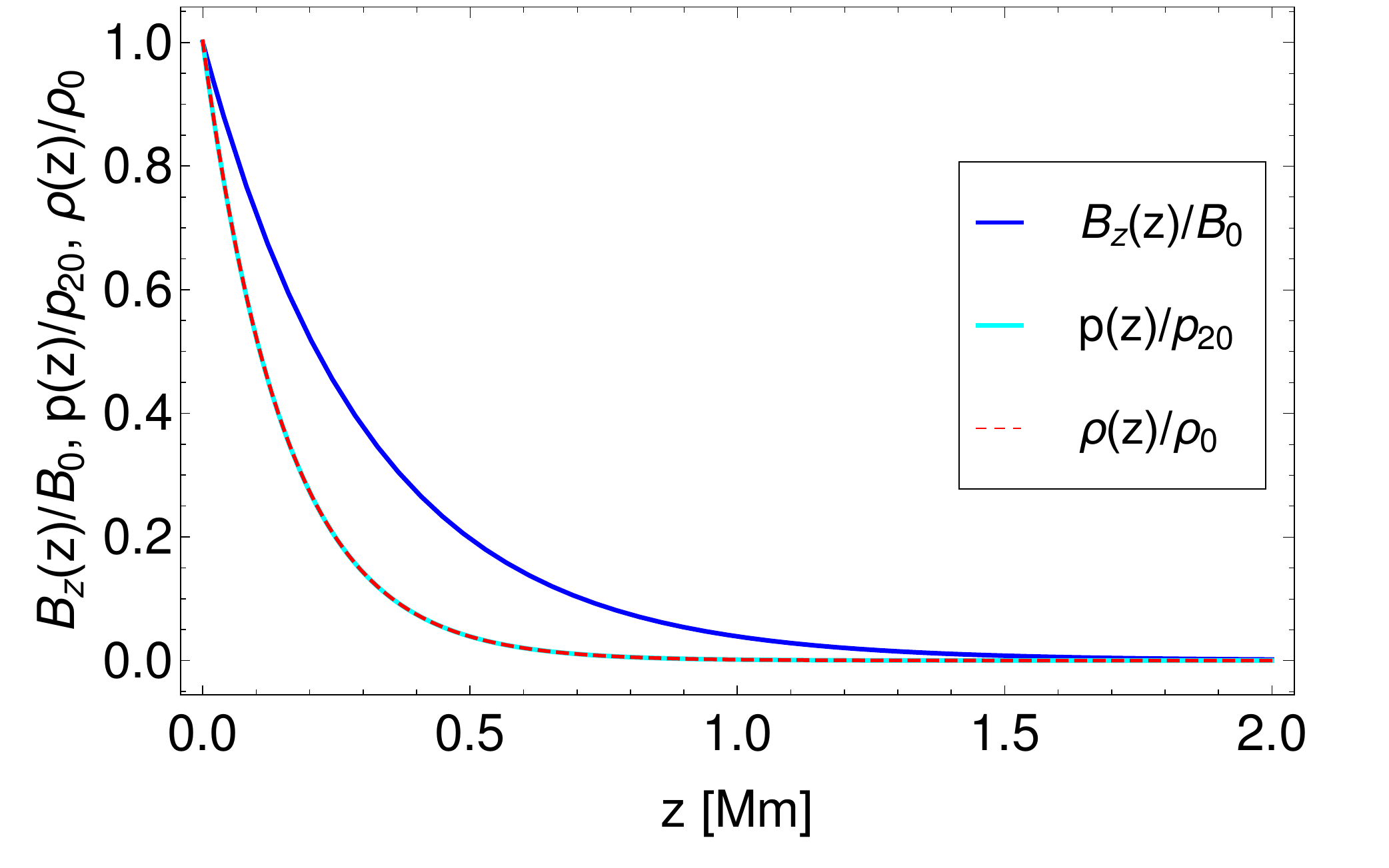}
\caption{The vertical distribution of $B_z$, $p$ and $\rho$, normalized w.r.t. the values at the flux tube center, $B_0$, $p_{20}$ and $\rho_0$ respectively, obtained from Coulomb function open field model for the parameter set of run $C4$ in Table \ref{gse_parameter_tab}. The horizontal axis is scaled in the units of Mm. The values of the scale factors are $B_0=1$ kG, $p_{20}=1.03 \times 10^5$ dyne cm$^{-2}$ and $\rho_0=2.44 \times 10^{-7}$ g cm$^{-3}$.} 
\label{gse_tdz_plot}
\end{figure}

This magnetohydrostatic Coulomb function helical flux tube model consists of the free parameters $R$ and $B_0$ and its functional dependence through $a(R)$, $\alpha(R)$, $\kappa(R)$, $b(R,B_0)$, $\psi_b(R,B_0)$ and $\bar{p}(R,B_0)$. We choose the parameter range, $1$ kG $\leq B_0 \leq$ $1.5$ kG and $100$ km $\leq R \leq$ $180$ km, consistent with the observations of MBP size and field strength distributions \citep{2009A&A...498..289U, 2013A&A...554A..65U}. In Table \ref{gse_parameter_tab}, we show the solutions for combinations of the free parameters $\lbrace R, B_0\rbrace$, where we notice the following trends:
\begin{itemize}
\item The boundary flux $\psi_b$ decreases with $R$ for same $B_0$, and  with $B_0$ for same $R$ within the parameter space of runs $C1-C21$.
\item Due to the pressure balance at the boundary of the flux tube, $\bar{p}$ increases with $R$ for same $B_0$, but there is no fixed trend  with $B_0$ for same $R$ within the parameter space of runs $C1-C21$.
\end{itemize}
As example, we show the solution of $\psi^O_C$, and the magnetic and thermodynamic structure of the flux tube for run $C4$. The radial variation of the solution of $\psi^O_C$, magnetic components and pressure inside the flux tube are shown in the Figs. \ref{gse_Ar_plot}, \ref{gse_B_plot} and \ref{gse_pr_plot} respectively. Examples of 3D configuration of the magnetic field lines for open and closed field are shown in the Figs. \ref{gse_openfield_topology} and {\ref{gse_closedfield_topology}} for runs $C4$ and $C10$. 2D vertical projection of the magnetic field lines for $\psi^O_C$ inside the flux tube along $r-z$ plane is shown in the Fig. \ref{gse_Acp_plot}. The density inside the flux tube is constant along the radial direction but it decreases along $z$ whereas the temperature varies along $r$ direction and is nearly constant along $z$ direction at the axis. The vertical variation of $B_z$, $p$ and $\rho$ are shown in the Fig. \ref{gse_tdz_plot}. Conclusions from figures and tables are discussed in $\S$\ref{discussions}.

\begin{table}[hbt!]
\centering
\begin{tabular}{|c | c  c  c  c  c  c  c c|}\hline \hline
Run \# & $B_0$ [kG]  & $R$ [km] & $\psi_b$ [$10^{-3}$] & $a$  & $\alpha$ [$10^{-2}$] & $\kappa$ [$10^{6}$] & $b$ [$10^{-3}$] & $\bar{p}$  \\ \hline
$C1$	&	1	&	100	&	2.57	&	9.390	&	2.74	&	4.85	&	1.350	&	0.109	\\
$C2$	&	1.2	&	100	&	2.22	&	9.390	&	2.74	&	4.85	&	1.170	&	0.105	\\
$C3$	&	1.5	&	100	&	1.80	&	9.390	&	2.74	&	4.85	&	0.949	&	0.104	\\
$C4$	&	1	&	120	&	1.92	&	9.388	&	2.54	&	6.99	&	1.008	&	0.159	\\
$C5$	&	1.2	&	120	&	1.62	&	9.388	&	2.54	&	6.99	&	0.849	&	0.165	\\
$C6$	&	1.5	&	120	&	1.31	&	9.388	&	2.54	&	6.99	&	0.692	&	0.163	\\
$C7$	&	1	&	130	&	1.69	&	9.389	&	2.43	&	8.21	&	0.880	&	0.184	\\
$C8$	&	1.2	&	130	&	1.42	&	9.389	&	2.43	&	8.21	&	0.744	&	0.181	\\
$C9$	&	1.5	&	130	&	1.15	&	9.389	&	2.43	&	8.21	&	0.603	&	0.182	\\
$C10$	&	1	&	140	&	1.50	&	9.383	&	2.31	&	9.52	&	0.783	&	0.205	\\
$C11$	&	1.2	&	140	&	1.25	&	9.383	&	2.31	&	9.52	&	0.661	&	0.204	\\
$C12$	&	1.5	&	140	&	1.00	&	9.383	&	2.31	&	9.52	&	0.535	&	0.208	\\
$C13$	&	1	&	150	&	1.38	&	9.378	&	2.18	&	10.93	&	0.723	&	0.233	\\
$C14$	&	1.2	&	150	&	1.16	&	9.378	&	2.18	&	10.93	&	0.606	&	0.235	\\
$C15$	&	1.5	&	150	&	0.94	&	9.378	&	2.18	&	10.93	&	0.492	&	0.237	\\
$C16$	&	1	&	160	&	1.31	&	9.388	&	1.98	&	12.43	&	0.665	&	0.276	\\
$C17$	&	1.2	&	160	&	1.11	&	9.388	&	1.98	&	12.43	&	0.577	&	0.276	\\
$C18$	&	1.5	&	160	&	0.89	&	9.388	&	1.98	&	12.43	&	0.465	&	0.279	\\
$C19$	&	1	&	180	&	1.14	&	9.395	&	1.72	&	15.73	&	0.587	&	0.402	\\
$C20$	&	1.2	&	180	&	0.96	&	9.395	&	1.72	&	15.73	&	0.497	&	0.407	\\
$C21$	&	1.5	&	180	&	0.78	&	9.395	&	1.72	&	15.73	&	0.405	&	0.409	\\
\hline   
\end{tabular}
\caption{Numerical values of the different parameters obtained from the Coulomb function open field flux tube model for different combinations of $R$ and $B_0$ are shown; the units of the various quantities are in the square brackets at the top.}
\label{gse_parameter_tab}
\end{table}

\subsection{Self similar model}

The self-similar model we developed consists of the dimensionless parameters $\psi_b, \bar{B'}_{z 0}$, $\bar{f}$ and $\bar{\chi}$ which are the functions of the input parameter set $\lbrace\Psi_b,\ B_0,\ B'_{z0},\ p_c$, $\chi \rbrace$. The self-similar flux tube solutions are spanned by these parameters but the structures remain similar. We use the values of these input parameters in the range, $\Psi_b=10^{17}$--$10^{18}$ Mx \citep{1998SoPh..183..283Z, 1999ApJ...511..932H, 2011IAUS..274..140G}, $B_0=1$--$2$ kG \citep{1998SoPh..183..283Z}, $B'_{z0}$ in the range $1$--$2$ G-km$^{-1}$ \citep{1974SoPh...36...29W, 1990A&A...228..246P, 1993A&A...279..243B}, $p_c<p_0$ [\cite{2010A&A...515A.107S} and SM$18$], which are observed for small scale magnetic structures in the photosphere. The generalized Gaussian profile reduces to the Gaussian profile for $n_G=2$, and it has been shown in \S \ref{ss_sol} that, for the power law profile, the flux function converges to finite value, at infinite radius, only for $n_P>2$. We study the different cases for $n_G=2$--$3$, $n_P=3$--$4$ and $\bar{\chi}=0.01$--$100$ for different combinations of the other parameter sets $\lbrace \bar{\Psi}_b, \bar{B'}_{z 0}$, $\bar{f} \rbrace$,  which are shown in runs $S1$--$S19$ of Table \ref{mpara_ss}. For the parameter set of runs $S1$--$S19$, we find the following results:

\begin{itemize}

\item For same $\Psi_b$ and $B_0$, with the increase of $n_G$ and $n_P$, $R_G$ and $R_P$ decrease respectively. 

\item  For same $n_G$, $n_P$ and $\Psi_b$, with the increase of $B_0$, $R_G$ and $R_P$ decrease; whereas for same $n_G$, $n_P$ and $B_0$; $R_G$ and $R_P$ increase with the increasing of $\Psi_b$.    

\item For a fixed parameter set $\lbrace \psi_b$, $\bar{B'}_{z 0}$ $\bar{f}$, $\bar{\chi} \rbrace$, we notice that $R_G>R_P$ for $n_G=2$ and $n_P=3$, but for values $n_G \geq 2.5$ and $n_P \geq 3.5$, $R_G<R_P$; this means that the radii of the flux tubes for the power law profiles falls off more quickly than those of the generalized Gaussian profiles for higher values of $n_G$ and $n_P$. 
\end{itemize}

\begin{figure}[h!]\hspace{8 mm}
\includegraphics[scale=0.38]{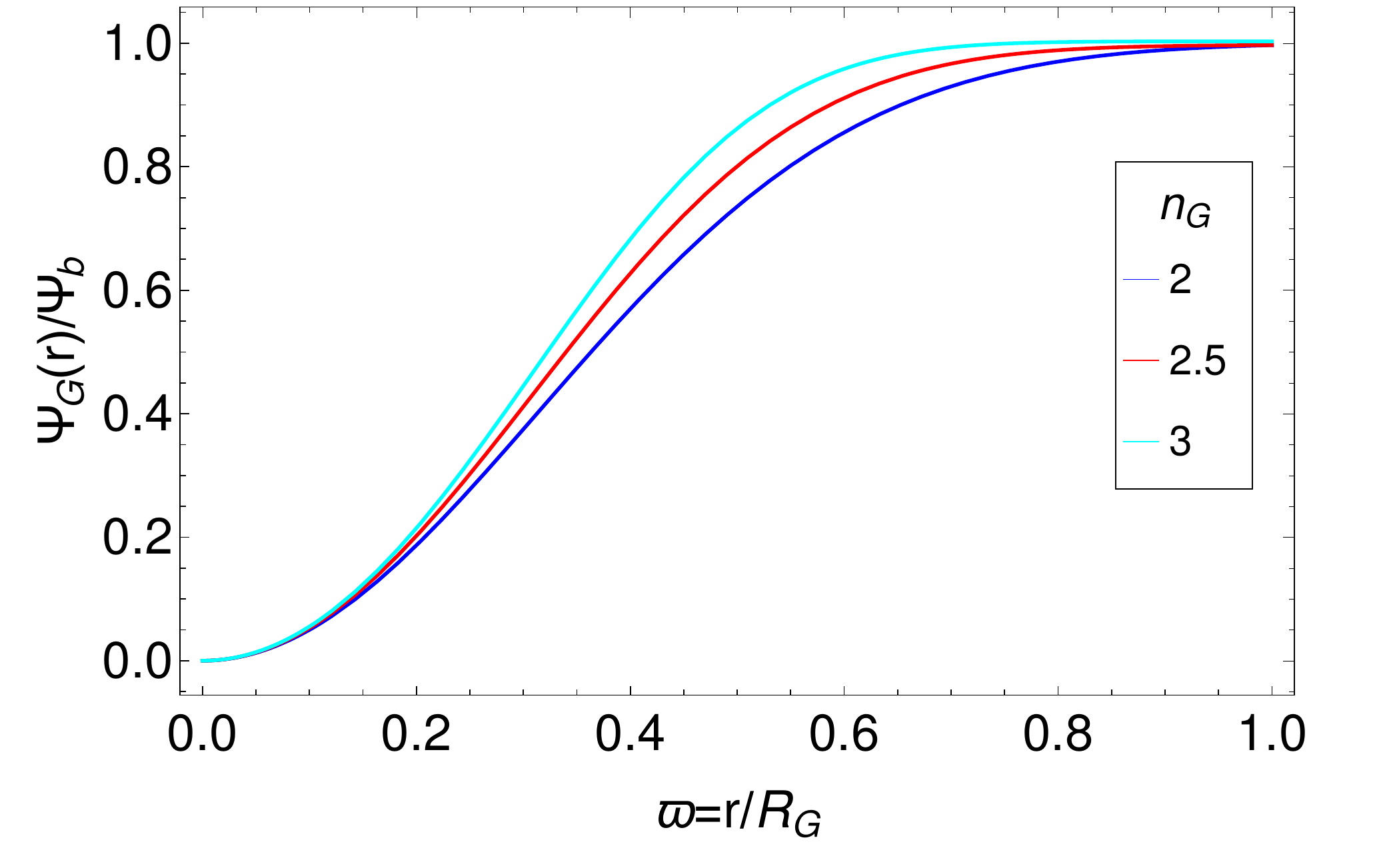}
\includegraphics[scale=0.38]{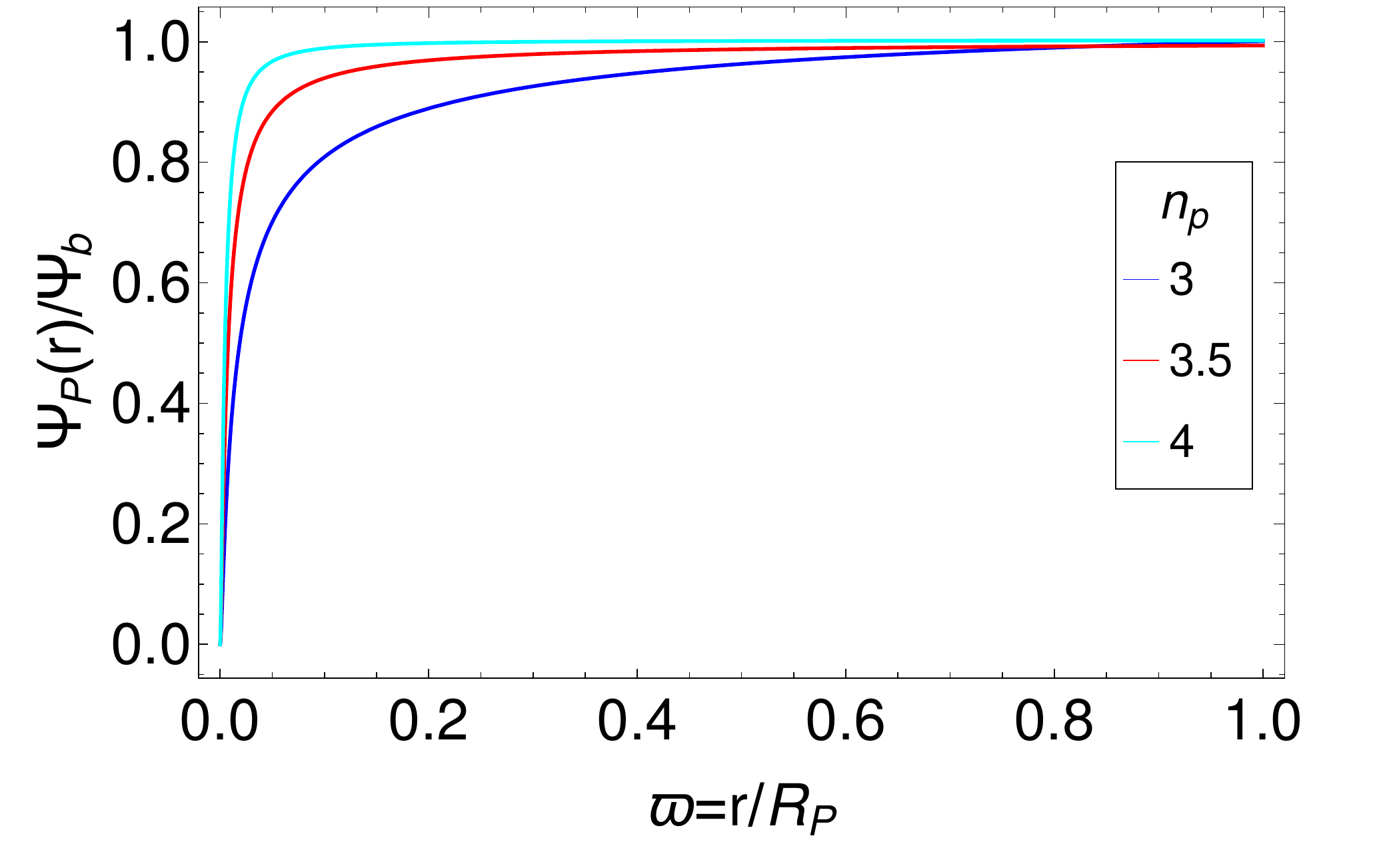}
\caption{The radial variation of the flux function, normalized with respect to the maximum values for different values of $n$ for generalized Gaussian ({\it left}) and power law ({\it right}) shape functions for the parameter set of run $S1$ in Table \ref{mpara_ss}. The horizontal axes are scaled with the total radius $R$.}
\label{ss_psir_plot}
\end{figure}

\begin{figure}[]\hspace{8 mm}
\includegraphics[scale=0.52]{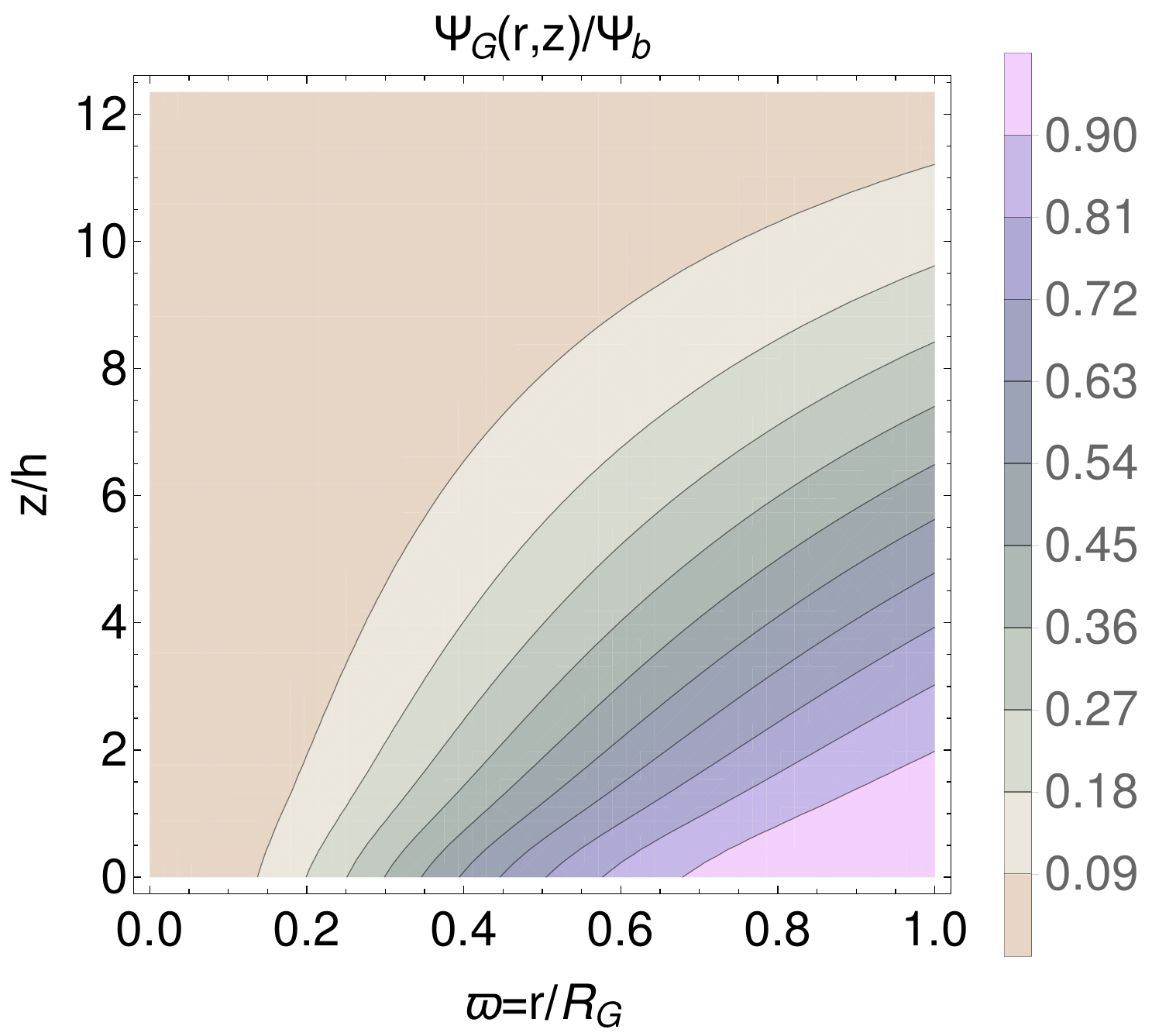}
\includegraphics[scale=0.5]{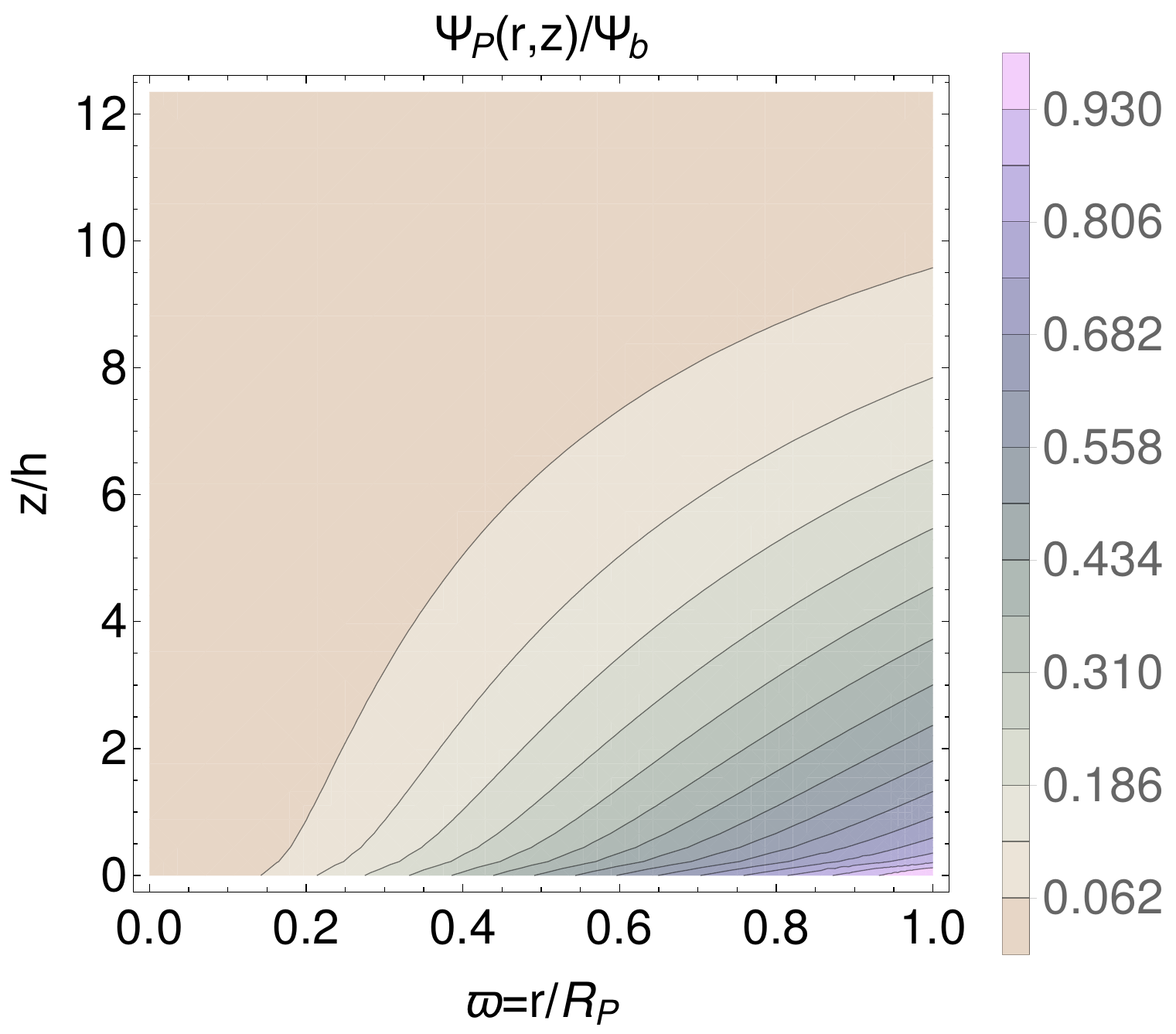}
\caption{Contour plots of the flux functions for Gaussian ({\it left}) and power law ({\it right}) profiles for $n_P=3$ for the parameter set of run $S1$ in Table \ref{mpara_ss}. The horizontal axes are scaled with the total radii $R_G=214$ and $R_P=261$ km, and the vertical axes are scaled with the pressure scale height $h=162$ km. The contours are normalized with respect to the maximum value of the flux function.}
\label{ss_psicp}
\end{figure}

\begin{figure}[h!]\hspace{10 mm}
\includegraphics[scale=0.55]{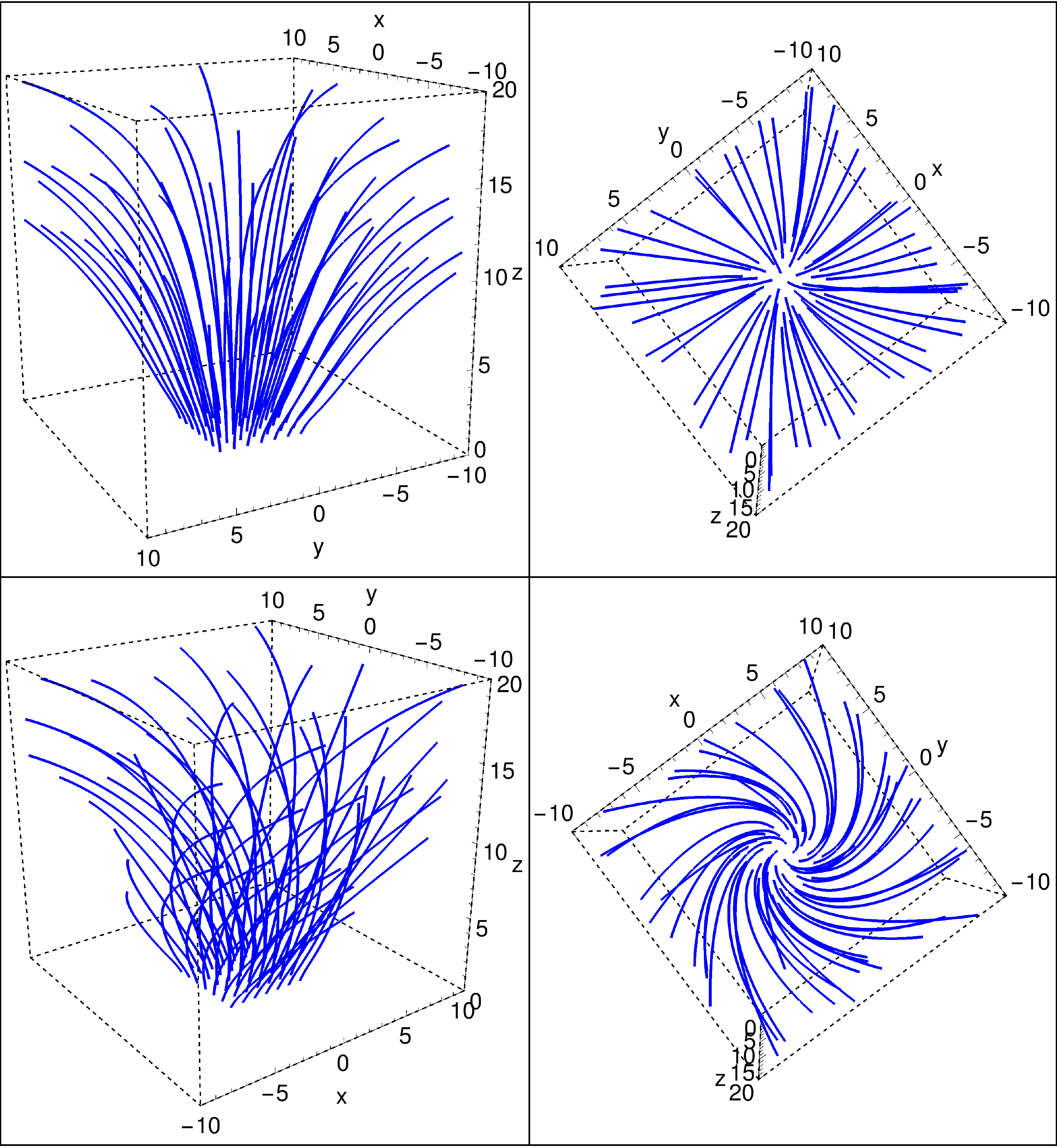}
\caption{The 3D configuration of 50 different open field lines inside the flux tube obtained from self-similar model for Gaussian profile. The left and the right columns show the side and the top view of the configurations. The domain of the simulation box is $-10 \leq x \leq 10$, $-10 \leq y \leq 10$ and $0 \leq z \leq 20$ where the $x$, $y$ and $z$ axes are scaled in units of $100$ km. The field line configurations for the bottom and the top rows are simulated for the parameter values of Table \ref{mpara_ss} corresponding to runs $S1$ and $S2$ respectively.}
\label{gausstopo_topology}
\end{figure}

\begin{figure}[h!]\hspace{10 mm}
\includegraphics[scale=0.55]{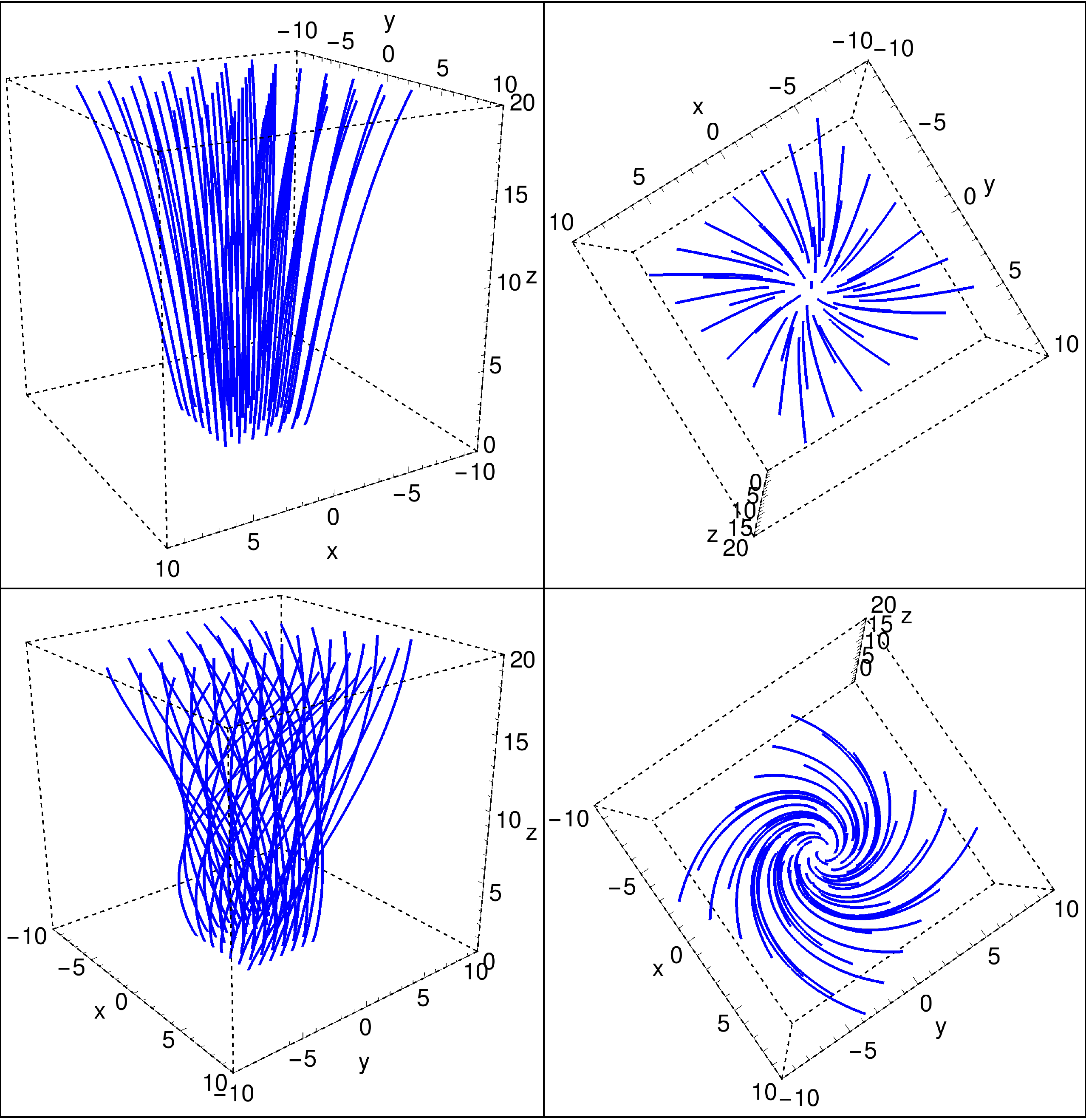}
\caption{The 3D configuration of 50 different open field lines inside the flux tube obtained from self-similar model for power law profile with $n_P=3$. The left and the right columns show the side and the top view of the configurations. The domain of the simulation box is $-10 \leq x \leq 10$, $-10 \leq y \leq 10$ and $0 \leq z \leq 20$ where the $x$, $y$ and $z$ axes are scaled in units of $100$ km. The field line configurations for the bottom and the top rows are simulated for the parameter sets of runs $S1$ and $S2$ respectively corresponding to Table \ref{mpara_ss}.}
\label{powerlawtopo_topology}
\end{figure}

\begin{figure}[h!]
\includegraphics[scale=0.3]{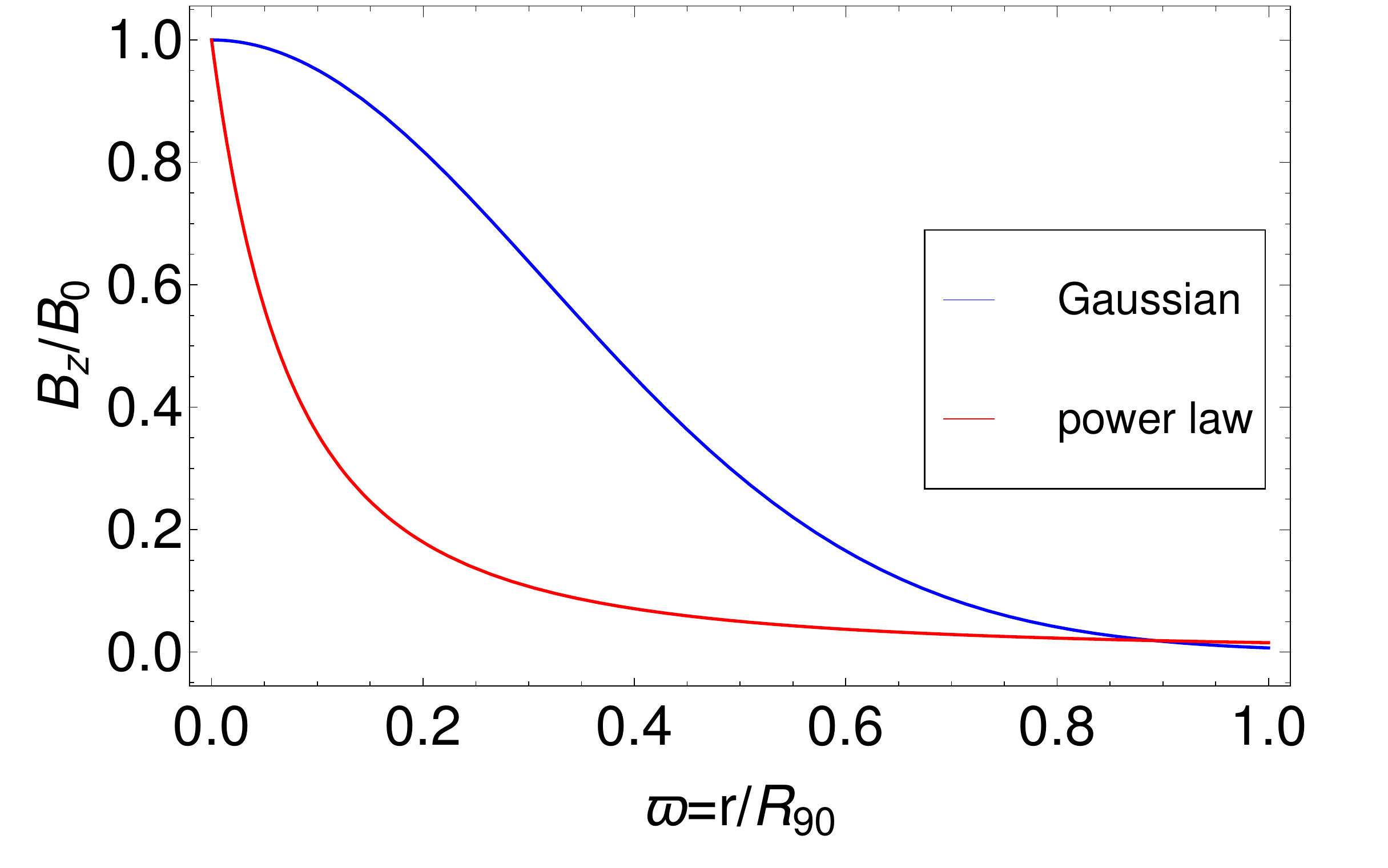}
\includegraphics[scale=0.3]{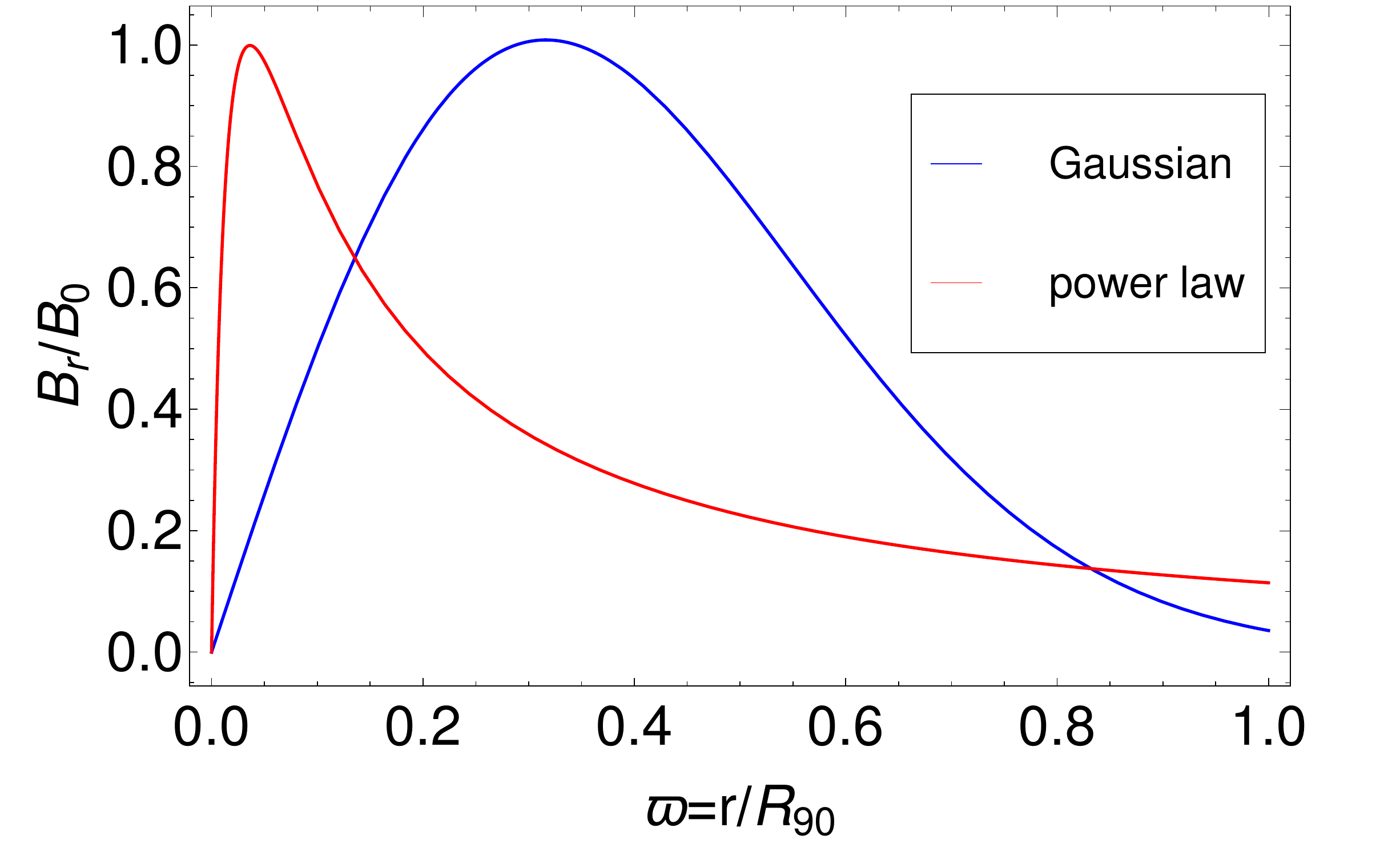}

\includegraphics[scale=0.3]{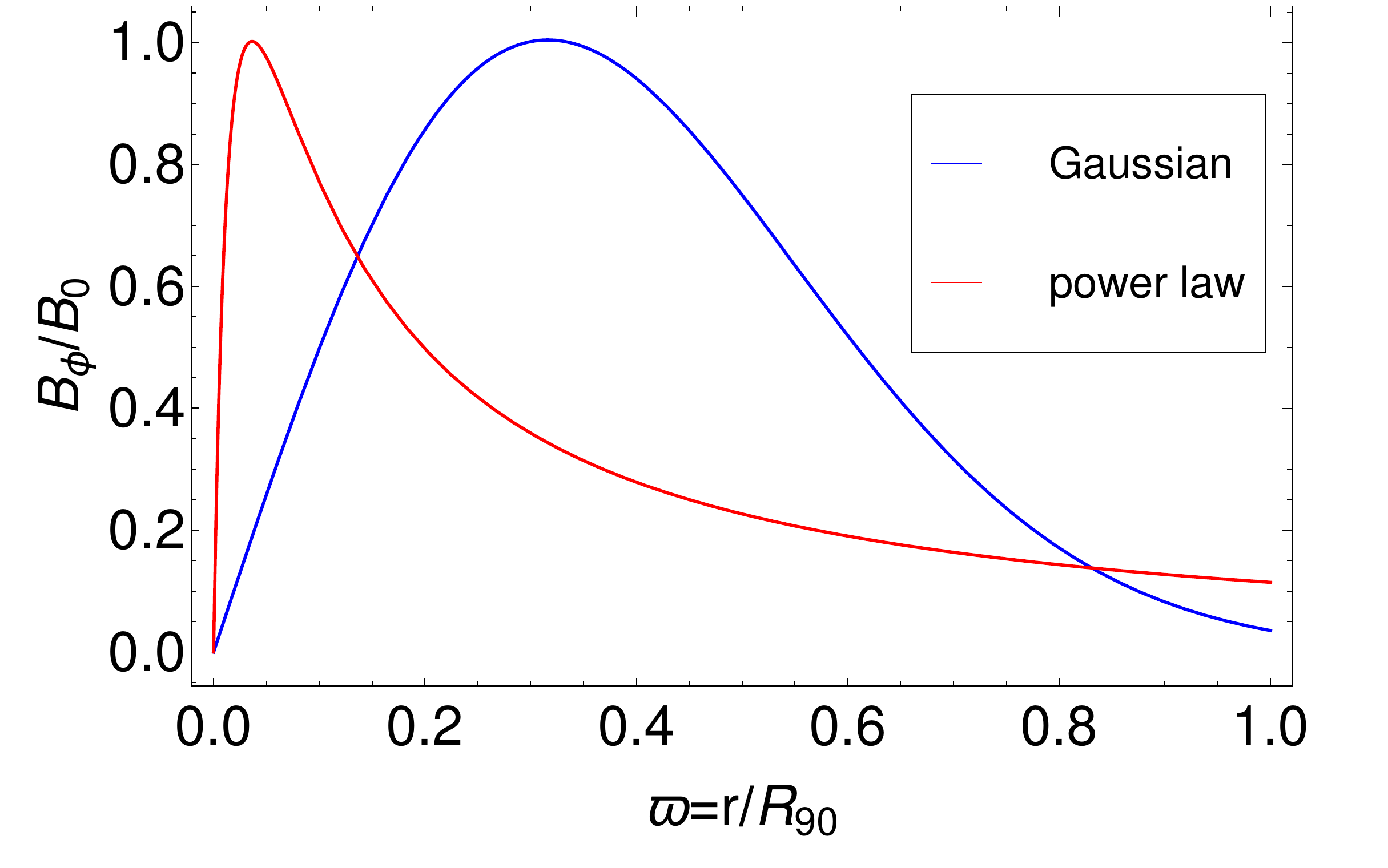}
\includegraphics[scale=0.3]{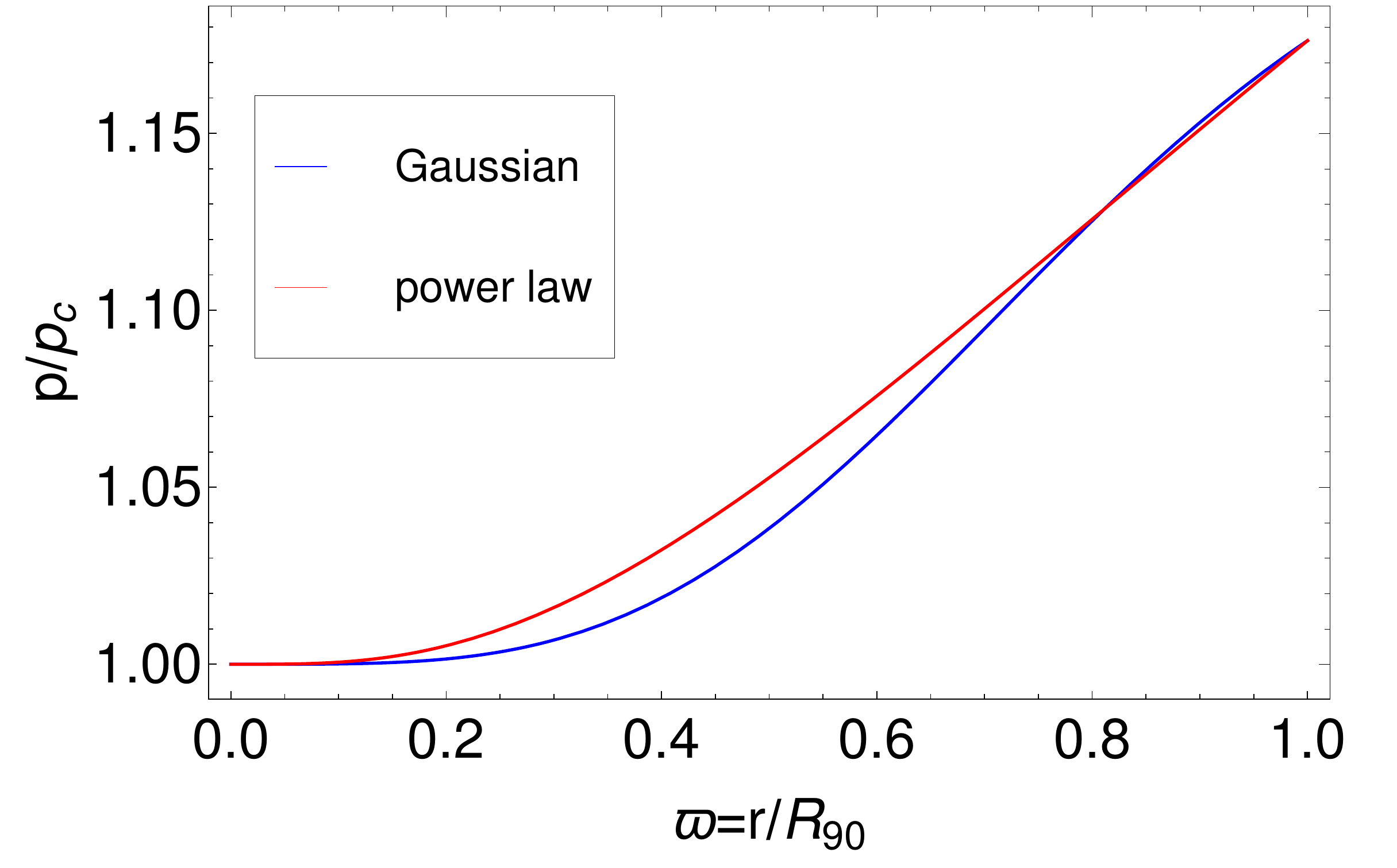}
\caption{The radial distribution of the magnetic field components $B_z$, $B_r$, $B_\phi$ and gas pressure $p$ normalized w.r.t. the values at the flux tube center, $B_0$, $p_c$, for Gaussian and power law shape functions for the parameter set of run $S1$ in Table \ref{mpara_ss}. The horizontal axes are scaled with the total radius of the flux tube $R$ and the values of the scale factors are $B_0=1$ kG and $p_c=10^5$ dyne cm$^{-2}$.}
\label{ss_B_plot}
\end{figure}

\begin{figure}[h!]\hspace{35 mm}
\includegraphics[scale=0.4]{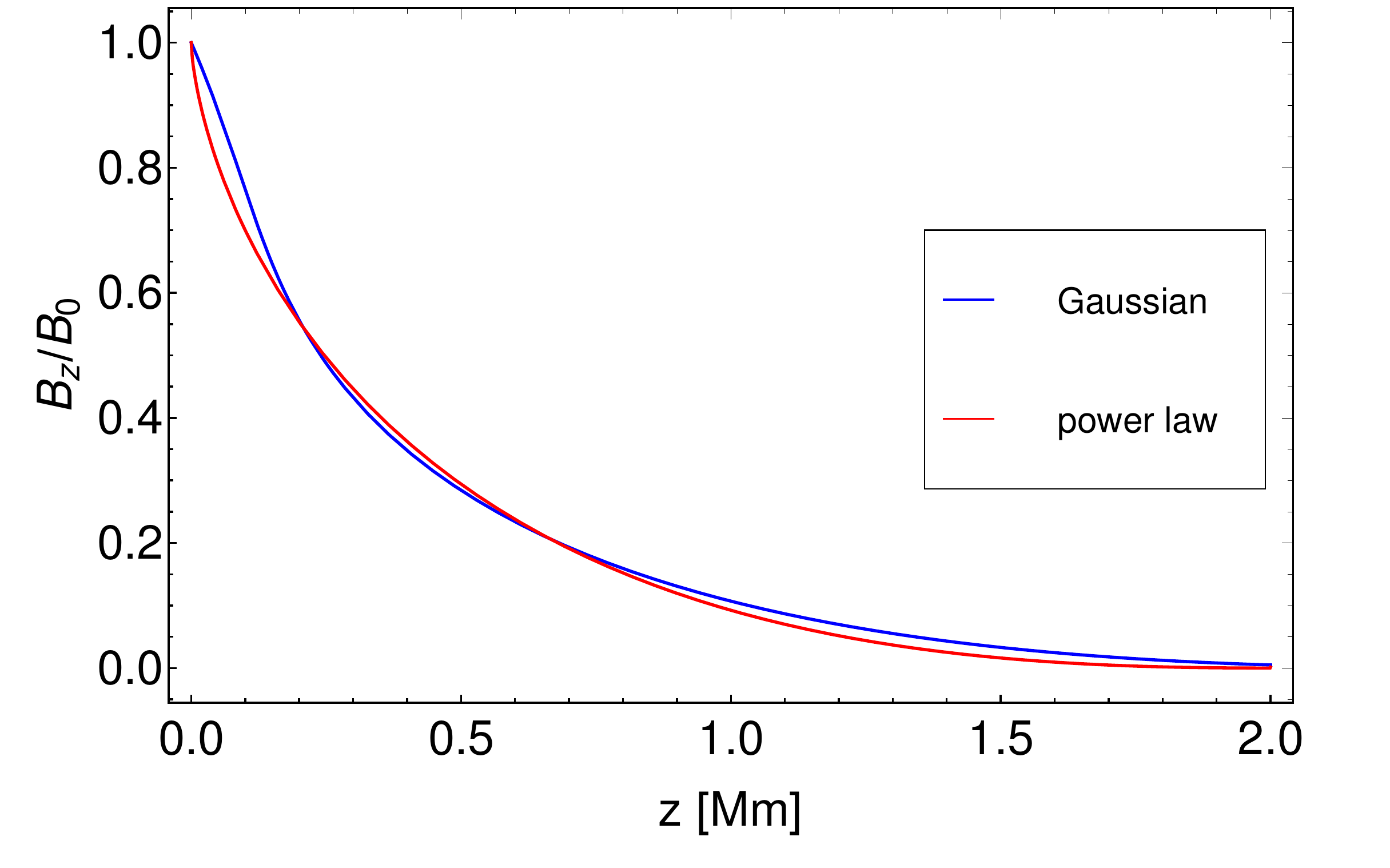}
\caption{The vertical distribution of $B_z$ at the axis of the flux tube, obtained from the self-similar model for the Gaussian and power law profiles for the parameter set of run $S1$ in Table \ref{mpara_ss}.}
\label{ss_Bzz}
\end{figure}

\begin{figure}[h!]\hspace{35 mm}
\includegraphics[scale=0.5]{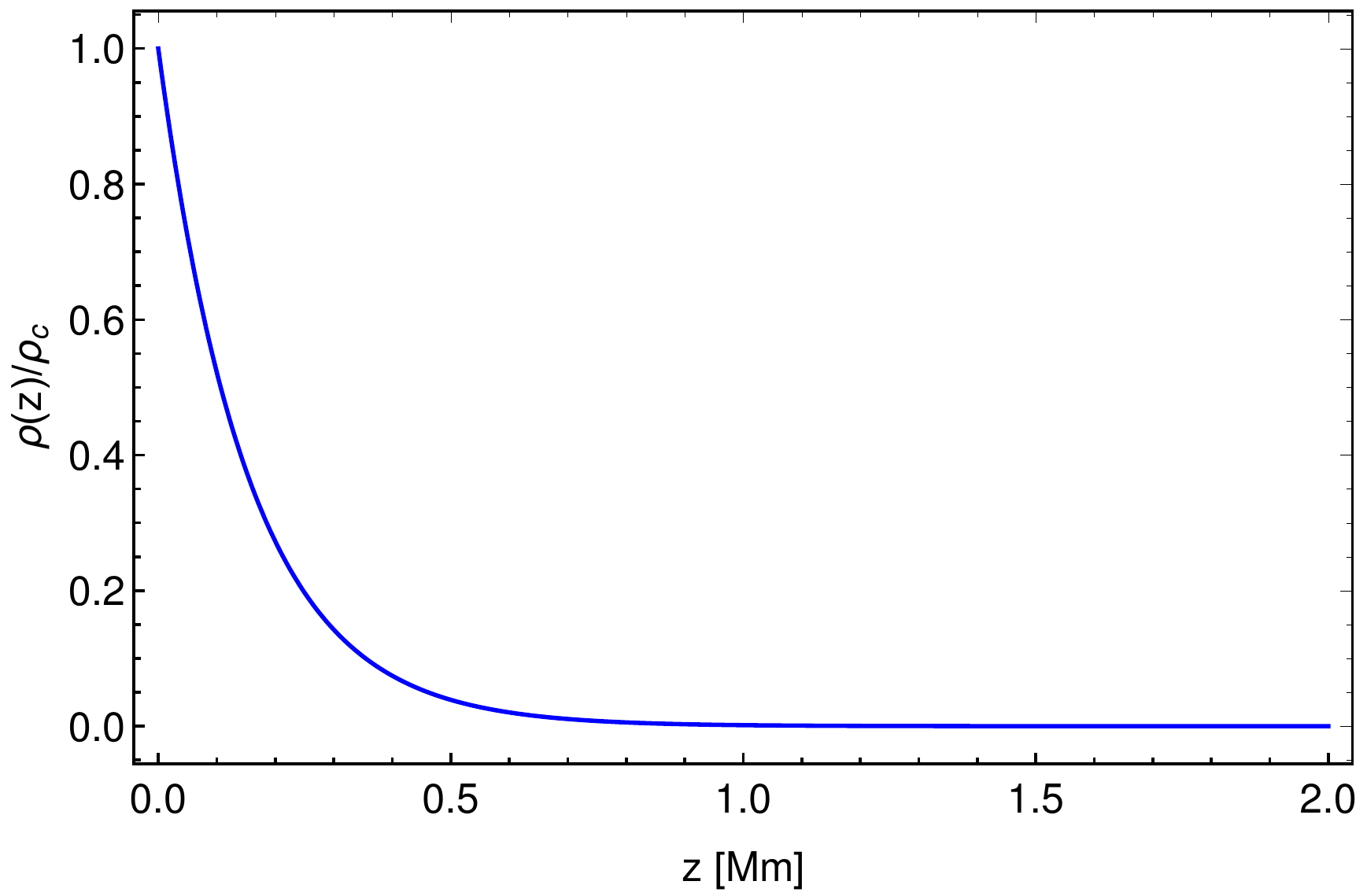}
\caption{The vertical distribution of density, $\rho(z)$ obtained from the self-similar model for the parameter set of run $S1$ in Table \ref{mpara_ss}, which is normalized w.r.t. $z=0$ value, $\rho_c$, for both Gaussian and power law profiles. The horizontal axis is scaled in units of Mm. The value of scale factor $\rho_c=2.37 \times 10^{-7}$ g cm$^{-3}$.}
\label{ss_rho_plot}
\end{figure}

\begin{figure}[h!]
\begin{center}
\includegraphics[scale=0.5]{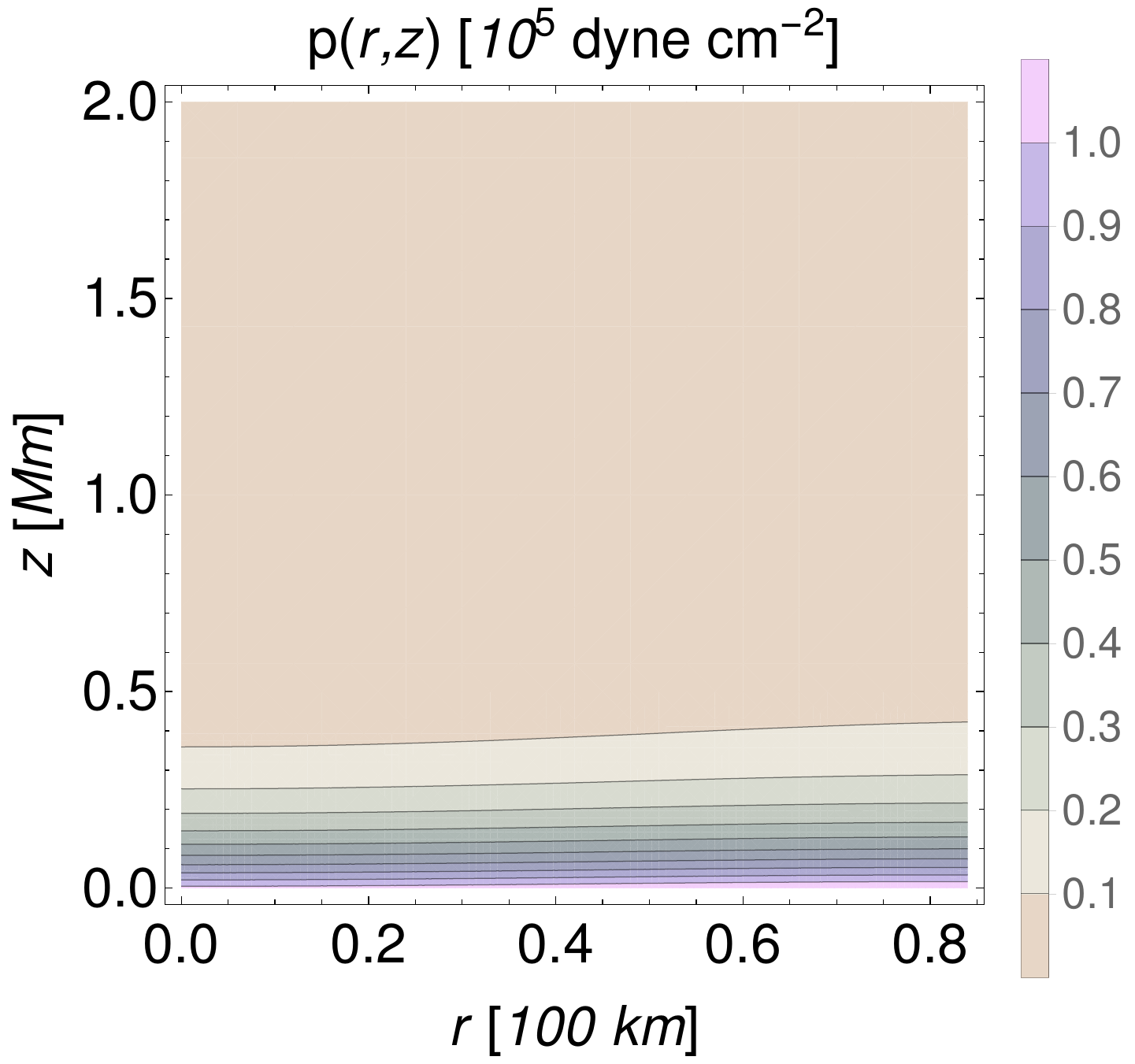}
\includegraphics[scale=0.5]{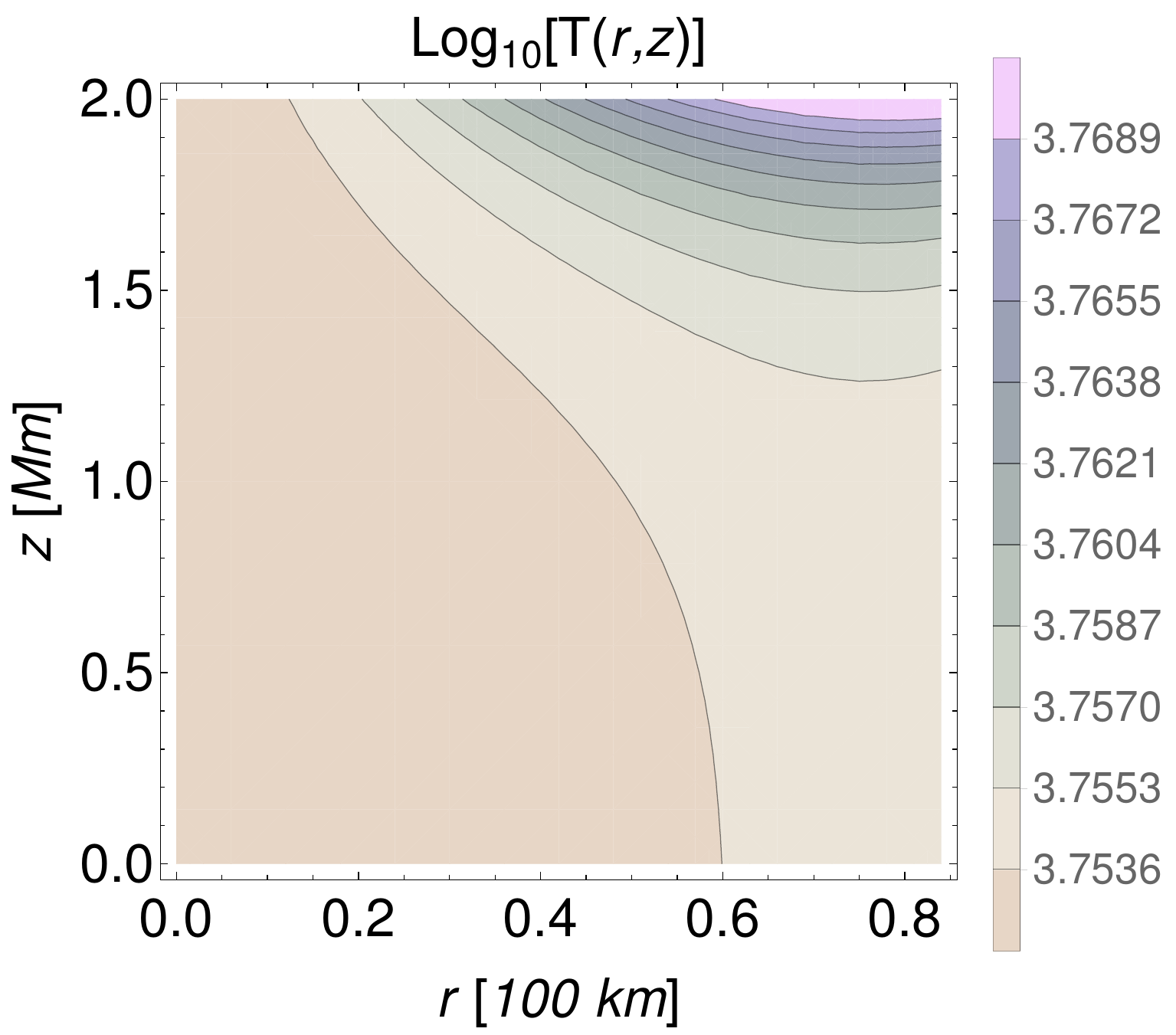}
\caption{The 2D variation of $p$ ({\it left}) and $T$ ({\it right}) in the  $r-z$ plane for $r_b=84$ km obtained from the Coulomb function model for the parameter set of run $C4$ in Table \ref{gse_parameter_tab}. The horizontal axes are scaled in the units of $100$ km and the vertical axes are scaled in the units of Mm.} 
\label{gse_tdcp_plot}
\end{center}
\end{figure}

\begin{figure}[h!]
\begin{center}
\includegraphics[scale=0.52]{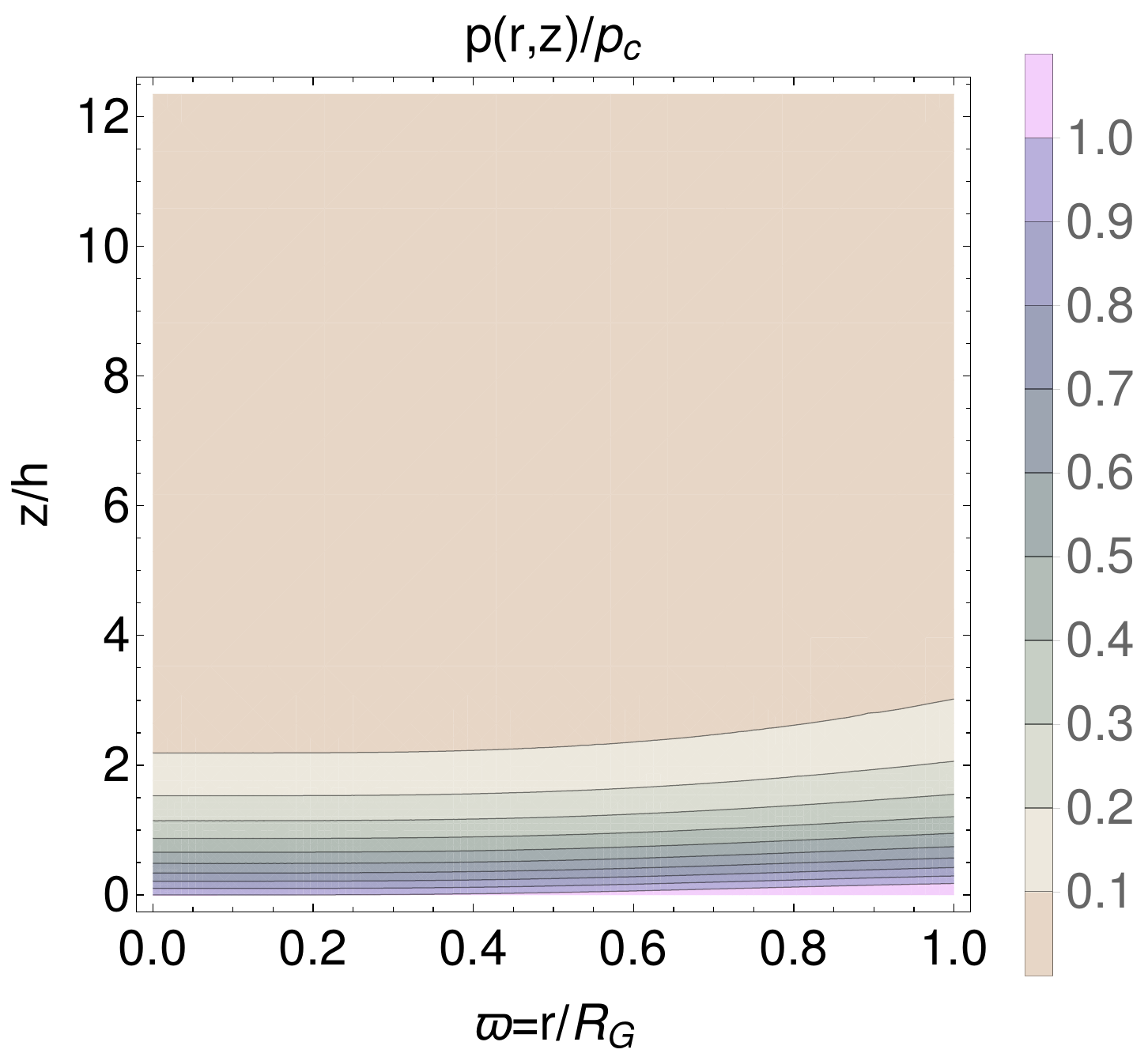}
\includegraphics[scale=0.5]{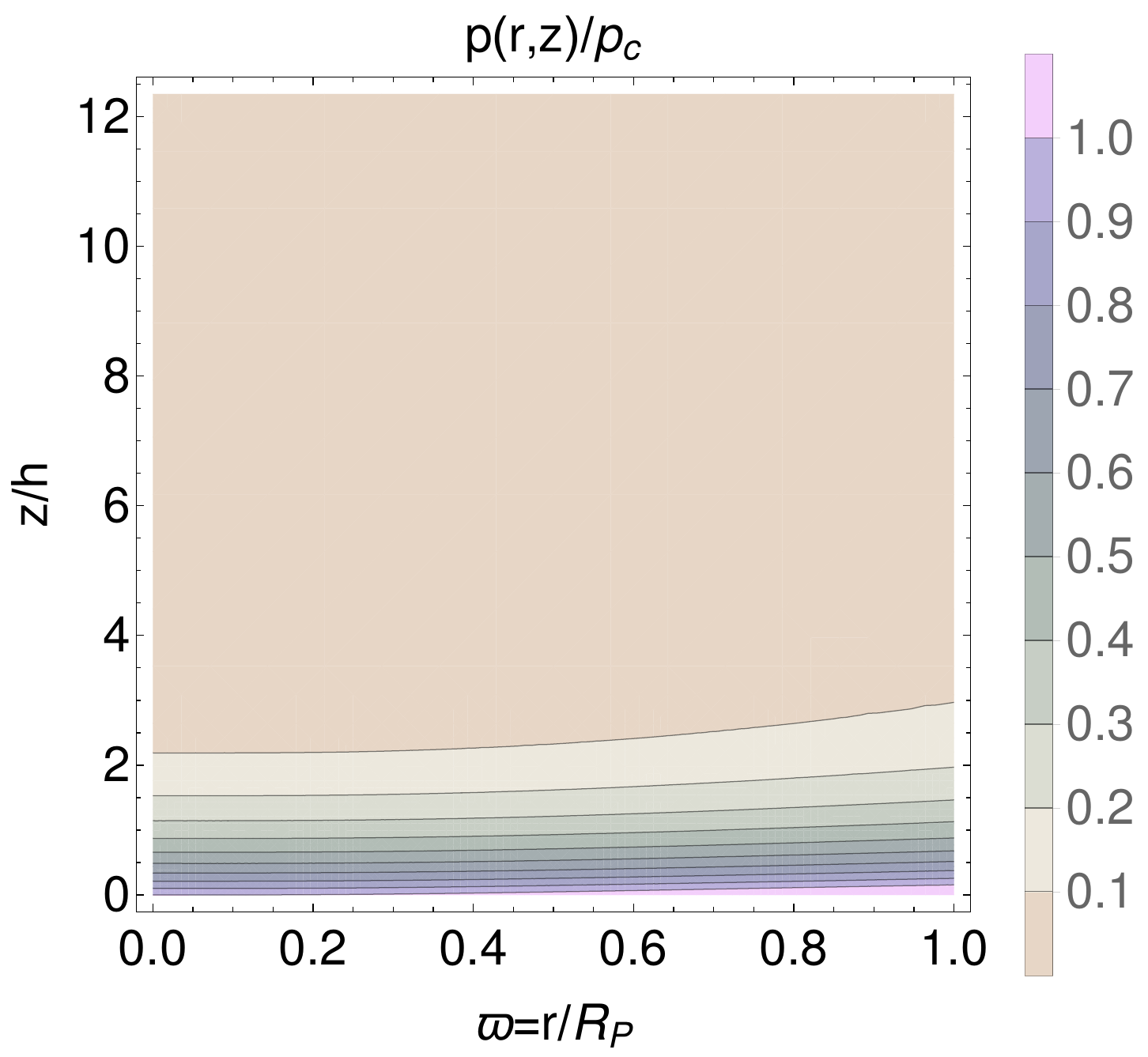}
\caption{The 2D variation of $p$  in the $r-z$ plane obtained from the self-similar model for Gaussian ({\it left}) and power law ({\it right}) profiles with $n_P=3$, for the parameter set of run $S1$ in Table \ref{mpara_ss}. The horizontal axes are scaled with the total radii $R_G=150$ and $R_P=130$ km, and the vertical axes are scaled with the pressure scale height, $h=162$ km.} 
\label{ss_pcp_plot}
\end{center}
\end{figure}

\begin{figure}[h!]
\begin{center}
\includegraphics[scale=0.52]{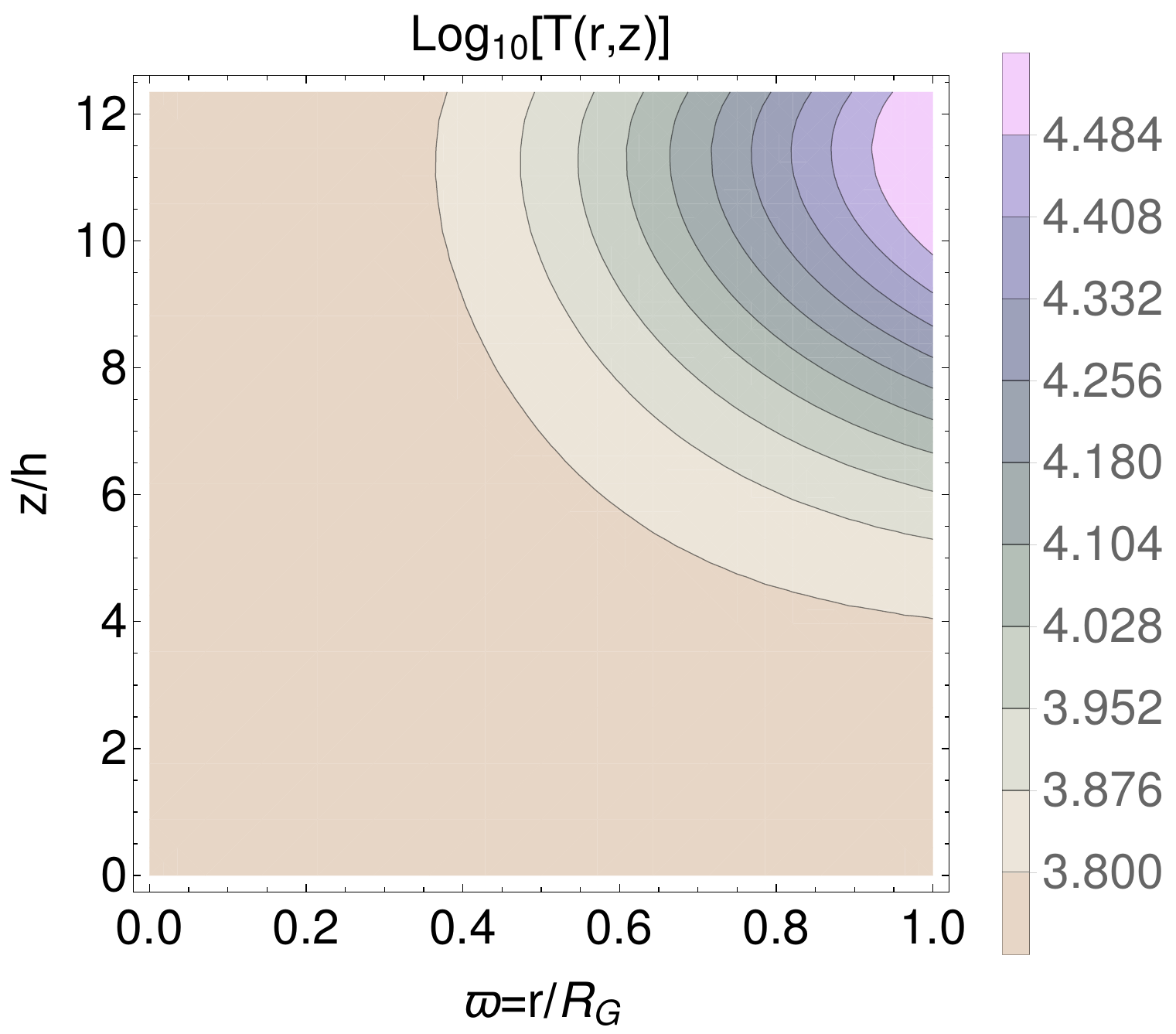}
\includegraphics[scale=0.5]{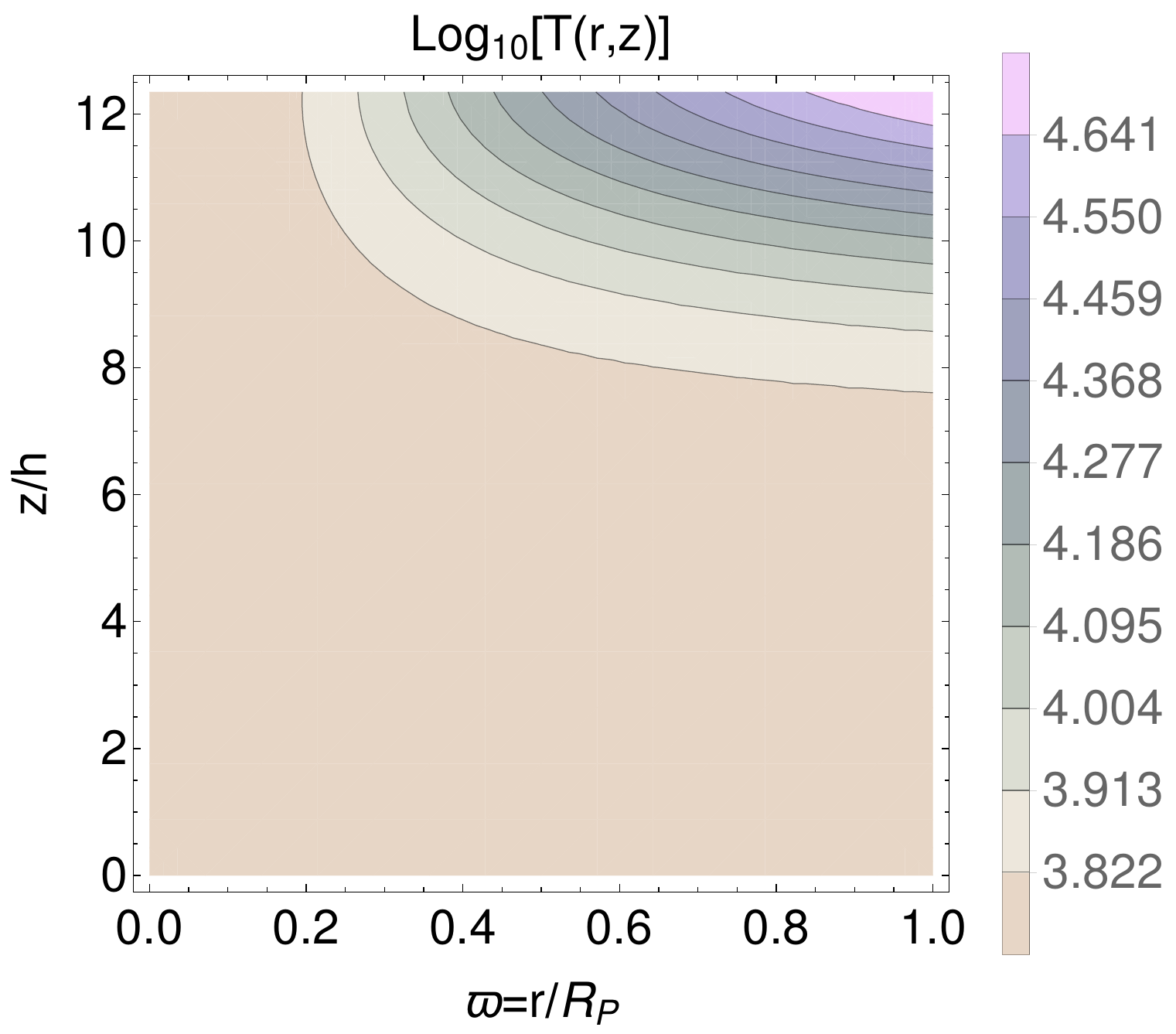}
\caption{The 2D variation of $T$  in the $r-z$ plane for Gaussian ({\it left}) and power law ({\it right}) profiles with $n_P=3$ obtained from the self-similar model for the parameter set $S1$ in Table \ref{mpara_ss}. The horizontal axes are scaled with the total radii $R_G=150$ km and $R_P=130$ km, and the vertical axes are scaled with the pressure scale height, $h=162$ km.} 
\label{ss_Tcp_plot}
\end{center}
\end{figure}

As an example, we show the solution of $\psi^O_S$ and the magnetic and thermodynamic structures for run $S1$ of Table \ref{mpara_ss}. The values of the magnetic and thermodynamic quantities obtained from the self-similar model are reported in Table \ref{td_tab}, for both the Gaussian and power law shape function profiles.  The radial variation of the generalized Gaussian and power law flux function are shown in the Fig. \ref{ss_psir_plot} for different values of $n_G$ and $n_P$, and the variation along the $r-z$ plane is shown in Fig. \ref{ss_psicp}. The $3$D configuration of the field lines for the generalized Gaussian and power law self-similar models are shown in the Figs. \ref{gausstopo_topology} and \ref{powerlawtopo_topology} for the parameter sets for runs $S1$ and $S2$ given in Table \ref{mpara_ss}. The radial and vertical distribution of the magnetic field components are shown in Figs. \ref{ss_B_plot} and \ref{ss_Bzz} respectively for both the Gaussian and power law models, whereas the density inside the flux tube does not vary along $r-$direction but decreases along the $z-$direction which is shown in the Fig. \ref{ss_rho_plot}. The variation of $p$ and $T$ in the $r-z$ plane obtained from the self-similar model are shown in Figs. \ref{ss_pcp_plot} and \ref{ss_Tcp_plot} for Gaussian and power law shape function profiles. Conclusions drawn from Figures [\ref{gse_openfield_topology}, \ref{gse_closedfield_topology}, \ref{gse_tdz_plot}, \ref{gausstopo_topology}, \ref{powerlawtopo_topology}, \ref{ss_rho_plot}, \ref{gse_tdcp_plot}, \ref{ss_pcp_plot}, \ref{ss_Tcp_plot}, \ref{magcanopy}] and Tables [\ref{gse_parameter_tab}, \ref{mpara_ss}, \ref{td_tab}] are discussed in $\S$\ref{discussions}. 

\begin{center}
\begin{table}[hbt!]
\begin{tabular}{|c |c  c c  c c c c c c c c c c c|}\hline \hline
Run \# & $\Psi_b$   & $B_0$  & $p_c$  & $B_{z0}'$ & $\chi$   & $f$ & $n_G$ & $n_P$ & $\psi_b$ &$\bar{f}$ & $\bar{B}'_{z0}$& $\bar{\chi}$ & $R_G$ & $R_P$ \\ 
& $[10^{17}$ Mx]  & [kG] &  [$10^5$ dyne cm$^{-2}$] &  [G km$^{-1}$] &  [cm$^{-2}$]  &  [$10^{-30}$ cm$^{-4}$]&  & & &  & &  & [km] & [km]\\ \hline

$S1$ & $1$ & $1$ & $1$ & $1$ & $10^{-16}$ & $4.56$ & $2$ & $3$ & $0.01$ & $456$ & $1$ & $1$ & $214$ & $261$\\

$S2$ & $1$ & $1$ & $1$ & $1$ & $10^{-14}$ & $4.56$ & $2$ & $3$ & $0.01$ & $456$ & $1$ & $100$ & $214$ & $261$\\

$S3$ & $1$ & $1$ & $0.8$ & $1.5$ & $10^{-18}$ & $8.56$ & $2.5$ & $3.5$ & $0.01$ & $856$ & $1.5$ & $0.01$ & $196$ & $138$\\

$S4$ & $1$ & $1$ & $0.5$ & $2$ & $10^{-16}$ & $14.56$ & $3$ & $4$ & $1$ & $1456$ & $2$ & $1$ & $186$ & $100$\\

$S5$ & $1$ & $2$ & $1$ & $1$ & $10^{-14}$ & $4.56$ & $2$ & $3$ & $0.005$ & $456$ & $0.5$ & $100$ & $151$ & $184$\\

$S6$ & $1$ & $2$ & $0.8$ & $1.5$ & $10^{-18}$ & $8.56$ & $2.5$ & $3.5$ & $0.005$ & $856$ & $0.75$ & $0.01$ & $139$ & $98$\\

$S7$ & $1$ & $2$ & $0.5$ & $2$ & $10^{-16}$ & $14.56$ & $3$ & $4$ & $0.005$ & $1456$ & $0.25$ & $1$ & $131$ & $71$\\

$S8$ & $5$ & $1$ & $1$ & $1$ & $10^{-14}$ & $0.182$ & $2$ & $3$ & $0.005$ & $18.2$ & $1$ & $100$ & $479$ & $584$\\

$S9$ & $5$ & $1$ & $0.8$ & $1.5$ & $10^{-18}$ & $0.342$ & $2.5$ & $3.5$ & $0.005$ & $34.2$ & $1.5$ & $0.01$ & $439$ & $308$\\

$S10$ & $5$ & $1$ & $0.5$ & $2$ & $10^{-16}$ & $0.582$ & $3$ & $4$ & $0.005$ & $58.2$ & $2$ & $1$ & $416$ & $225$\\

$S11$ & $5$ & $2$ & $1$ & $1$ & $10^{-14}$ & $0.182$ & $2$ & $3$ & $0.025$ & $18.2$ & $0.5$ & $100$ & $339$ & $413$\\

$S12$ & $5$ & $2$ & $0.8$ & $1.5$ & $10^{-18}$ & $0.342$ & $2.5$ & $3.5$ & $0.025$ & $34.2$ & $0.75$ & $0.01$ & $310$ & $218$\\

$S13$ & $5$ & $2$ & $0.5$ & $2$ & $10^{-16}$ & $0.582$ & $3$ & $4$ & $0.025$ & $58.2$ & $0.25$ & $1$ & $294$ & $159$\\

$S14$ & $10$ & $1$ & $1$ & $1$ & $10^{-14}$ & $0.0456$ & $2$ & $3$ & $0.1$ & $4.56$ & $1$ & $100$ & $678$ & $826$\\

$S15$ & $10$ & $1$ & $0.8$ & $1.5$ & $10^{-18}$ & $0.0856$ & $2.5$ & $3.5$ & $0.1$ & $8.56$ & $1.5$ & $0.01$ & $621$ & $436$\\

$S16$ & $10$ & $1$ & $0.5$ & $2$ & $10^{-16}$ & $0.145$ & $3$ & $4$ & $0.1$ & $14.56$ & $2$ & $1$ & $589$ & $318$\\

$S17$ & $10$ & $2$ & $1$ & $1$ & $10^{-14}$ & $0.0456$ & $2$ & $3$ & $0.05$ & $4.56$ & $0.5$ & $100$ & $479$ & $584$\\

$S18$ & $10$ & $2$ & $0.8$ & $1.5$ & $10^{-18}$ & $0.0856$ & $2.5$ & $3.5$ & $0.05$ & $8.56$ & $0.75$ & $0.01$ & $439$ & $308$\\

$S19$ & $10$ & $2$ & $2$ & $0.5$ & $10^{-16}$ & $0.145$ & $3$ & $4$ & $0.05$ & $14.56$ & $0.25$ & $1$ & $416$ & $225$\\ \hline
\end{tabular}
\caption{Different combinations of the input parameters and the dimensionless parameters for the self-similar model where $R_G$ and $R_P$ represents the radii of the flux tubes for generalized Gaussian and power law profiles respectively.}
\label{mpara_ss}
\end{table}
\end{center}

\section{Comparing our models with observations}\label{compare}

We compare our models with the observations reported by the high resolution and high cadence instruments. The small scale magnetic structures in the solar photosphere are often found in the forms of the magnetic bright points (MBPs) which are small scale magnetic flux tubes with open field lines \citep{1995ApJ...454..531B, 2007ApJ...666L.137C, 2010ApJ...723L.164L}. Therefore the MBPs are the best candidates to compare our open field flux tube models with the observations. MBPs can be identified by spectro polarimetric measurements or they can be seen by the G-band filtergrams \citep{2009A&A...498..289U, 2013A&A...554A..65U, 2016SoPh..291.1089Y}. Next, we compare the observed magnetic field strength, size and the thermodynamic quantities of the MBPs with that obtained from our models. The MBPs are observed as a region of the unipolar flux concentration, therefore, in the Coulomb function model, we construct a cylindrical boundary of cut-off radius $r_b$ inside the total simulation domain, where the line of sight magnetic field $B_z$ vanishes. The magnetic field strength inside the cylinder of the cut-off radius is always positive. From the recent observations by \cite{2009A&A...498..289U, 2013A&A...554A..65U}, it has been reported that the MBPs number distribution for the size, peaks in the range $160$-$200$ km and the magnetic field strength is at $\sim 1.4$ kG. From the Fig. \ref{gse_B_plot} we see that the $B_z$ vanishes at $r_b=84$ km, where $R=120$ km is the entire radial simulation domain. We choose the parameter range, $1$ kG $\leq B_0\leq$ $1.5$ kG and $100$ km $\leq R\leq$ $180$ km, for which the magnetic and thermodynamic quantities obtained from our model is in reasonable agreement with the solar atmosphere \citep{1981ApJS...45..635V}, and the selection of the $\lbrace R, B_0 \rbrace$ parameter space is also consistent with the observations of MBP size and field strength distributions \citep{2009A&A...498..289U, 2013A&A...554A..65U}. The values of the magnetic and thermodynamic quantities obtained from the Coulomb function open field model is reported in the Table \ref{td_tab}. For the self-similar model, the choice of the parameter space is consistent with the MBPs. We take the flux value in the range of $10^{17}$--$10^{18}$ Mx which is the typical flux value for MBPs \citep{1998SoPh..183..283Z, 1999ApJ...511..932H, 2011IAUS..274..140G}. According to the previous studies by \cite{2010A&A...515A.107S} and SM18, the gas pressure at the axis of MBP is lesser than its boundary gas pressure, so we have chosen the parameter $p_c< p_0$. The field strength of the magnetic footpoints observed in the photosphere for MBPs are $\sim1$ kG with a distribution peak at $1.3$ kG \citep{2013A&A...554A..65U}. Thus, we use the value of $B_0$ in the typical range of $1$--$2$ kG \citep{1998SoPh..183..283Z} in our model. The vertical gradient of the magnetic field strength at the photosphere is $\sim 1$ G-km$^{-1}$ \citep{1974SoPh...36...29W, 1990A&A...228..246P, 1993A&A...279..243B}. Hence we use the value of $B_{z0}'$ in the range of $1$--$2$ G-km$^{-1}$ in our model. We have reported the combinations of the free parameters and the corresponding input parameters in Table \ref{mpara_ss}. Within the parameter sets of runs $S1$--$S19$ in Table \ref{mpara_ss}, we notice that the minimum and maximum radii of the flux tubes  are $151$ and $678$ km respectively for Gaussian model, and $71$ and $826$ km respectively for the power law model, which are in the reasonable agreement  with the observations of MBP size distributions \citep{2009A&A...498..289U}. The values of the magnetic and thermodynamic quantities obtained from the self-similar model are reported in the Table \ref{td_tab}, which is also in a reasonable agreement with the solar atmosphere reported by \cite{1981ApJS...45..635V}.

\begin{figure}[h!]\hspace{35 mm}
\includegraphics[scale=0.6]{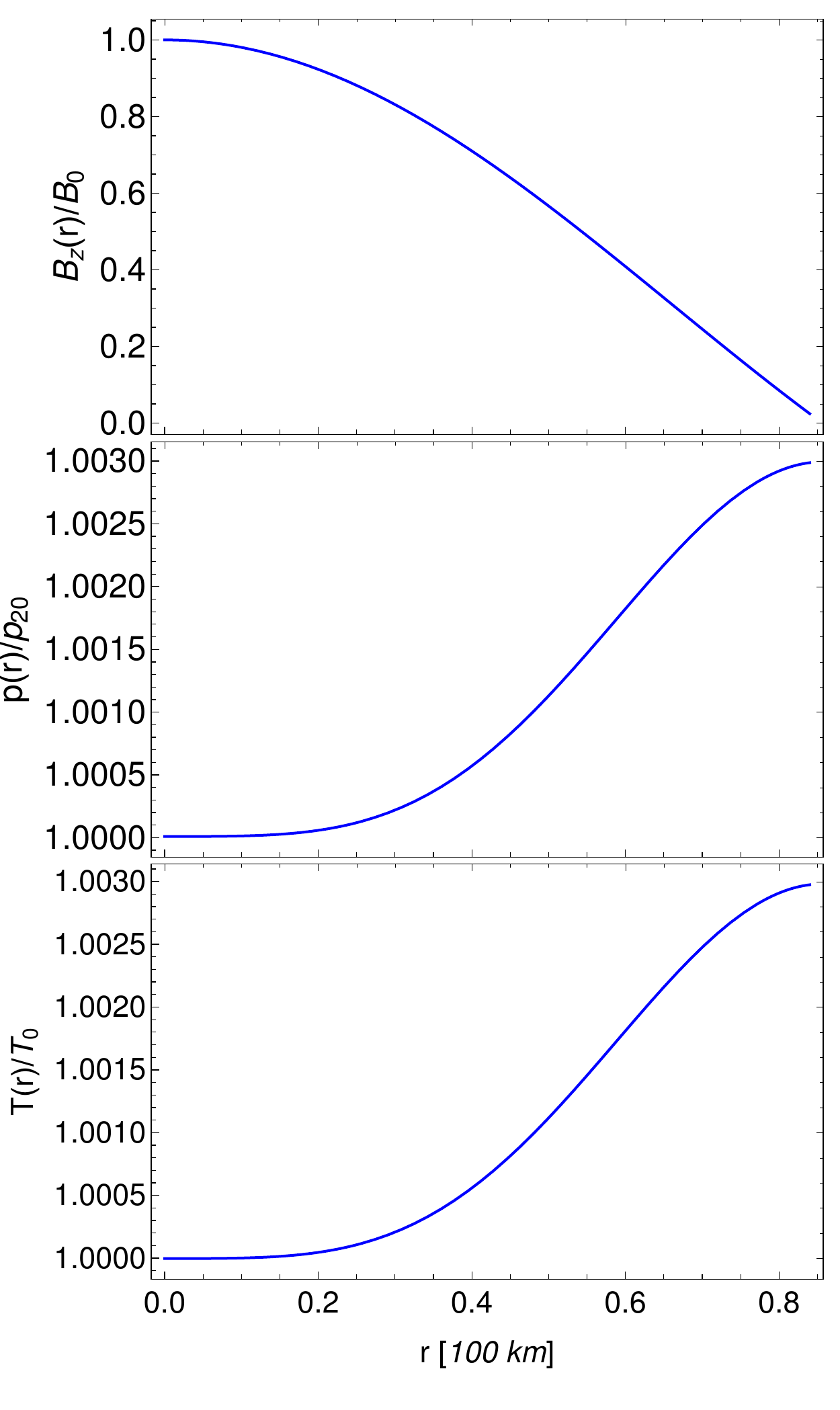}
\caption{The vertical distribution of $B_z$, $p$ and $T$, normalized w.r.t. the value at the center of the flux tube, $B_0$, $p_{20}$ and $T_0$ respectively from the axis of the flux tube to the MBP boundary for $r_b=84$ km at $z=0$, obtained from the Coulomb function open flux tube model for the parameter set of run $C4$ in Table \ref{gse_parameter_tab}. The horizontal axis is scaled in the units of $100$ km and the values of the scale factors are $B_0=1$ kG, $p_{20}=1.03 \times 10^5$ dyne cm$^{-2}$ and $T_0=5656$ K.} 
\label{gse_tdr_plot}
\end{figure}

\begin{table*}[h!]
  \centering
\begin{tabular}{|c | c  c  c  c  c  c| }\hline \hline
Models & $r$ & $z$ [Mm] & $B_z$[G] & $p$ [dyne cm$^{-2}$] & $\rho$ [g cm$^{-3}$] & $T$ [K] \\ \hline    
 & $0$ & $0$ & $1000$ & $1.030 \times 10^5$ & $2.44 \times 10^{-7}$ &  $5656$ \\
Coulomb function open field  & $0$ & $2$ & $2.61$ & $0.234$ & $5.56 \times 10^{-13}$ &  $5656$ \\
& $r_b$ & $0$ & $0$ & $1.04 \times 10^5$ & $2.44 \times 10^{-7}$ &  $5690$ \\ 
& $r_b$ & $2$ & $0$ & $0.2445$ & $5.56 \times 10^{-13}$ &  $5890$ \\  \hline \hline

& $0$ & $0$ & $10^3$ & $1.0 \times 10^5$ & $2.37 \times 10^{-7}$ &  $5630$ \\
Generalized Gaussian & $0$ & $2$ & $3.44$ & $0.227$ & $5.44 \times 10^{-13}$ &  $5630$ \\
& $R_G$ & $0$ & $6.73$ & $1.17 \times 10^5$ & $2.37 \times 10^{-7}$ &  $6620$ \\ 
& $R_G$ & $2$ & $2.2$  & $1.54$ & $5.44 \times 10^{-13}$ &  $38000$ \\
\hline \hline

& $0$ & $0$ & $10^3$ & $1.0 \times 10^5$ & $2.37 \times 10^{-7}$ &  $5630$ \\
Power law & $0$ & $2$ & $75$  & $0.227$ & $5.44 \times 10^{-13}$ &  $5630$ \\
& $R_P$ & $0$ &  $50$ & $1.17 \times 10^5$ & $2.37 \times 10^{-7}$ &  $6620$ \\ 
& $R_P$ & $2$ & $19$ &$1.75$ & $5.44 \times 10^{-13}$ &  $43000$ \\
\hline 

\end{tabular}
\caption{The values of the magnetic field strength and thermodynamic quantities obtained from the Coulomb function open field flux tube model for the parameter set of run $C4$ in Table \ref{gse_parameter_tab}, where $r_b=84$ km; and the self-similar model with Gaussian profile, where $R_G=214$ km, and power law profile, with $n_P=3$, where $R_P=261$ km, for the parameter set of run $S1$ corresponding to Table \ref{mpara_ss} are shown.}
\label{td_tab}
\end{table*}

\begin{table*}
  \resizebox{0.97 \textwidth}{!}{  
\begin{tabular}{|c | c |}\hline \hline
Functions  & Formulae for the Coulomb function model \\ \hline & \\
$s(\varpi)$  &  $\displaystyle{c (F_0\big(-\alpha-\kappa^2,\sqrt{a}\varpi^2\big)+F^*_0\big(-\alpha-\kappa^2,\sqrt{a}\varpi^2\big)}$ \\ 
 \hline & \\
$Z(\bar{z})$ & $\displaystyle{\exp\bigg(-\frac{2\sqrt{2}\kappa a^{1/4}\bar{z}}{\tau}\bigg)}$ \\
 \hline & \\
$\psi_p(\varpi)$ &  $\displaystyle{\frac{i\sqrt{a}b\varpi^2}{4\psi_b}\bigg[  e^{i \sqrt{a}\varpi^2}\sum_{n=0}^{\infty} \frac{F_2^1\big(n+2,1;n+2-i\alpha;\frac{1}{2}\big)(-i\sqrt{a}\varpi^2)^n}{(n+1-i\alpha)n!} -e^{-i \sqrt{a}\varpi^2}\sum_{n=0}^{\infty} \frac{F_2^1\big(n+2,1;n+2+i\alpha;\frac{1}{2}\big)(i\sqrt{a}\varpi^2)^n}{(n+1+i\alpha)n!} \bigg]}$  \\\hline & \\
$\psi_C(\varpi,\bar{z})$  & $s(\varpi) Z(\bar{z})+\psi_p(\varpi)$ \\ \hline & \\
$B_r(\varpi,\bar{z})$ & $\displaystyle{\frac{B_0\sqrt{2}\psi_b \kappa}{a^{1/4}}s(\varpi)Z(\bar{z})}$ \\ \hline & \\
$B_z(\varpi,\bar{z})$ & $\displaystyle{\frac{B_0 \psi_b}{2\sqrt{a}\varpi}[s'(\varpi)Z(\bar{z})+\psi'_p(\varpi)]}$ \\  \hline & \\
$B_\phi(\varpi,\bar{z})$ & $\displaystyle{\frac{B_0\sqrt{2}\alpha^{1/2}\psi_b}{a^{1/4}}[s(\varpi)Z(\bar{z})+\psi_p(\varpi)]}$ \\  \hline & \\
$p(\varpi,\bar{z})$ & $\displaystyle{  B_0^2\bigg[\bigg(\frac{\psi_b^2 s^2(\varpi)}{8\pi}+\bar{p}_{20}\bigg)Z^2(\bar{z})+\bigg(\frac{\psi_b^2 s(\varpi)\psi_p(\varpi)}{4\pi}+\frac{b\psi_b s(\varpi)}{2\sqrt{2a}}\bigg)Z(\bar{z})+\bigg(\frac{\psi_b^2 \psi_p^2}{8\pi}+\frac{b \psi_b \psi_p}{2\sqrt{2a}}\bigg)\bigg]}$ \\ \hline & \\
$\rho(\bar{z})$ & $\displaystyle{\frac{4\sqrt{2}\kappa a^{1/4}p_{20}}{g R} Z^2(\bar{z})}$ \\ \hline & \\
$T(\varpi,\bar{z})$ & $\displaystyle{\frac{\bar{\mu}B_0^2 g R}{4\sqrt{2} R_g \kappa a^{1/4} p_{20}}\bigg[\bigg(\frac{\psi_b^2 s^2(\varpi)}{8\pi}+\bar{p}_{20}\bigg)+\bigg(\frac{\psi_b^2 s(\varpi)\psi_p(\varpi)}{4\pi}+\frac{b\psi_b s(\varpi)}{2\sqrt{2a}}\bigg)\frac{1}{Z(\bar{z})}+\bigg(\frac{\psi_b^2 \psi_p^2}{8\pi}+\frac{b \psi_b \psi_p}{2\sqrt{2a}}\bigg)\frac{1}{Z^2(\bar{z})}\bigg]}$  \\ & \\ \hline  \hline 
Functions  & Formulae for the self-similar model \\  \hline & \\
$\xi(\varpi,\bar{z})$ & $\displaystyle{\sqrt{\frac{\tau}{\psi_b D_0}}\varpi y(\bar{z})}$\\ \hline & \\
$\psi_G(\varpi,\bar{z})$ & $\displaystyle{1-\frac{\Gamma(2/n_G,\xi^{n_G})}{\Gamma(2/n_G)}; \quad (n_G>0)}$\\ \hline  & \\
$\psi_P(\varpi,\bar{z})$ & $\displaystyle{1-(1+\xi)^{1-n_P}\big(1+\xi(n_P-1)\big); \quad (n_P>2)}$ \\ \hline & \\
$B_r(\varpi,\bar{z})$ & $\displaystyle{-\frac{B_0 \varpi}{D_0}y(\bar{z}) y'(\bar{z})D_X(\xi)}$ \\ \hline & \\
$B_z(r,z)$ & $\displaystyle{\frac{B_0}{D_0}y^2(z) D_X(\xi)}$ \\ \hline & \\
$B_\phi(\varpi,\bar{z})$ & $\displaystyle{\frac{\sqrt{\bar{\chi}}B_0}{D_0} \varpi y^2(\bar{z}) D_X(\xi)}$ \\  \hline & \\
$p(\varpi,\bar{z})$ & $\displaystyle{B_0^2\bigg[\frac{\bar{f}}{2} \psi^2+\bar{p}_c e^{-2\bar{k}\bar{z}}\bigg]}$ \\  \hline & \\
$\rho(\bar{z})$ & $\displaystyle{\frac{2\bar{k}\bar{z}\bar{p}_c B_0^2}{g z_0} e^{-2\bar{k}\bar{z}}}$ \\ \hline & \\
$T(\varpi,\bar{z})$ & $\displaystyle{\frac{\bar{\mu}g z_0}{2 R_g\bar{k}\bar{z}\bar{p}_c}\bigg(\frac{\bar{f}}{2}\psi^2 e^{2\bar{k}\bar{z}}+\bar{p}_c\bigg)}$ \\  \hline  
\end{tabular}
}
\caption{ A formulary of different functions obtained for the Coulomb function helical flux tube and self-similar model. Here, $s(\varpi)$, $Z(\bar{z})$ and $\psi_p(\varpi)$ are given by eqns (\ref{S2}, \ref{Z}, \ref{Ap2}), and $\displaystyle{\bar{p}_{20}=p_{20}/B_0^2}$. $\displaystyle{\xi}$ is the self-similar parameter where $y(\bar{z})$ is obtained by solving eqn (\ref{ss_y_eq}) and $\bar{p}_0=p_0/B_0^2$, $\bar{p}_c=p_c/B_0^2$, $\bar{k}=k z_0$. The value of the constants are $\bar{\mu}=1.12$, $g=2.74 \times 10^4$ cm s$^{-2}$, $k=3.4 \times 10^{-8}$ cm$^{-1}$ and $z_0=10^8$ cm.}
\label{func_models}
\end{table*}

\section{Discussion of the models}\label{discussions}

We discuss the findings of our simulations below:
\begin{itemize}
\item The Coulomb function model is easier to implement for the numerical studies as it consists of two free parameters ($R$, $B_0$); on the other hand, the self-similar model consists of five free parameters ($\Psi_b$, $B_0$,  $p_c$, $B'_{z0}$ and $\chi$). From Table \ref{td_tab}, we see that the rise of the gas pressure along the radial direction from axis to the boundary is higher for the self-similar model than the Coulomb function model at higher $z$. The density within the flux tube does not vary with $r$; hence, the rise of the temperature from axis to the boundary at higher $z$ is also higher for the self-similar model relative to the Coulomb function model. For the Coulomb function model, the radial boundary of the flux tube is defined where $B_z$ vanishes; on the other hand, for the self-similar model, $B_z(R) \neq 0$, whereas $B_z$ reduces along the radial direction from axis to the boundary for the Gaussian model faster than the Power law model. 

\item The radial size and the magnetic field strength at the center of the flux tube are the free parameters in the Coulomb function model. The magnetic and thermodynamic structure of the flux tube remain similar for different values of the free parameters, whereas the magnitude of the magnetic and thermodynamic quantities vary. We have explored the parameter space and notice that, in the domain of $100$ km $\leq R \leq$ $180$ km, and $1$ kG $\leq B_0 \leq$ $1.5$ kG, the magnetic and thermodynamic quantities are in reasonable agreement with the solar atmosphere \citep{1981ApJS...45..635V}, which also validate the MBP size and magnetic field strength distribution \citep{2009A&A...498..289U, 2013A&A...554A..65U}. For the self-similar model, the radial sizes of the flux tubes depend on the choice of the dimensionless input parameters $\lbrace \psi_b, \bar{f}, \bar{B'_{z0}}, \bar{\chi} \rbrace$. In the domain of the selected parameter space (see Table \ref{mpara_ss}), the maximum and minimum radii of the flux tubes are $678$ and $151$ km obtained from the Gaussian model, whereas for the power law model with $n_P=3$, the maximum and minimum values of the radii are calculated to be $826$ and $184$ km respectively which are also in reasonable agreement with the observation of MBP size distribution by \cite{2009A&A...498..289U}.

\item For the Coulomb function model, we notice that the value of $\alpha$ decreases with $R$ (see Table \ref{gse_parameter_tab}), which lowers the poloidal current $I_p$ and the twist of the field lines. The $3$D geometry of the field lines for different twists are shown in the Figs. \ref{gse_openfield_topology} and \ref{gse_closedfield_topology} for open and closed field Coulomb function models respectively. In the self-similar model, the twist of the field lines increases with $\bar{\chi}$ and are shown in Figs. \ref{gausstopo_topology} and \ref{powerlawtopo_topology} for Gaussian and power law profiles respectively, which follows from eqn (\ref{flux_functionGP}).

\item The gas pressure for both Coulomb function and self-similar models increases along radial direction from axis to the boundary, whereas it decreases along the vertical height from photosphere to the transition region (see Figs. \ref{gse_tdcp_plot} and \ref{ss_pcp_plot}) which is similar to the result obtained by \cite{2010A&A...515A.107S} for MBPs, where the gas pressure inside the MBPs increases radially though the change is not significant, and decreases vertically. \cite{2013MNRAS.435..689G, 2014ApJ...789...42G} have studied for the cases of single and multiple flux tubes, where the internal gas pressure is nearly same along the radial distance but decreases with height. The density within the flux tube does not change radially but it decreases along $z$, for both Coulomb function and self-similar models (see Figs. \ref{gse_tdz_plot} and \ref{ss_rho_plot}).  Our model predicts that the atmosphere inside the flux tube is nearly plane parallel which is comparable to the model obtained for MBPs by \cite{2010A&A...515A.107S}.

\item In the solar atmosphere, the temperature in the transition region rises perhaps because the shock dissipation of waves play a dominant role, which is not included in our model. We have also not implemented the temperature profile by \cite{1981ApJS...45..635V} (VAL model); however our model is self consistent, obtained by solving the GSE without shock heating. Therefore, we do not see the drastic rise of the temperature with height. Our vertical simulation domain is restricted from the photosphere to the transition region where our input external atmosphere model is valid. Both the flux tube models we built are non-isothermal where the temperature increases along the radial direction for both Coulomb function open field and self-similar models. The vertical variation of the temperature is constant at the axis but it increases with height away from the axis for the Coulomb function open field and self-similar models (see Figs. \ref{gse_tdcp_plot} and \ref{ss_Tcp_plot}).

\item \cite{2014A&A...565A..84H}, \cite{2013MmSAI..84..369U}, \cite{2016arXiv161207887R} have reported the simulation results of MBPs by using MuRAM and Copenhagen-Stagger code where the obtained values of magnetic field strength, pressure, density and temperature inside the flux tube are in reasonable agreement with our predictions.

\item The 2D simulations of the propagation of linear and non-linear magneto acoustic wave through an open magnetic flux tube, embedded in the solar atmosphere from photosphere to corona were carried out by \cite{2011ApJ...740L..46F}. We can incorporate our solutions as the background condition for such numerical studies of waves and their kinematic properties taking realistic inputs of field strength and pressure distribution observed in the solar atmosphere.

\item The Coulomb function model gives both open and closed field flux tube solutions, which can be co-added to build the canopy structure. A cartoon diagram of the magnetic canopy is shown in Fig. \ref{magcanopy}, where the closed field lines ({\it red}), $\Psi^C_C$, are present between the open field flux tubes and obtained from the Coulomb function, where the open field lines ({\it blue}), $\Psi^O_C$ and $\Psi^O_S$, of the neighboring flux tubes merge to each other to form a canopy structure. This is similar to structures assumed in the numerical simulations by \cite{2014ApJ...789...42G}, constructed by a different self-similar flux tube solution. We can use our solutions for inputs to simulations to build such canopy structures. The self-similar flux tube model gives an open field structure of the flux tube which is embedded in a continuous magnetic medium and span upto  infinity in the radial direction. The magnetic and thermodynamic quantities we estimated from both Coulomb function and self-similar models are nearly similar, whereas there are some differences in the structures of the magnetic and thermodynamic profiles.

\end{itemize}

Future advancement of the observations of magnetic and thermodynamic structures of the MBPs will provide better selection of the parameter inputs and discriminate between our models. 

\begin{figure}[h!]
\begin{center}
\includegraphics[scale=0.6]{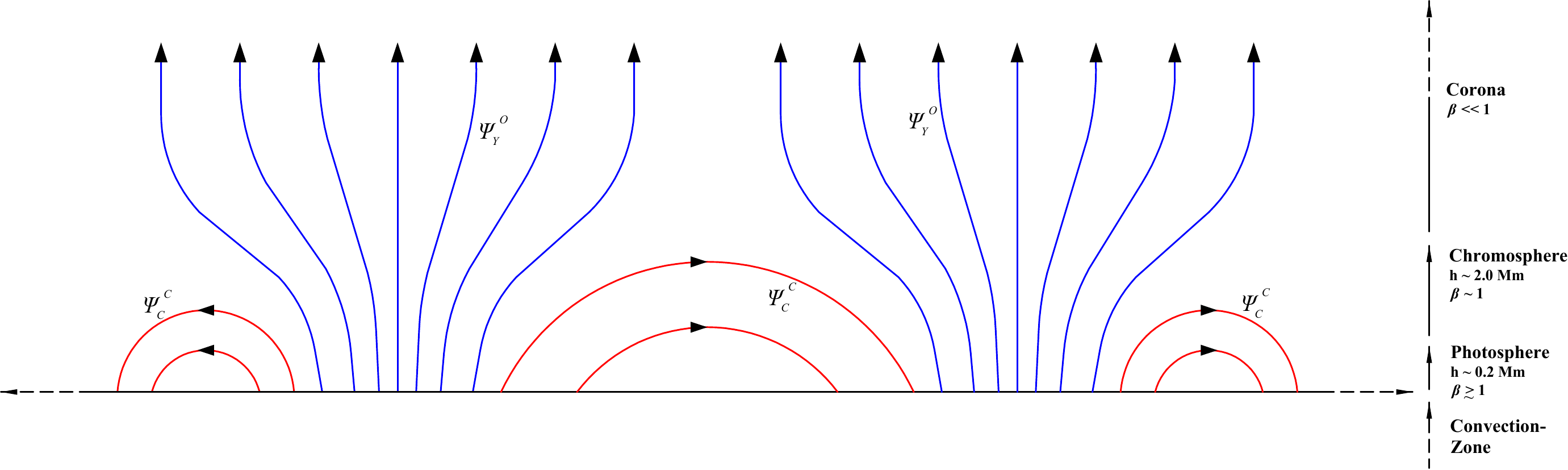}
\caption{A cartoon diagram of magnetic canopy structure is shown, where the closed field lines ({\it red}), which is obtained by the Coulomb function closed field solution, $\Psi^C_C$, rise and fall back in the photosphere, present between two open field flux tubes. The open field lines ({\it blue}), obtained by Coulomb function open field and self-similar solution, $\Psi^O_Y$ ($Y=C$, for Coulomb function and $Y=S$ for self-similar models), of two neighboring flux tubes merge together to form magnetic canopy structure [An improvised  version of the illustration in \cite{2006ASPC..354..259J}].}
\label{magcanopy}
\end{center}
\end{figure}

\section{Summary and Conclusions}\label{summary}

In this work, we have constructed two different models of flux tubes with twisted magnetic fields which are the Coulomb function helical flux tube model and self-similar model by solving GSE semi-analytically. We tabulate the expressions of magnetic and thermodynamic functions for Coulomb and self-similar models in Table \ref{func_models}, and highlight the novel features of this work below.

\begin{enumerate}
\item By incorporating the form of gas pressure and poloidal current we have solved GSE to obtain the flux function for the Coulomb function model. The solution of the Coulomb function model is the combination of a homogeneous part and a particular part. The homogeneous part with closed geometry is separable with a Coulomb function in $r$ whereas the $z$ part decreases exponentially with height, and  the particular part with open geometry is a power series of $r$ which is independent of $z$. 

\item Using appropriate BCs and employing the presence of the sheet current at the boundary of the flux tube, we have determined the parameters $a(R)$, $\alpha(R)$, $\kappa(R)$, $b(R,B_0)$ , $\psi_b(R,B_0)$ and $\bar{p}(R,B_0)$ in terms of the input parameters $\lbrace R, B_0 \rbrace$, which are the free parameters in the model, and $k$ is calculated from the pressure values at photosphere and transition region obtained from \cite{2008ApJS..175..229A} model. The values of the parameters for Coulomb function model are listed in Table \ref{gse_parameter_tab}.

\item In the Coulomb function model, the solution consisting of homogeneous and particular parts together represents an open field flux tube solution, where the field lines rise from the photosphere. The  homogeneous solution depicts a closed field flux tube model which is discussed in SM18. The values of the magnetic field strength and thermodynamic quantities inside the flux tube are calculated and are summarized in Table \ref{td_tab}. The 3D visualization of both open and closed field lines are shown in the Figs. \ref{gse_openfield_topology} and \ref{gse_closedfield_topology} for the parameter set of run no. $C4$ and $C10$ corresponding to Table \ref{gse_parameter_tab}.

\item In the self-similar model, we have employed an extra term $p_c \exp(-2kz)$ with $p_1$ in eqn (\ref{ss_p1}), to maintain the hydrostatic pressure balance under the influence of stratified solar gravity, and taken two options for the shape functions, $D_X(\xi)$ from eqn (\ref{DgP}), which is the extension of previous models by ST$58$; \cite{ 1971SoPh...16..398Y, 1979SoPh...64..261O, 1980SoPh...77..63O}. We have incorporated the resulting two different shape functions, generalized Gaussian and power law profiles, to obtain open field flux tube solutions. We have taken a range of the parameters $\Psi_b$, $B_0$, $p_c$, $B'_{z0}$ and $\chi$ (see Table \ref{mpara_ss}), that are consistent with the solar atmosphere to study the structure and the properties of the flux tubes. The size of the flux tubes and the magnitude of the thermodynamic and magnetic field strengths depend on the choice of the input parameters, but the magnetic and thermodynamic structures remain similar. We have calculated the magnetic field strength and the thermodynamic quantities inside the flux tube which are given in Table \ref{td_tab}, for the parameter set of run no. $S1$ corresponding to Table \ref{mpara_ss}. 

\item Preliminary calculations using the constraint of relative helicity based on the formulations given in \citep{2014ApJ...786...81P, 2016ApJ...817...12P} and applying the constrained energy minimization principle \citep{2000JApA...21..299M, 1983PhFl...26.3540F, 1974PhRvL..33.1139T} indicate that stable configurations are possible for some regions in the parameter space of $\lbrace B_0, R \rbrace$. We plan a complete solution of this allowed region and test it with numerical simulations in a paper in preparation. \\

The flux tube models presented here give useful estimates of the magnitude and the distribution of the magnetic field strength and thermodynamic quantities from the photosphere to the transition region which can be verified by the future observations. Work on self-similar closed and twisted field structure is in progress. The solutions we obtained for different flux tubes can be used for the dynamical simulation of wave propagation through the flux tubes, which is important for studying the coronal heating by waves.

\end{enumerate}

We thank V. Fedun, R. Erd\'{e}lyi and S. Shelyag for useful discussions. We thank the anonymous referee for insightful comments and helpful suggestions. We also thank the support staff of the IIA HPC facility, VBO (IIA, Kavalur) staff for hospitality during our visits, and Saikat Das for help with Figs. \ref{tube} and \ref{magcanopy}.

\appendix

\section{Derivation of the explicit form of $B_z$ for Coulomb function model}
\label{app_Bz}
The homogeneous solution $s(\varpi)$ which is given by the eqn (\ref{S2}), can be represented in terms of the Whittaker--M function (SM18), where the Whittaker--M function can be expressed in terms of hypergeometric function by the standard relation \citep{2015msa...26...49}
\begin{align}\label{wfm}
M_{t,m}(\nu)=e^{-\nu/2} \nu^{m+\frac{1}{2}} F_1^1\bigg(\frac{1}{2}+m-t,1+2m,\nu\bigg),
\end{align}
where $F_1^1$ represents the hypergeometric function with the arguments $t,m$ and $\nu$. Taking the real part of $\psi_h(\varpi,{\bar z})$ from eqn (\ref{Ah}) and $\psi_p(\varpi)$ from (\ref{Ap2}), and using eqns [(\ref{Bz2}), (\ref{wfm})] we obtain 
\begin{align}\label{Bz3}
B_z(\varpi,{\bar z})=& B_0 \psi_b c \exp\bigg(-\frac{2\sqrt{2}\kappa a^{1/4} {\bar z}}{\tau}\bigg) \bigg[ 8(1+i\sqrt{a}\varpi^2)F_1^1(1+i\alpha,2,2i\sqrt{a}\varpi^2)-\\ \nonumber & 8\sqrt{a}\varpi^2F_1^1(1+i\alpha,3,2i\sqrt{a}\varpi^2)\bigg]+\frac{B_0\psi_b}{2\sqrt{a}}\bigg(\frac{\psi'_p(\varpi)+\psi^{'*}_p(\varpi)}{2\varpi}\bigg) 
\end{align}
where, $B_0 \equiv B_z(0,0)$, and  from eqn (\ref{Bz3}) we obtain, 
\begin{align}\label{B0}
1=8\psi_b c+\frac{\psi_b}{2\sqrt{a}}\bigg[\frac{\psi_p'(\varpi)+\psi^{'*}_p(\varpi)}{2\varpi}\bigg]_{\varpi=0},
\end{align}
where the identity $\displaystyle{F_1^1(1+i\alpha,2,0)}=1$ is applied. By expanding the last term on the RHS of eqn (\ref{B0}), we obtain
\begin{align}\label{B02}
\displaystyle{1=8 \psi_b c+\frac{ib}{2}\bigg[ \frac{F_2^1(1,-i\alpha,2-i\alpha,-1)}{1-i \alpha} - \frac{F_2^1(1,i\alpha,2+i\alpha,-1)}{1+i \alpha}\bigg]}
\end{align} 
and the expression for $c$ is given by
\begin{align}\label{C}
c=\frac{1}{8\psi_b}\bigg[1-\frac{ib}{2}\bigg(\frac{F_2^1(1,-i\alpha,2-i\alpha,-1)}{1-i \alpha} - \frac{F_2^1(1,i\alpha,2+i\alpha,-1)}{1+i \alpha}\bigg)\bigg].
\end{align}
Hence from eqn (\ref{Bz3}), we obtain the explicit form for $B_z(\varpi,\bar{z})$, given in eqn (\ref{coulomb_Bz_explicit}). 
\section{Explicit forms of the BCs for Coulomb function model}
\label{app_explicit_bcs}   
The explicit forms of the eqns (\ref{bc1}--\ref{bc4}) are 
\begin{align}
 \label{bc1_explicit}
& F_0(-\alpha-\kappa^2,\sqrt{a})+F^*_0(-\alpha-\kappa^2,\sqrt{a})=0, \\ \nonumber \label{bc2_explicit}
& \displaystyle{\bigg[\frac{1}{\varpi^2}\frac{{\rm d}}{{\rm d}\varpi} \bigg(\varpi^2e^{i\sqrt{a}\varpi^2}\sum_{n=0}^\infty \frac{F_2^1(n+2,1,n+2-i\alpha,1/2)(-i\sqrt{a}\varpi^2)^n}{(n+1-i\alpha)n!}  } \nonumber \\  & \displaystyle{-\varpi^2e^{-i\sqrt{a}\varpi^2}\sum_{n=0}^\infty \frac{F_2^1(n+2,1,n+2+i\alpha,1/2)(i\sqrt{a}\varpi^2)^n}{(n+1+i\alpha)n!}\bigg)\bigg]_{\varpi=1}=0, }   
 \\ \label{bc3_explicit} 
&\displaystyle{ \bigg[1-\frac{ib}{2}\bigg(\frac{F_2^1(1,-i\alpha,2-i\alpha,-1)}{1-i\alpha}-\frac{F_2^1(1,i\alpha,2+i\alpha,-1)}{1=i\alpha}\bigg)\bigg]^2 } \nonumber \\ 
& \displaystyle{\cdot \bigg( \frac{{\rm d}}{{\rm d}\varpi}\bigg[F_0(\alpha-\kappa^2,\sqrt{a}\varpi^2)+F^*_0(\alpha-\kappa^2,\sqrt{a}\varpi^2)\bigg]_{\varpi=1}\bigg)^2 =\bar{p} a, } 
\\ \label{bc4_explicit}
& \displaystyle{b=-\frac{\psi_b}{2}\bigg(1+\frac{6\alpha}{\sqrt{a}}\bigg),}\\ \nonumber \label{bc5_explicit}
& \displaystyle{\int_0^1\bigg[e^{-2\kappa_t}\bigg(1-\frac{i b}{2}\bigg[\frac{F_2^1(1,-i\alpha,2-i\alpha,-1)}{1-i\alpha}-\frac{F_2^1(1,i\alpha,2+i\alpha,-1)}{1+i\alpha}\bigg]\bigg)^2}\\
 \nonumber &  \displaystyle{\cdot \bigg(F_0(-\alpha-\kappa^2,\sqrt{a}\varpi^2)+F^*_0(-\alpha-\kappa^2,\sqrt{a}\varpi^2)\bigg)^2+ 8e^{-\kappa_t}\psi_b \big(\psi_p+\psi^*_p+\frac{2b}{\psi_b}\big)}\\ 
 \nonumber  &  \displaystyle{\cdot \bigg(1-\frac{ib}{2}\bigg[\frac{F_2^1(1,-i\alpha,2-i\alpha,-1)}{1-\alpha}-\frac{F_2^1(1,i\alpha,2+i\alpha,-1)}{1+\alpha}\bigg]\bigg)^2} \\  
 &  \displaystyle{\cdot \bigg(F_0(-\alpha-\kappa^2,\sqrt{a}\varpi^2)+F^*_0(-\alpha-\kappa^2,\sqrt{a}\varpi^2)\bigg)+\frac{\psi_b^2}{4}\big(\psi_p+\psi^*_p\big)^2+b\psi_b (\psi_p+\psi^*_p)\bigg]{\rm d}\varpi=\bar{p} }
\end{align}
respectively, where, $\psi_p(\varpi)$ is given by eqn (\ref{Ap2}), $\kappa_t= k z_t$, and $\psi_p(\varpi=1)=\psi_b$.

\bibliography{paper2}

\begin{thebibliography}{}
\expandafter\ifx\csname natexlab\endcsname\relax\def\natexlab#1{#1}\fi
\providecommand{\url}[1]{\href{#1}{#1}}

\bibitem[{{Abramowitz} \& {Stegun}(1972)}]{abramowitz+stegun}
{Abramowitz}, M., \& {Stegun}, I.~A. 1972, Handbook of Mathematical Functions
  with Formulas, Graphs, and Mathematical Tables, ninth dover printing, tenth
  gpo printing edn. (New York City: Dover)

\bibitem[{{Aschwanden} {et~al.}(2000){Aschwanden}, {Nightingale}, \&
  {Alexander}}]{2000ApJ...541.1059A}
{Aschwanden}, M.~J., {Nightingale}, R.~W., \& {Alexander}, D. 2000, \apj, 541,
  1059

\bibitem[{Atanasiu {et~al.}(2004)Atanasiu, Günter, Lackner, \&
  Miron}]{doi:10.1063/1.1756167}
Atanasiu, C.~V., Günter, S., Lackner, K., \& Miron, I.~G. 2004, Physics of
  Plasmas, 11, 3510.
\newblock \url{http://aip.scitation.org/doi/abs/10.1063/1.1756167}

\bibitem[{{Avrett} \& {Loeser}(2008)}]{2008ApJS..175..229A}
{Avrett}, E.~H., \& {Loeser}, R. 2008, \apjs, 175, 229

\bibitem[{{Balthasar} \& {Schmidt}(1993)}]{1993A&A...279..243B}
{Balthasar}, H., \& {Schmidt}, W. 1993, \aap, 279, 243

\bibitem[{{Berger} {et~al.}(1995){Berger}, {Schrijver}, {Shine}, {Tarbell},
  {Title}, \& {Scharmer}}]{1995ApJ...454..531B}
{Berger}, T.~E., {Schrijver}, C.~J., {Shine}, R.~A., {et~al.} 1995, \apj, 454,
  531

\bibitem[{{Centeno} {et~al.}(2007){Centeno}, {Socas-Navarro}, {Lites}, {Kubo},
  {Frank}, {Shine}, {Tarbell}, {Title}, {Ichimoto}, {Tsuneta}, {Katsukawa},
  {Suematsu}, {Shimizu}, \& {Nagata}}]{2007ApJ...666L.137C}
{Centeno}, R., {Socas-Navarro}, H., {Lites}, B., {et~al.} 2007, \apjl, 666,
  L137

\bibitem[{{Dixit} \& {Moll}(2015)}]{2015msa...26...49}
{Dixit}, A., \& {Moll}, V. 2015, \msa, 26, 49

\bibitem[{{Fedun} {et~al.}(2009){Fedun}, {Erd{\'e}lyi}, \&
  {Shelyag}}]{2009SoPh..258..219F}
{Fedun}, V., {Erd{\'e}lyi}, R., \& {Shelyag}, S. 2009, \solphys, 258, 219

\bibitem[{{Fedun} {et~al.}(2011){Fedun}, {Verth}, {Jess}, \&
  {Erd{\'e}lyi}}]{2011ApJ...740L..46F}
{Fedun}, V., {Verth}, G., {Jess}, D.~B., \& {Erd{\'e}lyi}, R. 2011, \apjl, 740,
  L46

\bibitem[{{Finn} \& {Antonsen}(1983)}]{1983PhFl...26.3540F}
{Finn}, J.~M., \& {Antonsen}, Jr., T.~M. 1983, Physics of Fluids, 26, 3540

\bibitem[{{Gent} {et~al.}(2014){Gent}, {Fedun}, \&
  {Erd{\'e}lyi}}]{2014ApJ...789...42G}
{Gent}, F.~A., {Fedun}, V., \& {Erd{\'e}lyi}, R. 2014, \apj, 789, 42

\bibitem[{{Gent} {et~al.}(2013){Gent}, {Fedun}, {Mumford}, \&
  {Erd{\'e}lyi}}]{2013MNRAS.435..689G}
{Gent}, F.~A., {Fedun}, V., {Mumford}, S.~J., \& {Erd{\'e}lyi}, R. 2013,
  \mnras, 435, 689

\bibitem[{{Grad} \& {Rubin}(1958)}]{1958conf...21...190}
{Grad}, H., \& {Rubin}, H. 1958, {Hydromagnetic Equlibria and Force-Free
  Fields} (U.S. Government Printing Office, International Conference on the
  Peaceful Uses of Atomic Energy, Geneva, 31, 190-197)

\bibitem[{{Guglielmino} {et~al.}(2011){Guglielmino}, {Pillet}, {del Toro
  Iniesta}, {Rubio}, {Zuccarello}, {Solanki}, \&
  {Solanki}}]{2011IAUS..274..140G}
{Guglielmino}, S.~L., {Pillet}, V.~M., {del Toro Iniesta}, J.~C., {et~al.}
  2011, in IAU Symposium, Vol. 274, Advances in Plasma Astrophysics, ed.
  A.~{Bonanno}, E.~{de Gouveia Dal Pino}, \& A.~G. {Kosovichev}, 140--142

\bibitem[{{Hagenaar} {et~al.}(1999){Hagenaar}, {Schrijver}, {Title}, \&
  {Shine}}]{1999ApJ...511..932H}
{Hagenaar}, H.~J., {Schrijver}, C.~J., {Title}, A.~M., \& {Shine}, R.~A. 1999,
  \apj, 511, 932

\bibitem[{{Hewitt} {et~al.}(2014){Hewitt}, {Shelyag}, {Mathioudakis}, \&
  {Keenan}}]{2014A&A...565A..84H}
{Hewitt}, R.~L., {Shelyag}, S., {Mathioudakis}, M., \& {Keenan}, F.~P. 2014,
  \aap, 565, A84

\bibitem[{{Judge}(2006)}]{2006ASPC..354..259J}
{Judge}, P. 2006, in Astronomical Society of the Pacific Conference Series,
  Vol. 354, Solar MHD Theory and Observations: A High Spatial Resolution
  Perspective, ed. J.~{Leibacher}, R.~F. {Stein}, \& H.~{Uitenbroek}, 259

\bibitem[{{Lagg} {et~al.}(2010){Lagg}, {Solanki}, {Riethm{\"u}ller},
  {Mart{\'{\i}}nez Pillet}, {Sch{\"u}ssler}, {Hirzberger}, {Feller}, {Borrero},
  {Schmidt}, {del Toro Iniesta}, {Bonet}, {Barthol}, {Berkefeld}, {Domingo},
  {Gandorfer}, {Kn{\"o}lker}, \& {Title}}]{2010ApJ...723L.164L}
{Lagg}, A., {Solanki}, S.~K., {Riethm{\"u}ller}, T.~L., {et~al.} 2010, \apjl,
  723, L164

\bibitem[{{Mangalam} \& {Krishan}(2000)}]{2000JApA...21..299M}
{Mangalam}, A., \& {Krishan}, V. 2000, Journal of Astrophysics and Astronomy,
  21, 299

\bibitem[{{Muller} \& {Mena}(1987)}]{muller1987}
{Muller}, R., \& {Mena}, B. 1987, \solphys, 112, 295

\bibitem[{{Muller} {et~al.}(1994){Muller}, {Roudier}, {Vigneau}, \&
  {Auffret}}]{1994A&A...283..232M}
{Muller}, R., {Roudier}, T., {Vigneau}, J., \& {Auffret}, H. 1994, \aap, 283,
  232

\bibitem[{{Murawski} {et~al.}(2015){Murawski}, {Solov'ev}, {Musielak},
  {Srivastava}, \& {Kra{\'s}kiewicz}}]{2015A&A...577A.126M}
{Murawski}, K., {Solov'ev}, A., {Musielak}, Z.~E., {Srivastava}, A.~K., \&
  {Kra{\'s}kiewicz}, J. 2015, \aap, 577, A126

\bibitem[{{Osherovitch}(1979)}]{1979SoPh...64..261O}
{Osherovitch}, V.~A. 1979, \solphys, 64, 261

\bibitem[{{Osherovitch}(1982)}]{1980SoPh...77..63O}
---. 1982, \solphys, 77, 63

\bibitem[{{Pahlke} \& {Wiehr}(1990)}]{1990A&A...228..246P}
{Pahlke}, K.-D., \& {Wiehr}, E. 1990, \aap, 228, 246

\bibitem[{{Parker}(1988)}]{1988ApJ...330..474P}
{Parker}, E.~N. 1988, \apj, 330, 474

\bibitem[{{Peter} {et~al.}(2005){Peter}, {Gudiksen}, \&
  {Nordlund}}]{2005ESASP.596E..14P}
{Peter}, H., {Gudiksen}, B.~V., \& {Nordlund}, A. 2005, in ESA Special
  Publication, Vol. 596, Chromospheric and Coronal Magnetic Fields, ed. D.~E.
  {Innes}, A.~{Lagg}, \& S.~A. {Solanki}, 14.1

\bibitem[{{Prasad} \& {Mangalam}(2016)}]{2016ApJ...817...12P}
{Prasad}, A., \& {Mangalam}, A. 2016, \apj, 817, 12

\bibitem[{{Prasad} {et~al.}(2014){Prasad}, {Mangalam}, \&
  {Ravindra}}]{2014ApJ...786...81P}
{Prasad}, A., {Mangalam}, A., \& {Ravindra}, B. 2014, \apj, 786, 81

\bibitem[{{Riethm{\"u}ller} \& {Solanki}(2016)}]{2016arXiv161207887R}
{Riethm{\"u}ller}, T.~L., \& {Solanki}, S.~K. 2016, ArXiv e-prints,
  arXiv:1612.07887

\bibitem[{{Ruzmaikin} \& {Berger}(1998)}]{1998A&A...337L...9R}
{Ruzmaikin}, A., \& {Berger}, M.~A. 1998, \aap, 337, L9

\bibitem[{Schl{\"u}ter \& Temesv{\'a}ry(1958)}]{Schluter_Temesvary_1958}
Schl{\"u}ter, A., \& Temesv{\'a}ry, S. 1958, Symposium - International
  Astronomical Union, Cambridge: Cambridge University Press, 6, 263

\bibitem[{{Sen} \& {Mangalam}(2018)}]{2018AdSpR..61..617S}
{Sen}, S., \& {Mangalam}, A. 2018, Advances in Space Research, 61, 617

\bibitem[{{Shafranov}(1958)}]{1958jetp...710...33}
{Shafranov}, V. 1958, \jetp, 33, 710

\bibitem[{{Shelyag} {et~al.}(2010){Shelyag}, {Mathioudakis}, {Keenan}, \&
  {Jess}}]{2010A&A...515A.107S}
{Shelyag}, S., {Mathioudakis}, M., {Keenan}, F.~P., \& {Jess}, D.~B. 2010,
  \aap, 515, A107

\bibitem[{{Solov'ev} \& {Kirichek}(2015)}]{2015AstL...41..211S}
{Solov'ev}, A.~A., \& {Kirichek}, E.~A. 2015, Astronomy Letters, 41, 211

\bibitem[{{Solov'ev}(1968)}]{1968JETP...26..2S}
{Solov'ev}, L. 1968, \textit{Journal of Experimental and Theoretical Physics},
  26, 626

\bibitem[{{Srivastava} {et~al.}(2017){Srivastava}, {Shetye}, {Murawski},
  {Doyle}, {Stangalini}, {Scullion}, {Ray}, {W{\'o}jcik}, \&
  {Dwivedi}}]{2017NatSR...743147S}
{Srivastava}, A.~K., {Shetye}, J., {Murawski}, K., {et~al.} 2017, Scientific
  Reports, 7, 43147

\bibitem[{{Steiner} {et~al.}(1986){Steiner}, {Pneuman}, \&
  {Stenflo}}]{1986A&A...170..126S}
{Steiner}, O., {Pneuman}, G.~W., \& {Stenflo}, J.~O. 1986, \aap, 170, 126

\bibitem[{{Stepanov}(1965)}]{1965IAUS...22..267S}
{Stepanov}, V.~E. 1965, in IAU Symposium, Vol.~22, Stellar and Solar Magnetic
  Fields, ed. R.~{Lust}, 267

\bibitem[{{Taylor}(1974)}]{1974PhRvL..33.1139T}
{Taylor}, J.~B. 1974, Physical Review Letters, 33, 1139

\bibitem[{{Thalmann} {et~al.}(2013){Thalmann}, {Tiwari}, \&
  {Wiegelmann}}]{2013ApJ...769...59T}
{Thalmann}, J.~K., {Tiwari}, S.~K., \& {Wiegelmann}, T. 2013, \apj, 769, 59

\bibitem[{{Uitenbroek} \& {Criscuoli}(2013)}]{2013MmSAI..84..369U}
{Uitenbroek}, H., \& {Criscuoli}, S. 2013, \memsai, 84, 369

\bibitem[{{Utz} {et~al.}(2009){Utz}, {Hanslmeier}, {M{\"o}stl}, {Muller},
  {Veronig}, \& {Muthsam}}]{2009A&A...498..289U}
{Utz}, D., {Hanslmeier}, A., {M{\"o}stl}, C., {et~al.} 2009, \aap, 498, 289

\bibitem[{{Utz} {et~al.}(2013){Utz}, {Jur{\v c}{\'a}k}, {Hanslmeier}, {Muller},
  {Veronig}, \& {K{\"u}hner}}]{2013A&A...554A..65U}
{Utz}, D., {Jur{\v c}{\'a}k}, J., {Hanslmeier}, A., {et~al.} 2013, \aap, 554,
  A65

\bibitem[{{van Ballegooijen}(1986)}]{1986ApJ...311.1001V}
{van Ballegooijen}, A.~A. 1986, \apj, 311, 1001

\bibitem[{{Vernazza} {et~al.}(1981){Vernazza}, {Avrett}, \&
  {Loeser}}]{1981ApJS...45..635V}
{Vernazza}, J.~E., {Avrett}, E.~H., \& {Loeser}, R. 1981, \apjs, 45, 635

\bibitem[{{Vigeesh} {et~al.}(2009){Vigeesh}, {Hasan}, \&
  {Steiner}}]{2009A&A...508..951V}
{Vigeesh}, G., {Hasan}, S.~S., \& {Steiner}, O. 2009, \aap, 508, 951

\bibitem[{{Wittmann}(1974)}]{1974SoPh...36...29W}
{Wittmann}, A. 1974, \solphys, 36, 29

\bibitem[{{Yang} {et~al.}(2016){Yang}, {Li}, {Ji}, {Feng}, {Deng}, {Wang}, \&
  {Lin}}]{2016SoPh..291.1089Y}
{Yang}, Y., {Li}, Q., {Ji}, K., {et~al.} 2016, \solphys, 291, 1089

\bibitem[{{Yun}(1971)}]{1971SoPh...16..398Y}
{Yun}, H.~S. 1971, \solphys, 16, 398

\bibitem[{{Zhang} {et~al.}(1998){Zhang}, {Scharmer}, {Lofdahl}, \&
  {Yi}}]{1998SoPh..183..283Z}
{Zhang}, H., {Scharmer}, G., {Lofdahl}, M., \& {Yi}, Z. 1998, \solphys, 183,
  283

\end{thebibliography}
\end{document}